\documentclass[prb,twocolumn,floatfix]{revtex4}

\usepackage{amsmath}
\usepackage[dvips]{graphicx}
\usepackage{subfigure}
\usepackage{latexsym}
\usepackage{bm}
\usepackage{wasysym}
\usepackage{amssymb}
\usepackage{longtable}

\def\etal{~\textit{et~al.}} 	
\def\ra{\rangle} 			
\def\la{\langle} 			
\def\z2{$\mathbb{Z}_2$}

\def\dd{\textrm{d}}

\newcommand{\ket}[1]{
\begin{tabular}{@{} r @{} c @{} l @{}}
$\Bigl|$ & #1 & $\Bigr\ra$
\end{tabular}}

\newcommand{\bra}[1]{
\begin{tabular}{@{} r @{} c @{} l @{}}
$\Bigl\la$ & #1 & $\Bigr|$
\end{tabular}}

\newcommand{\kagplaq}[1]{
\includegraphics[width=0.5cm,height=0.5cm,viewport=0 320 450 -150]{#1}}
\newcommand{\sixRLau}{\kagplaq{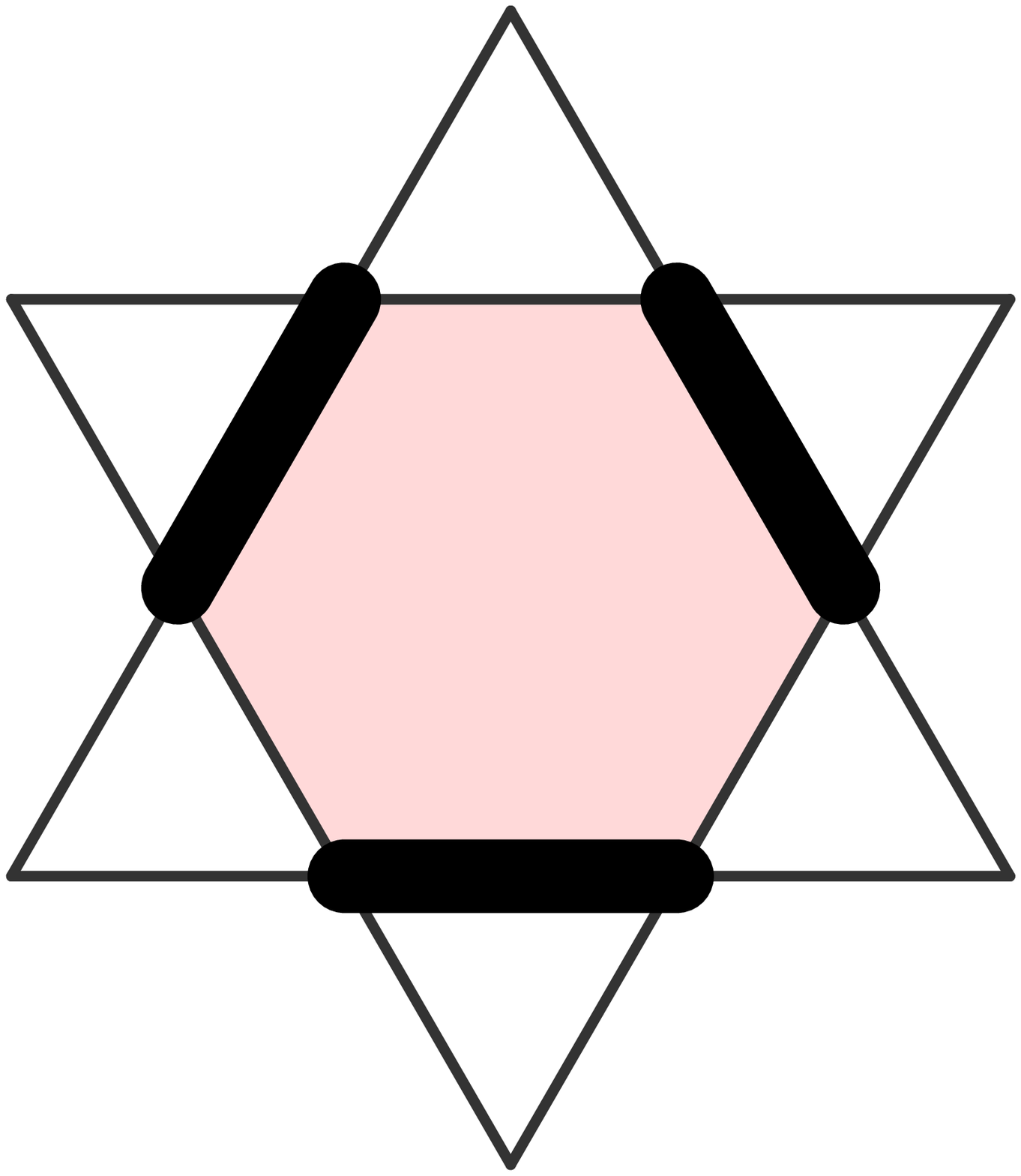}}
\newcommand{\sixRLad}{\kagplaq{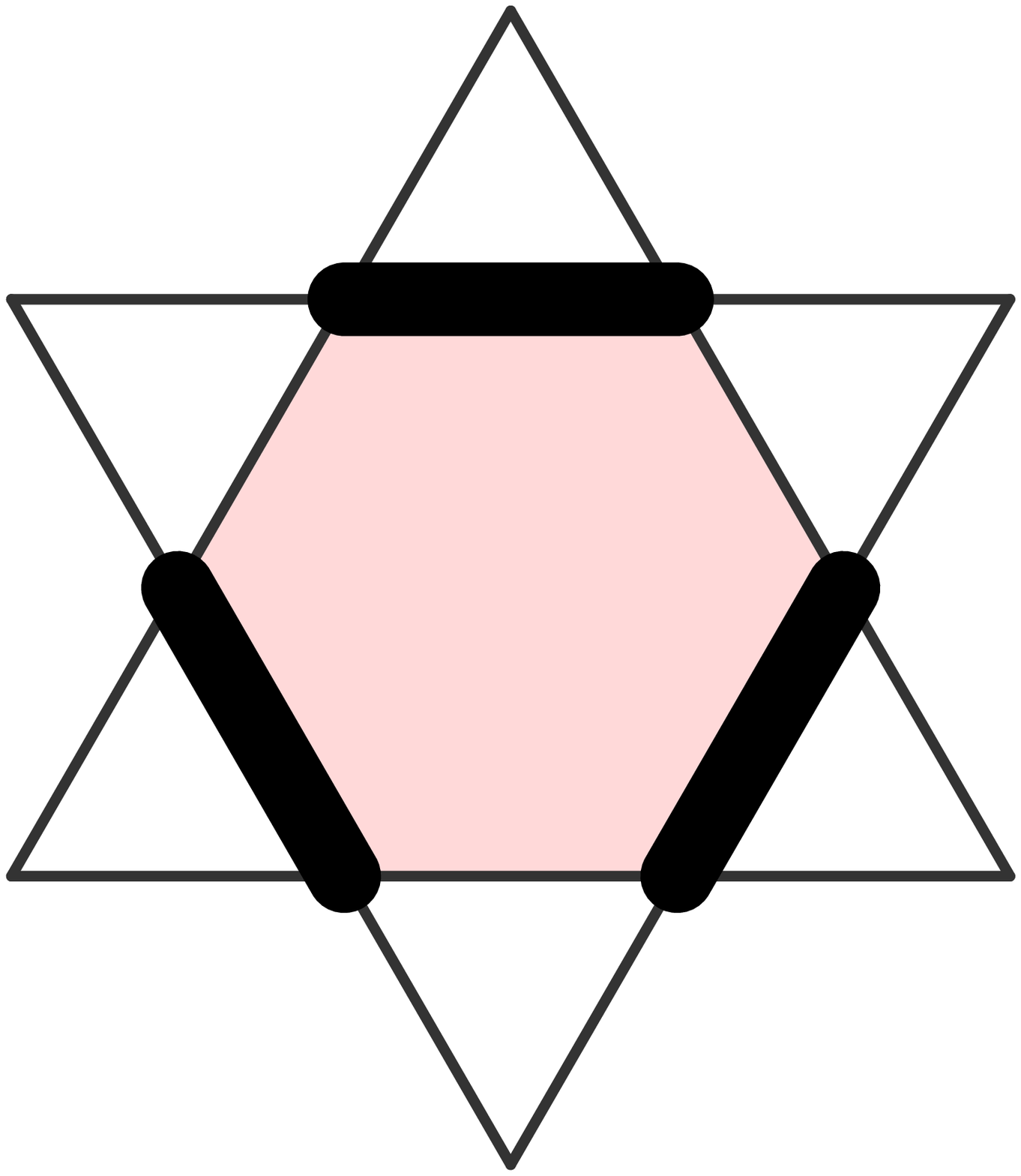}}
\newcommand{\eightRLau}{\kagplaq{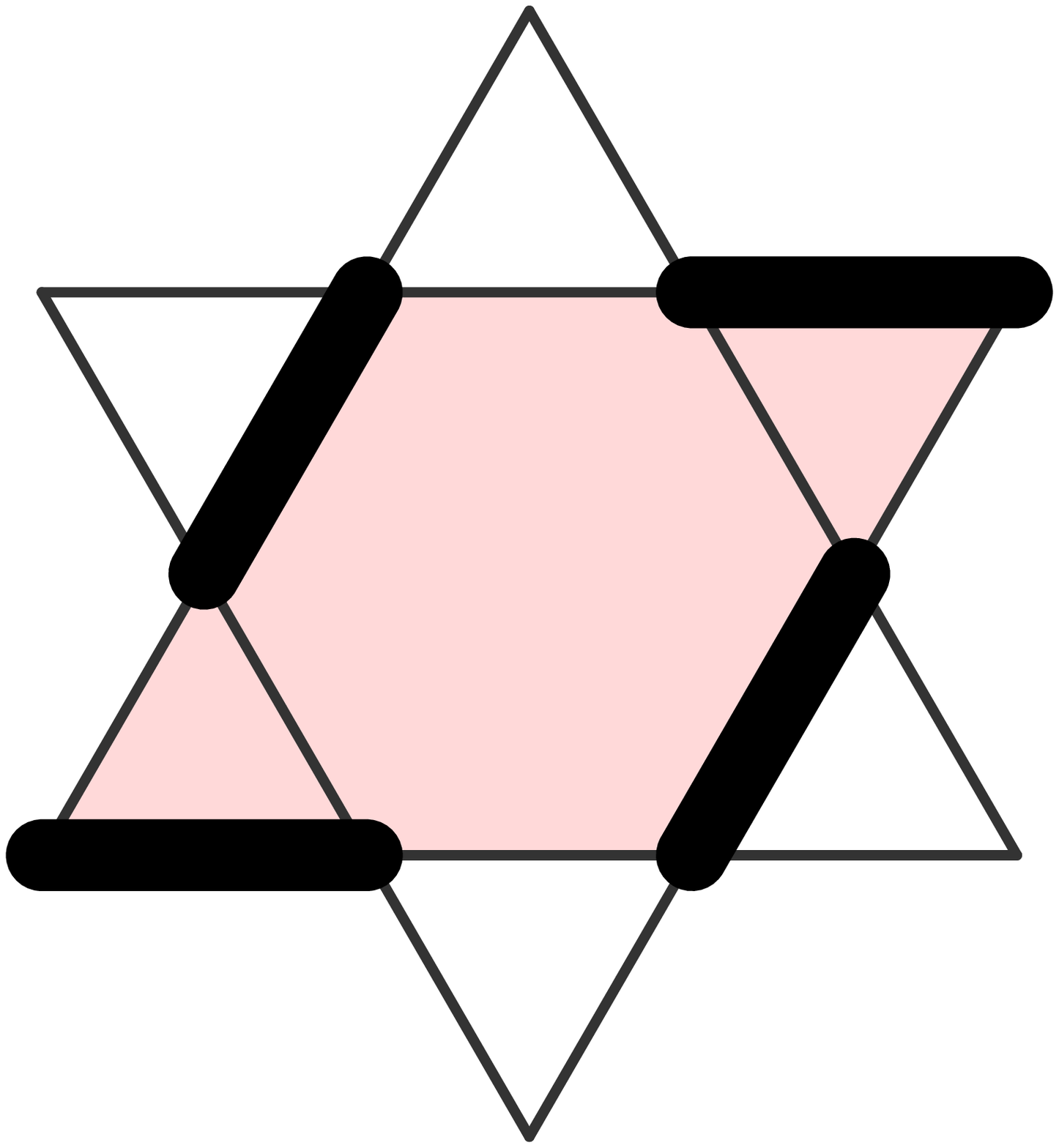}}
\newcommand{\eightRLad}{\kagplaq{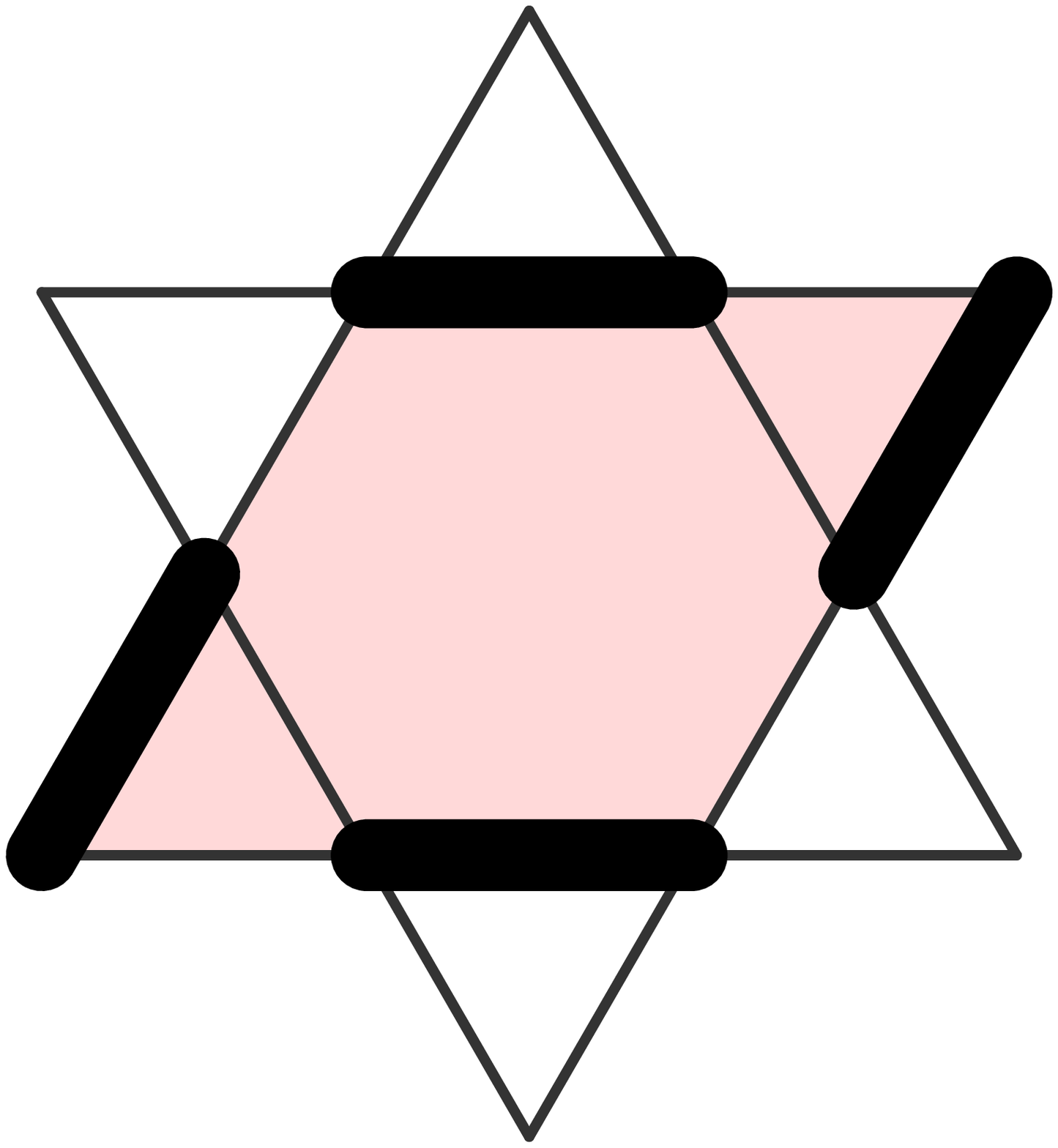}}
\newcommand{\eightRLbu}{\kagplaq{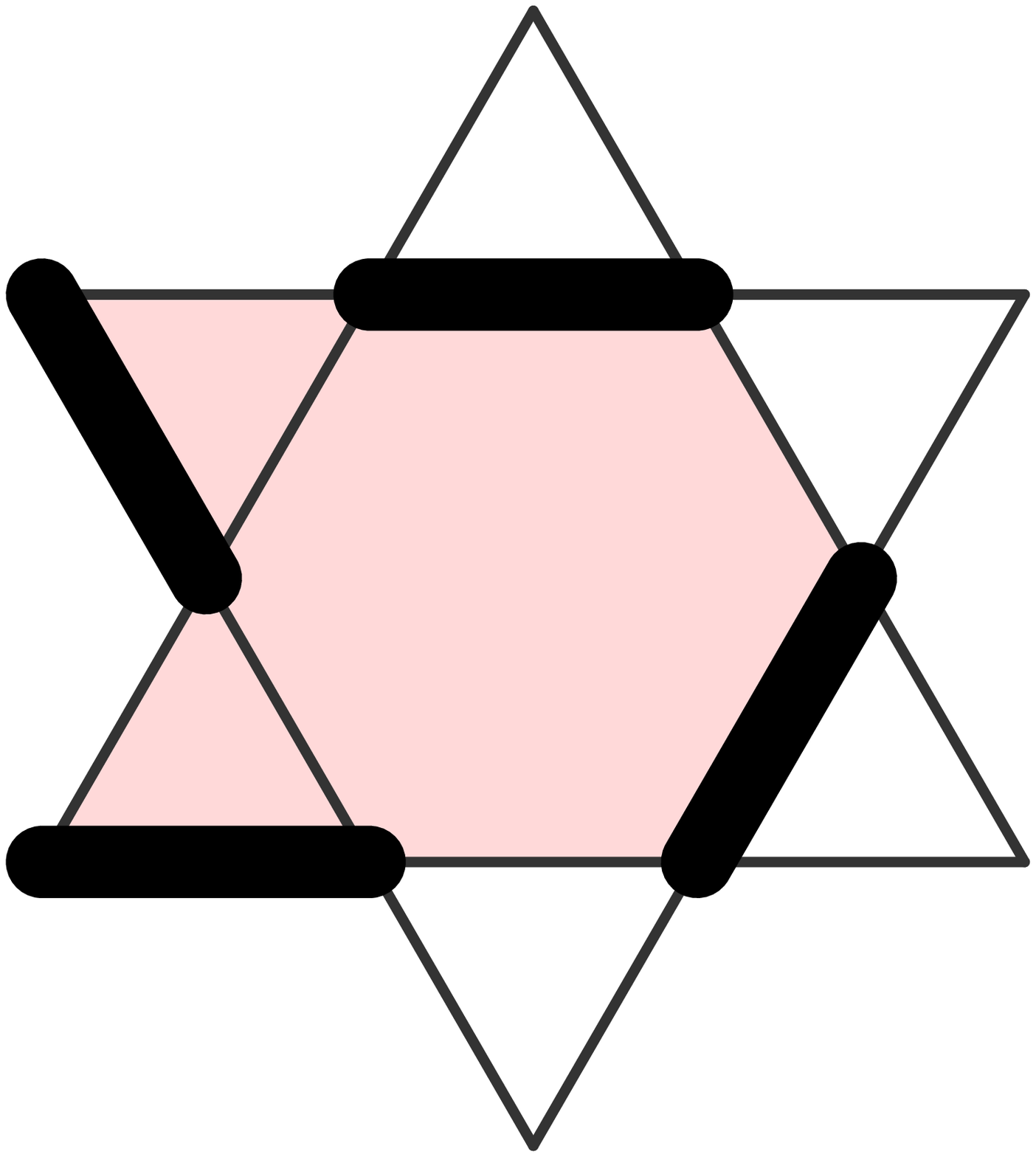}}
\newcommand{\eightRLbd}{\kagplaq{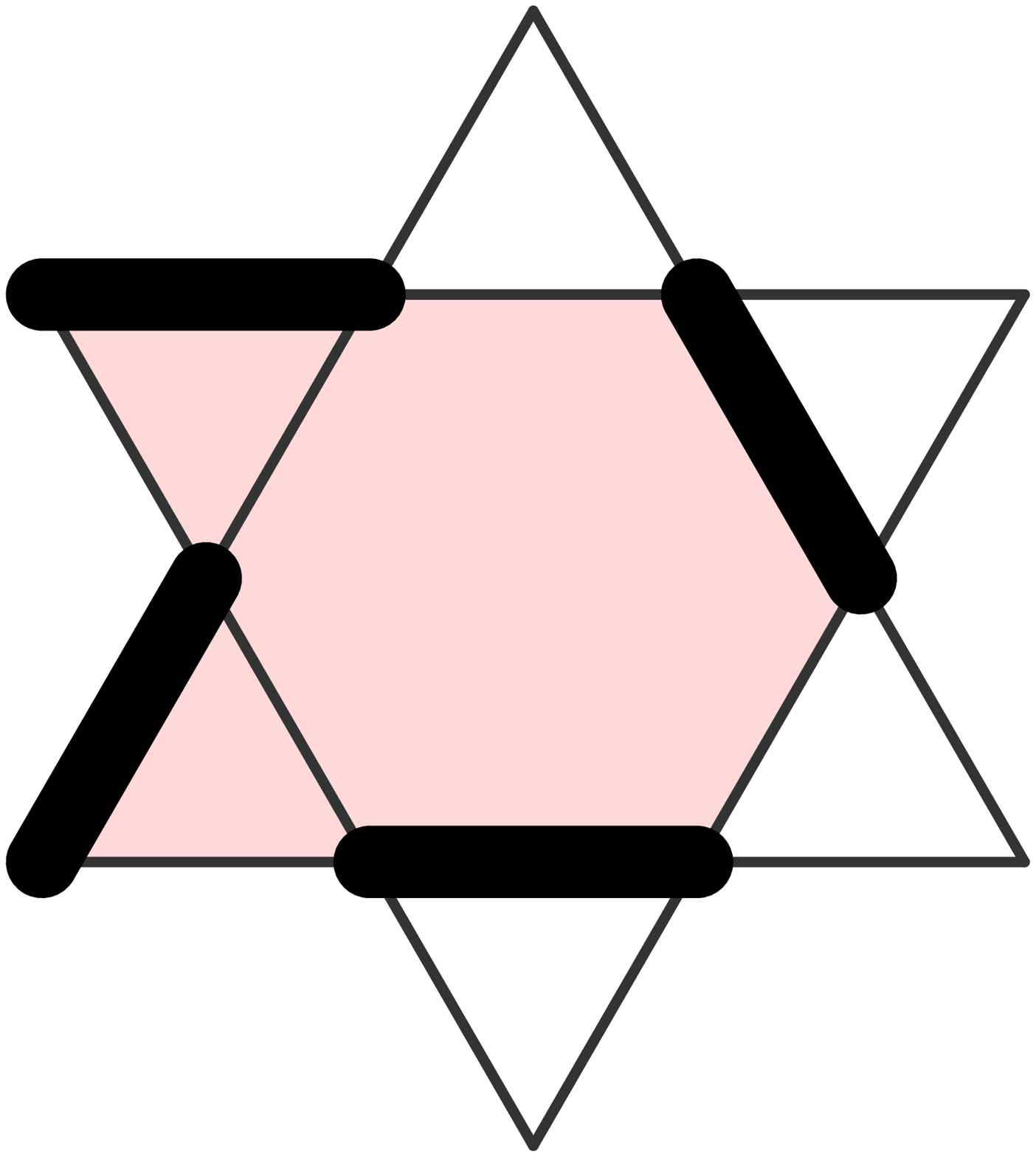}}
\newcommand{\eightRLcu}{\kagplaq{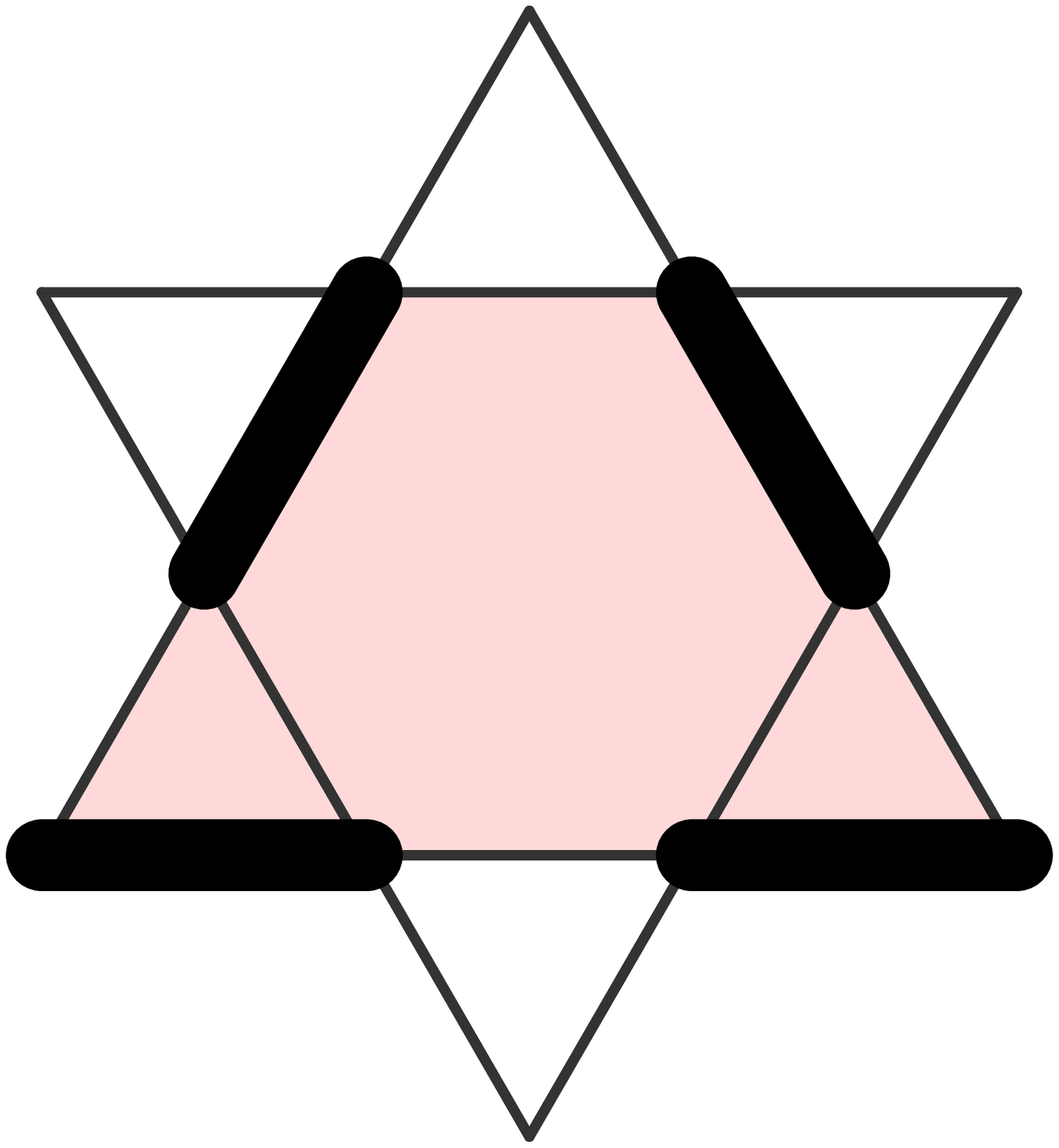}}
\newcommand{\eightRLcd}{\kagplaq{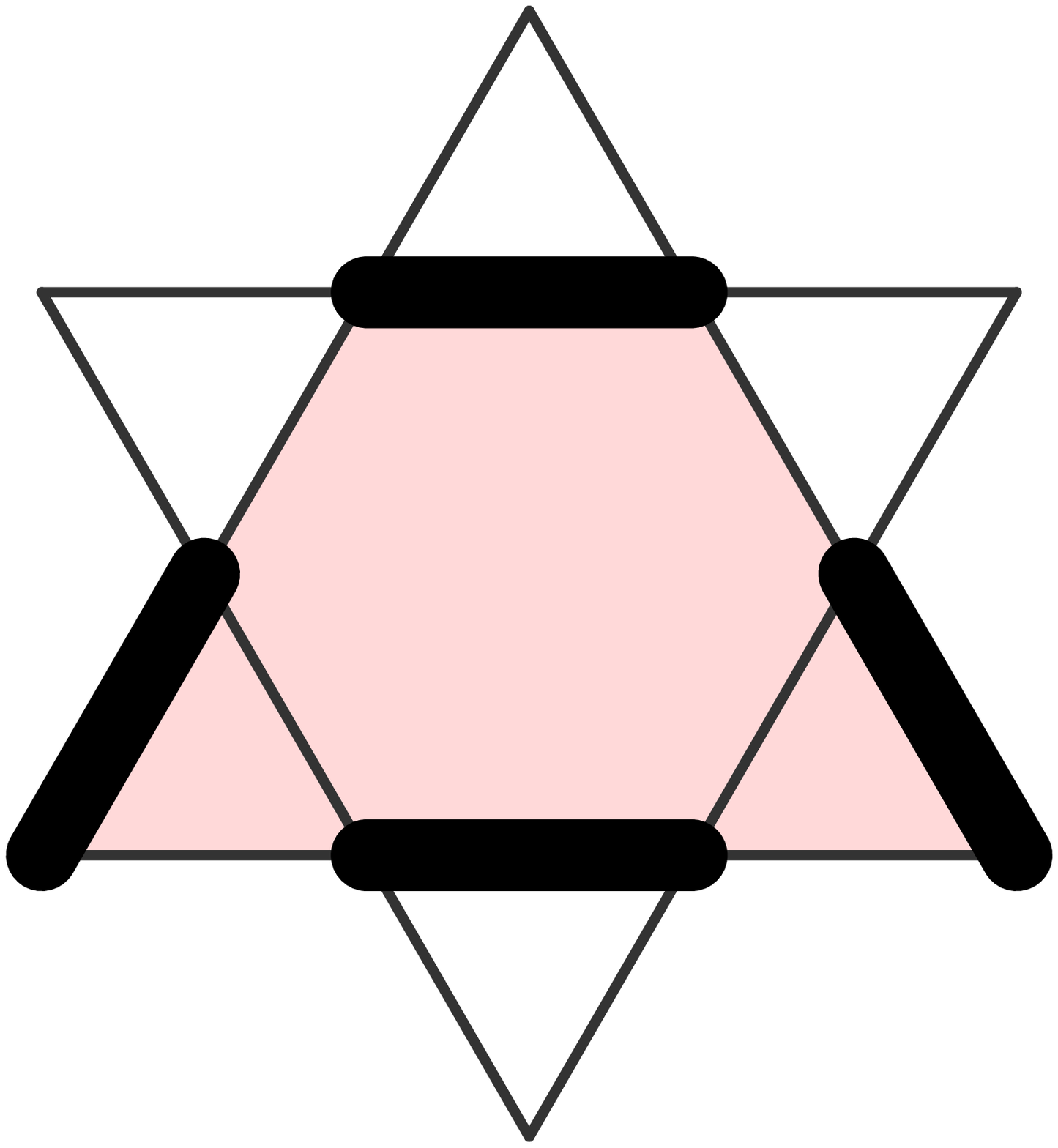}}
\newcommand{\tenRLau}{\kagplaq{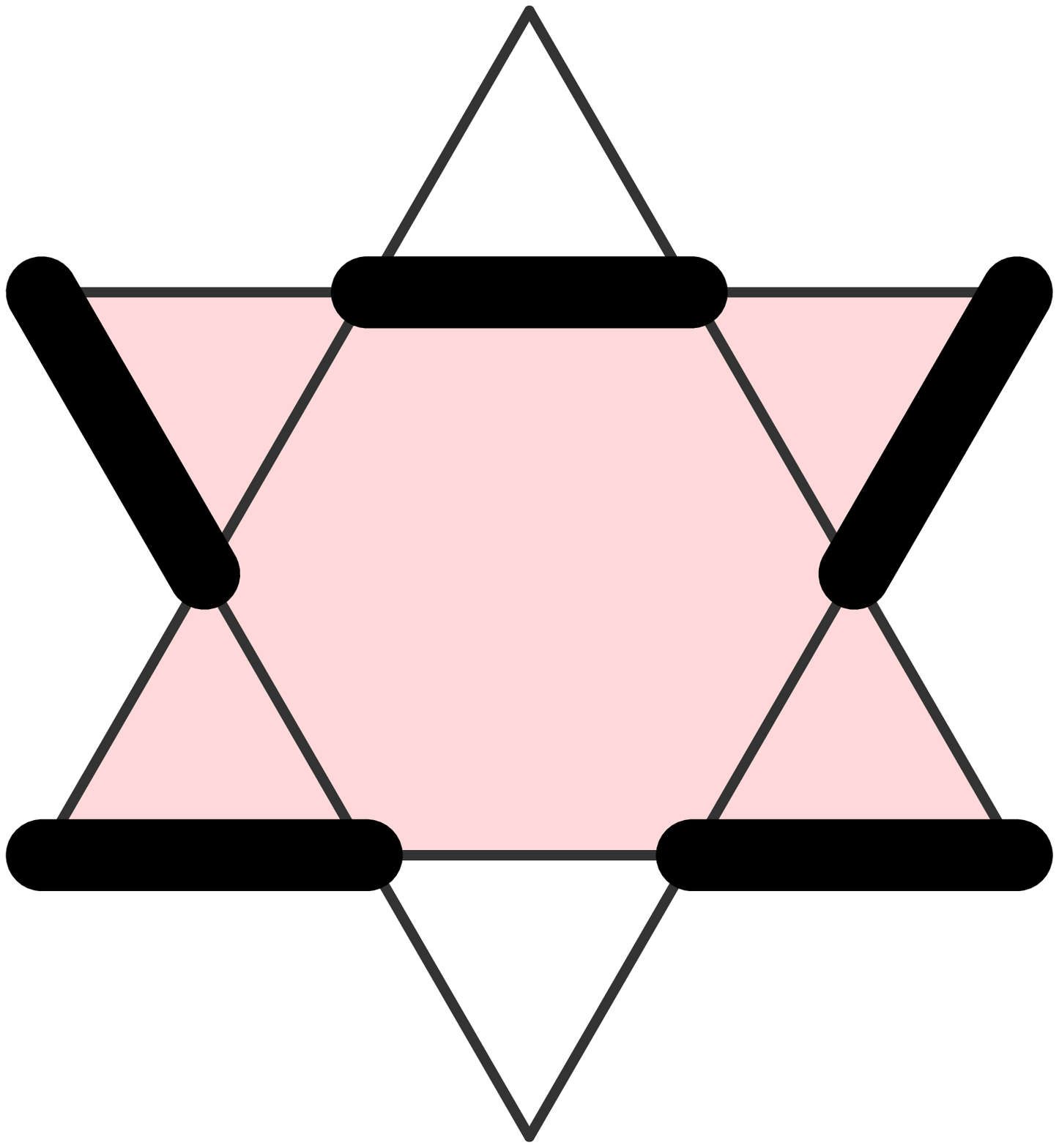}}
\newcommand{\tenRLad}{\kagplaq{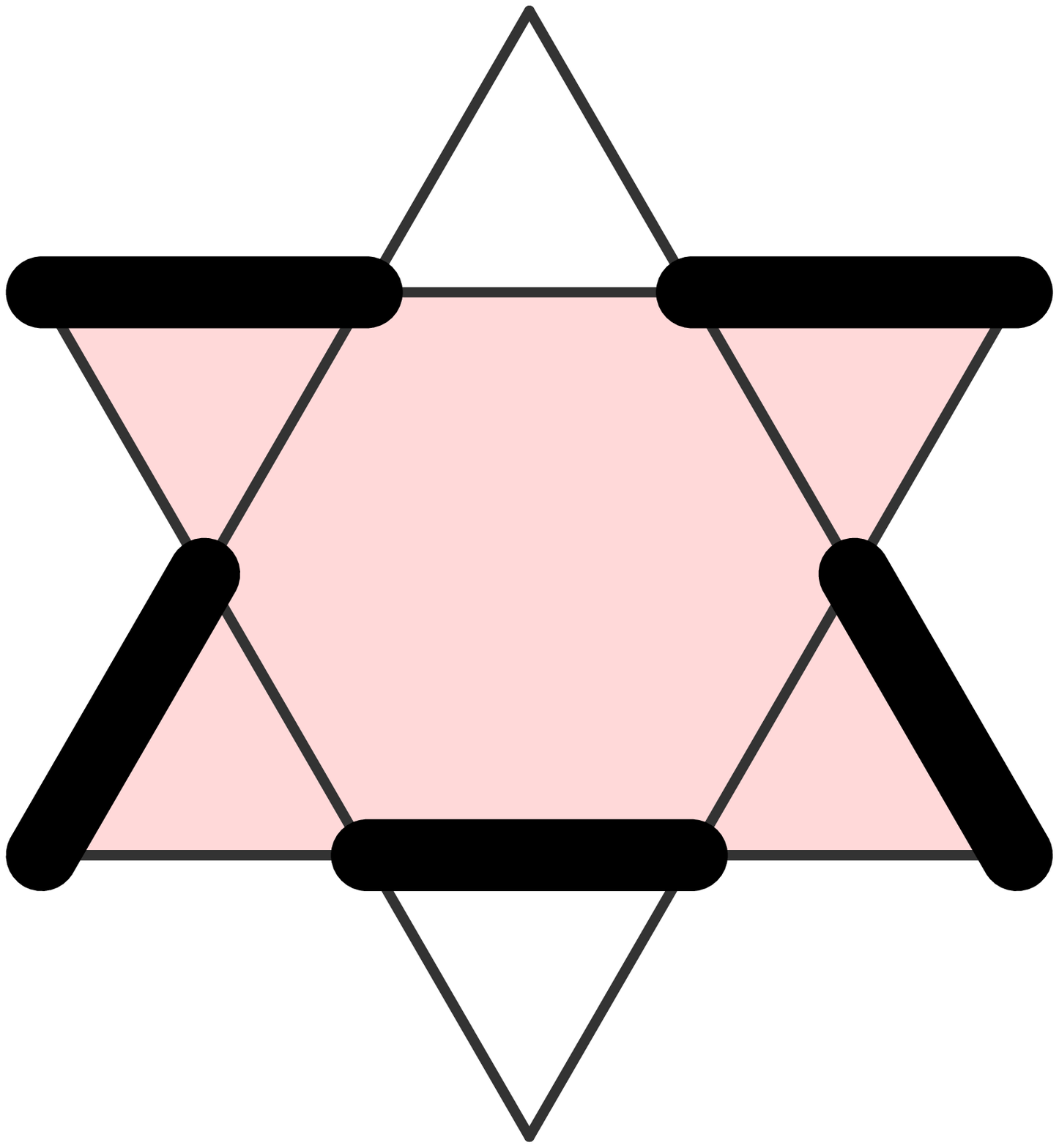}}
\newcommand{\tenRLbu}{\kagplaq{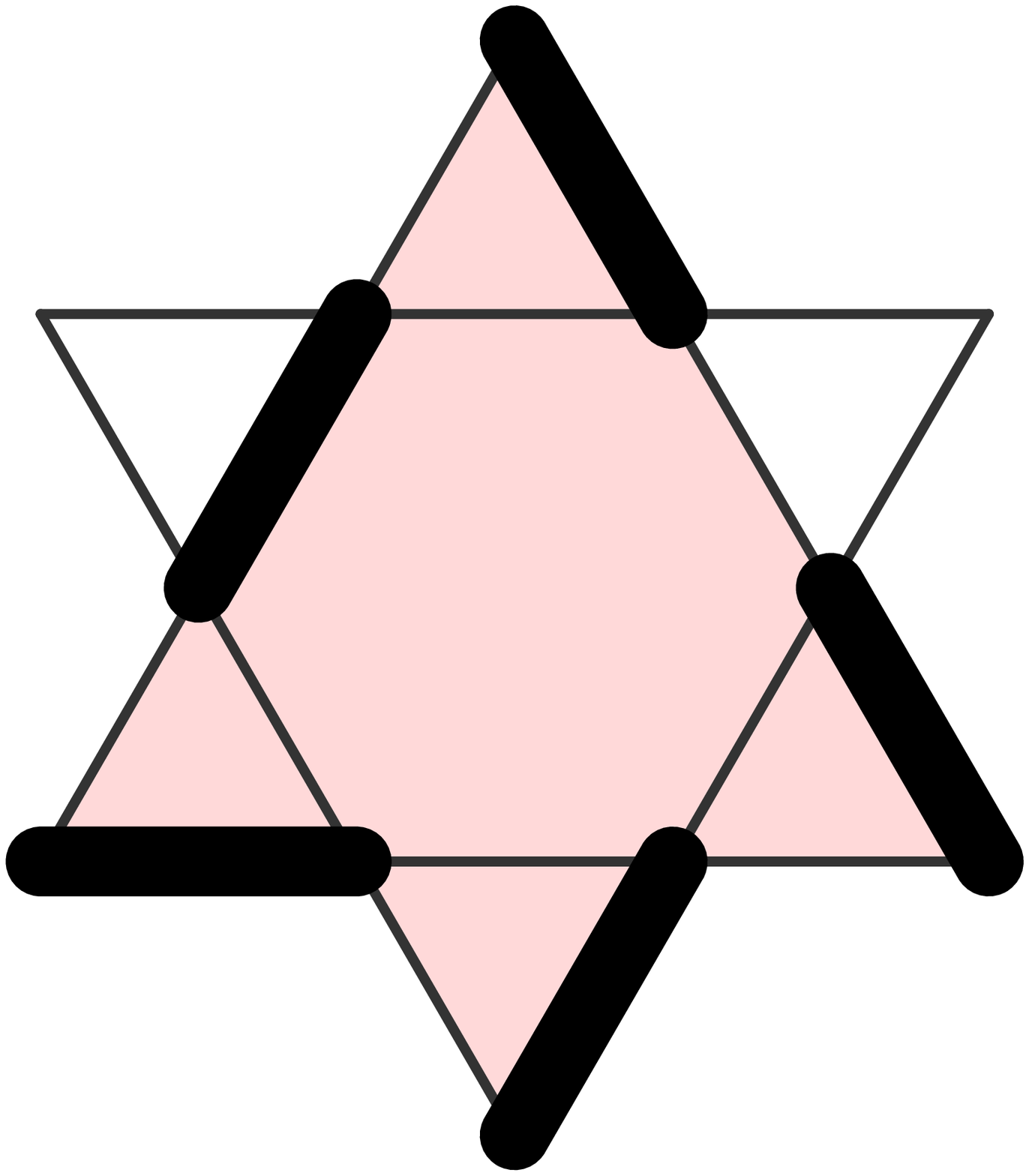}}
\newcommand{\tenRLbd}{\kagplaq{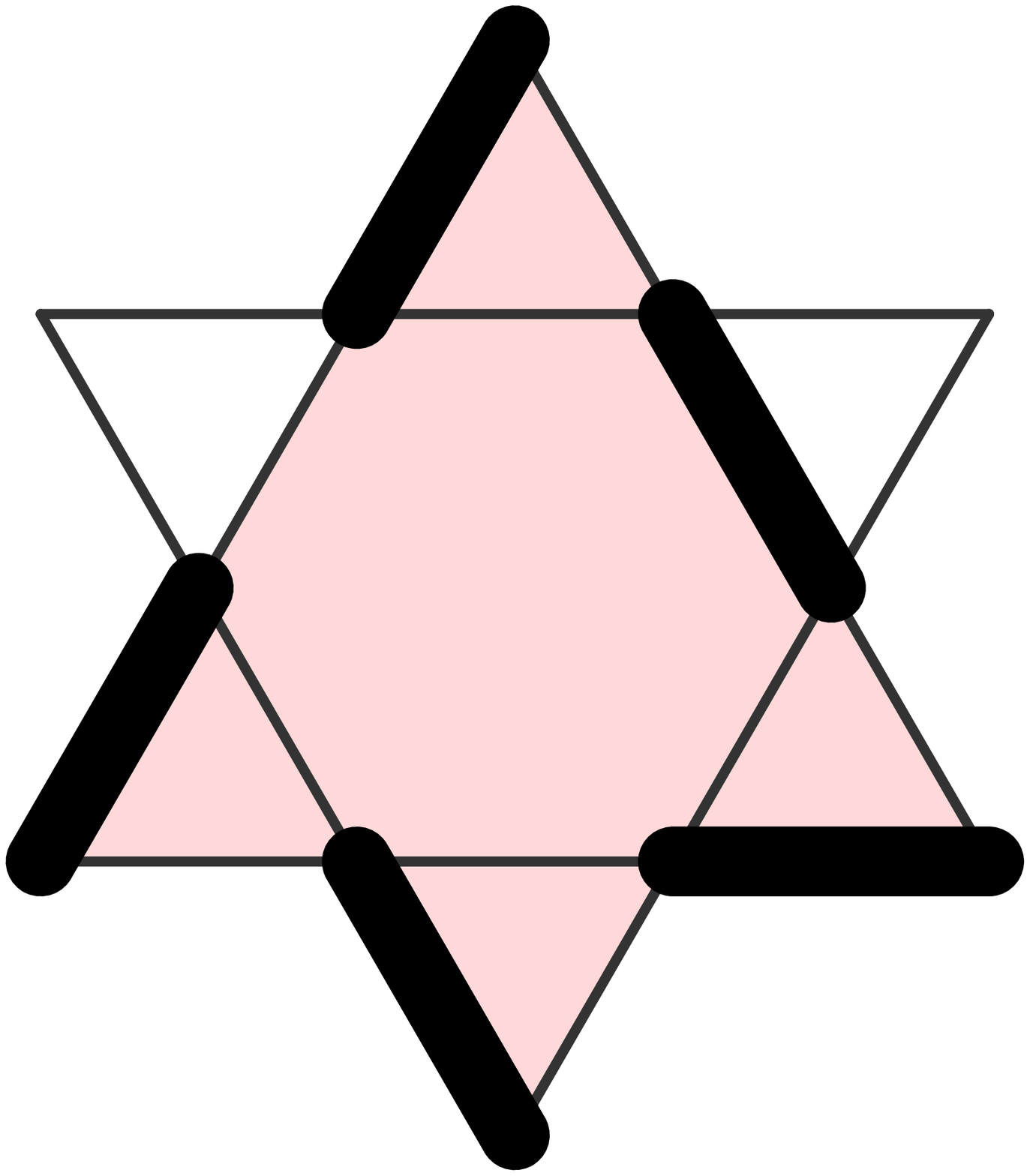}}
\newcommand{\tenRLcu}{\kagplaq{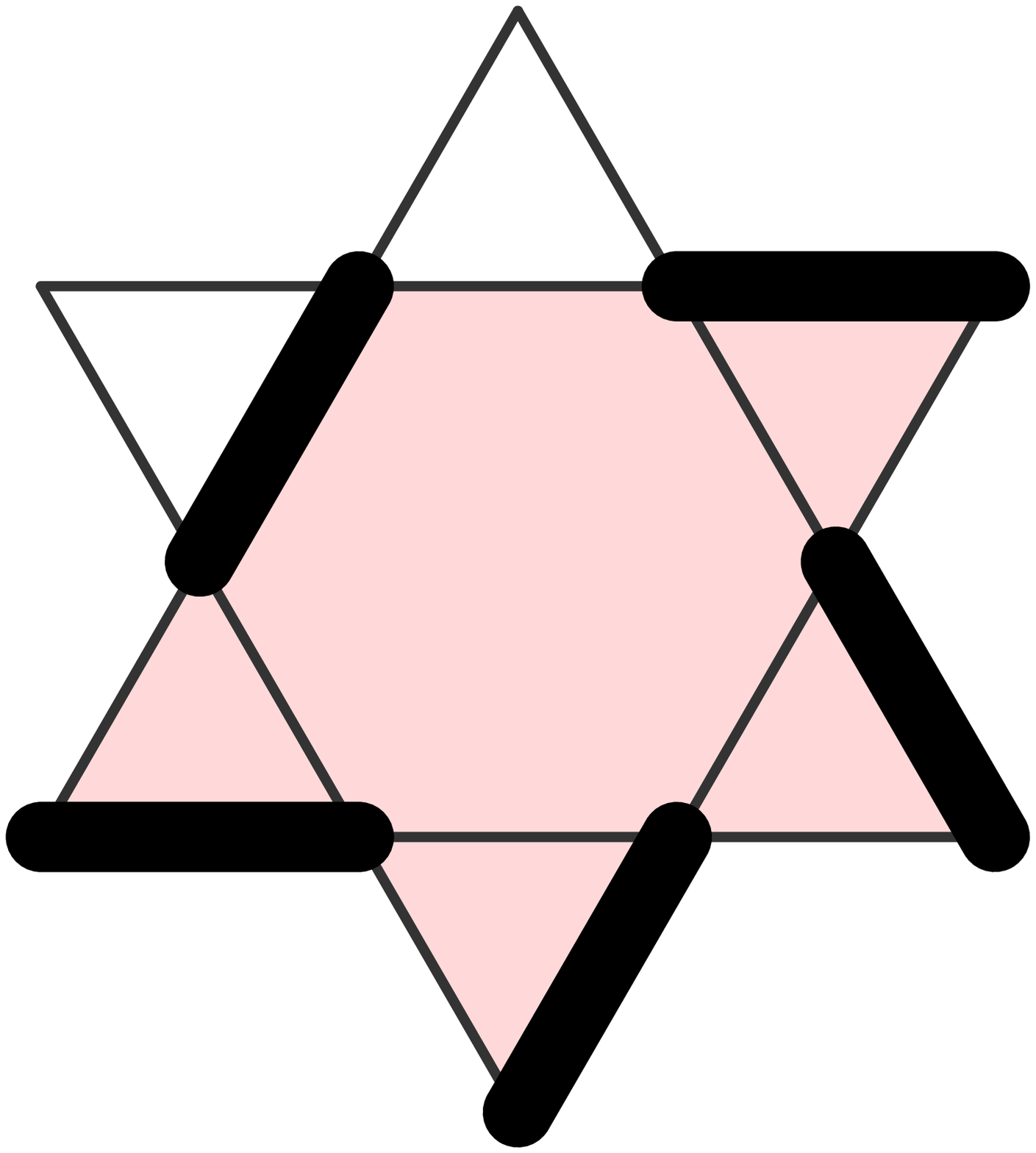}}
\newcommand{\tenRLcd}{\kagplaq{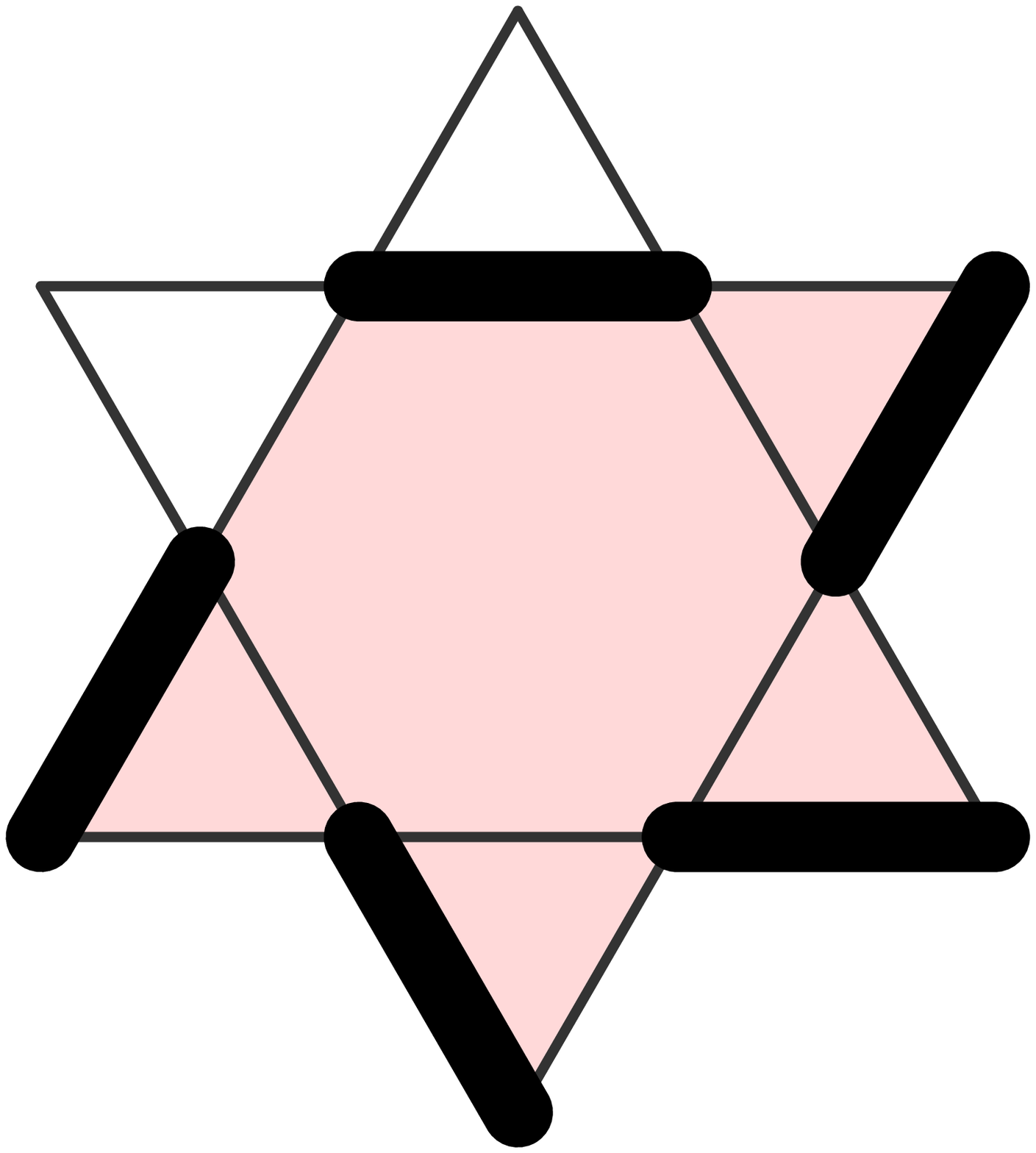}}
\newcommand{\twelveRLau}{\kagplaq{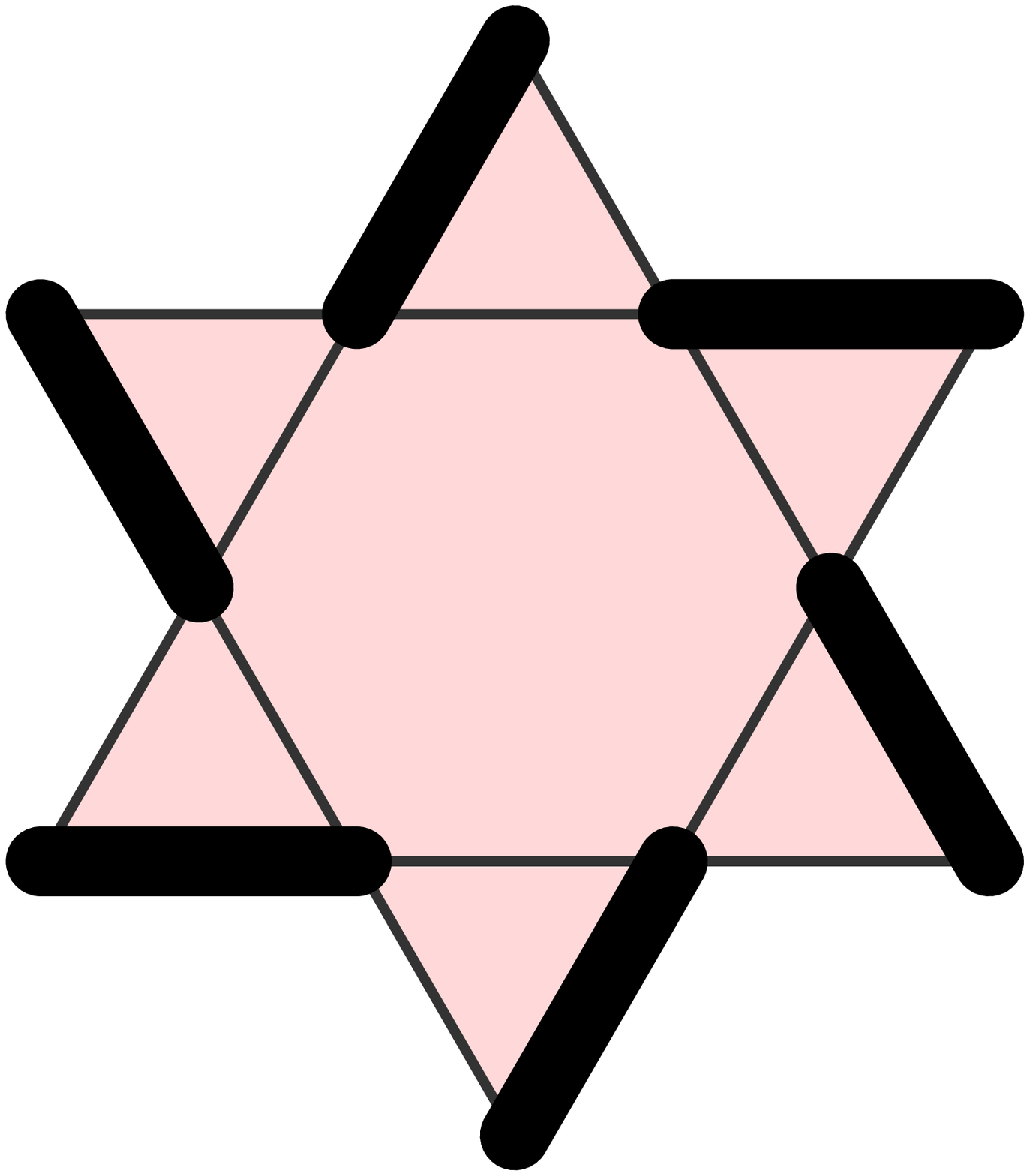}}
\newcommand{\twelveRLad}{\kagplaq{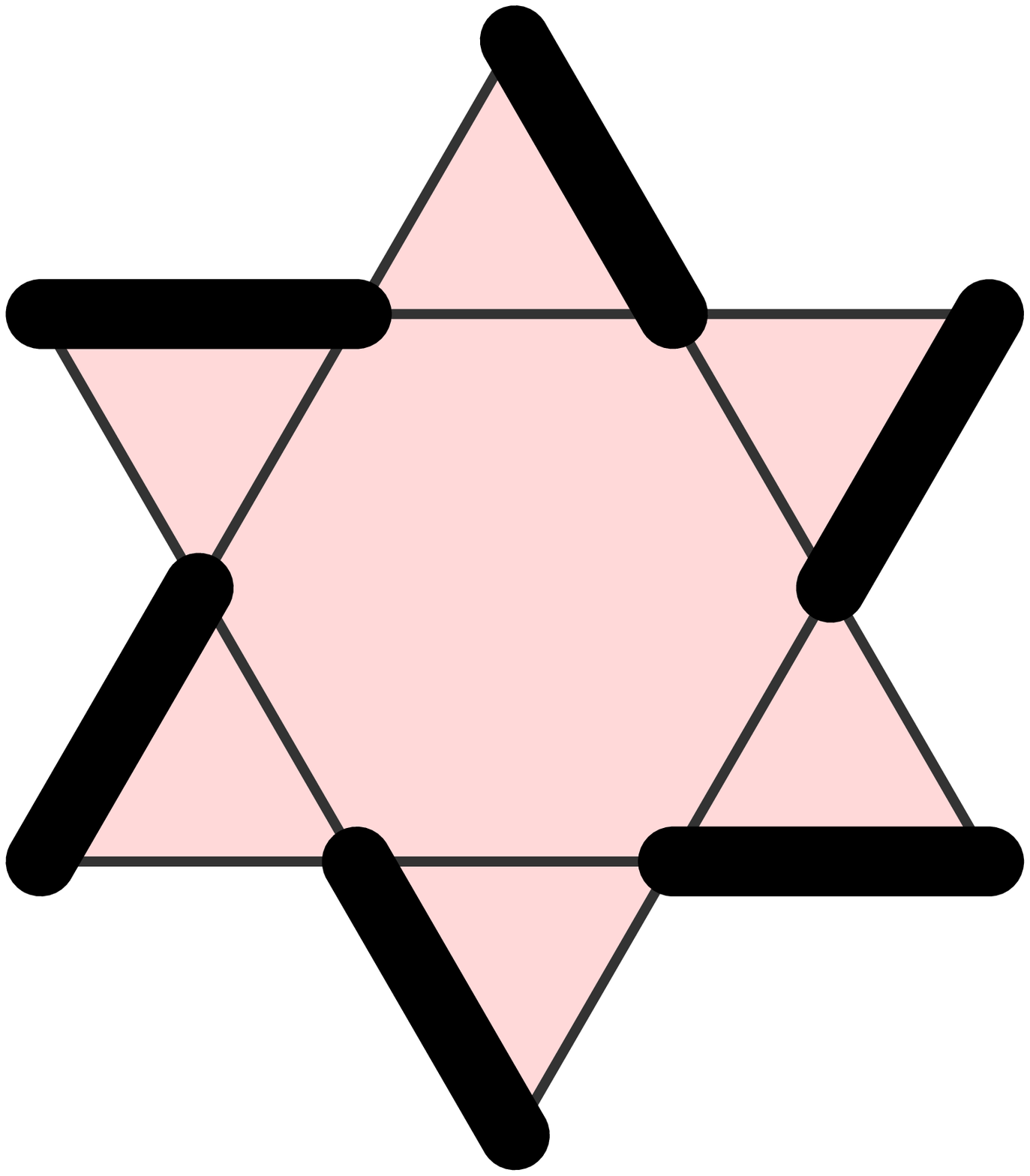}}

\begin{document}

\title{Physics of low energy singlet states of the Kagome lattice quantum Heisenberg antiferromagnet}
\author{P. Nikolic}
\author{T. Senthil}
\affiliation{Massachusetts Institute of Technology, 77 Massachusetts Ave.,  Cambridge, MA 02139}

\date{May 08, 2003}

\begin{abstract}
This paper is concerned with physics of the low energy singlet excitations found to exist below the spin gap in numerical studies of the Kagome lattice quantum Heisenberg antiferromagnet. Insight into the nature of these excitations is obtained by exploiting an approximate mapping to a fully frustrated transverse field Ising model on the dual dice lattice. This Ising model is shown to possess at least two phases - an ordered phase that also breaks translational symmetry with a large unit cell, and a paramagnetic phase. The former is argued to be a likely candidate for the ground state of the original Kagome magnet which thereby exhibits a specific pattern of dimer ordering with a large unit cell. Comparisons with available numerical results are made.

\end{abstract}

\maketitle

\section{Introduction}\label{intro}

Geometrically frustrated quantum magnets (GFQM) are a class of magnetic materials with various unusual properties at low temperatures. The combination of strong frustration and quantum effects makes them  potentially a source of new strongly correlated physics. One common possible consequence of frustration is the macroscopically large degeneracy of the classical ground-states, which makes them very sensitive to even weak thermal and quantum fluctuations. For example, the phenomenon of  order-from-disorder may occur: the fluctuations may lift the initial degeneracy and yield a symmetry broken ground-state. Alternatively, the initial degeneracy may be completely lifted, in which case the result is a spin-liquid state. In the traditional point of view, the liquid state is a result of fluctuations strong enough to destroy the long-range order of the classical ground-state. However, in several GFQM, the starting classical ground-state is typically already disordered, so that the liquid state, eventually shaped by the fluctuations, may be of a different kind. In general, one can hope to find various exotic phenomena in these systems.

The most studied GFQM are the Kagome and pyrochlore lattices of antiferromagnetically coupled nearest neighbor spins. They belong to a class of lattices which are composed of corner-sharing units, where every unit is a small frustrated spin system, like a triangle. Other examples include the triangular lattice frustrated by the higher order exchange processes, and the checkerboard lattice.

\begin{figure}
\subfigure[{}]{\includegraphics[width=2.1in,angle=90]{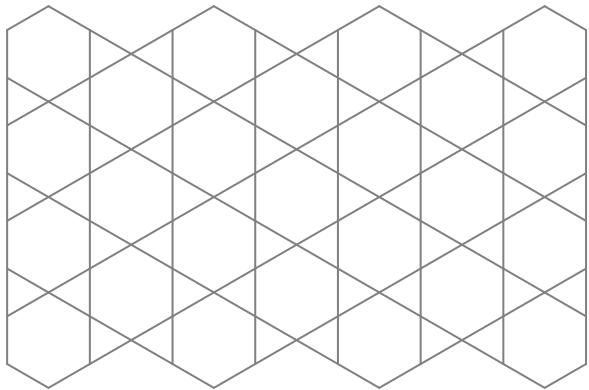}}
\subfigure[{}]{\includegraphics[width=2.1in,angle=90]{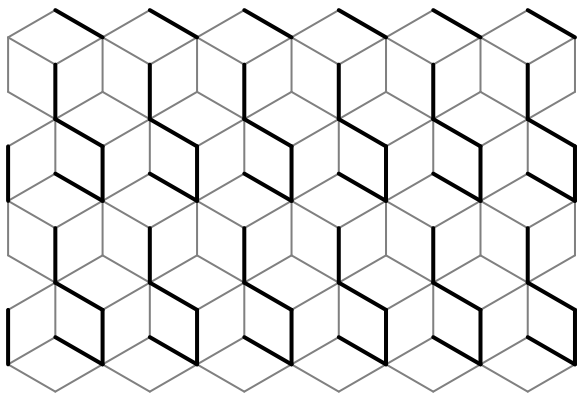}}
\vskip -2mm
\caption{\label{KDGrid}(a) Kagome lattice; (b) dice lattice (dual to Kagome). The effective theory for the low-energy singlet degrees of freedom will be described by a fully frustrated transverse-field quantum Ising model on the dice lattice. Frustration can be achieved in this model by giving opposite signs to the Ising spin interactions on the normal and thick bonds}
\end{figure}

This paper is concerned with the Kagome lattice spin-$1/2$ Heisenberg antiferromagnet (See Fig.~\ref{KDGrid}). Several groups have performed numerical calculations of spectra on finite clusters with up to 36 sites \cite{Elser,KNum}. The exact diagonalization \cite{KNum,Lhlong,Lhshort} showed that the ground-state of the 36-site Kagome lattice has short-ranged spin-spin and some other correlations. The spin-gap is finite, and the extrapolation to the thermodynamic limit suggests that it remains finite in the infinite lattice. It is, however, small, estimated around $J/20$, where $J$ is the exchange coupling. Remarkably however, at energies below the spin gap there appear to be a large number of {\em singlet} excitations. The origin and nature of these states is unclear (see Ref.\cite{mila,SyroMale} for some interesting approaches). The number of these states is found to be an exponential function of the sample size: $1.15^N$ below the spin-gap, for even number of sites $N$, and there is no clear gap that separates them from the ground-state. There are some indications that these low-energy singlet excitations may also occur in other frustrated quantum paramagnets. Indeed, evidence for such excitations in the multiple-ring exchange triangular lattice spin-$1/2$ quantum antiferromagnet has been reported in Ref. \cite{TNum}. This led to a suggestion that there was a new spin liquid phase characterized by having gapless singlet excitations, which was named the Type II spin liquid in Ref.~\cite{Lhlong,Lhshort}.

There also are interesting experimental results on $SrCr_{9p}Ga_{12-9p}O_{19}$ (or SCGO, a bilayer spin-$3/2$ Kagome magnet). Absence of long-range order at low temperatures \cite{SCGOmuon} accompanied however by a large entropy of low-lying excitations \cite{SCGOHeatCap} has been reported. Furthermore, the heat capacity is virtually independent of magnetic field, and is not thermally activated \cite{SCGOHeatCap}. These results suggest that the ground-state of SCGO is not magnetically ordered, and that the low-energy sector of excited states contains a (possibly gapless) band of singlet excitations, while spinful excitations are separated from the ground-state by a finite spin-gap. Despite the obvious differences between SCGO and the theoretical model of a single spin-$1/2$ Kagome layer, the similarities to the picture obtained from exact diagonalization of the latter are striking. (Other unusual phenomena such as spin-glass like behavior in clean samples \cite{SG,SCGOmuon} and its speculated coexistence with a spin-liquid component \cite{SCGOHeatCap} have also been suggested to happen in SCGO).  

These unusual properties of the Kagome Heisenberg magnet have motivated various theoretical efforts, but our understanding of the underlying physics is still very incomplete. Several controlled limits enable making some definite statements. One of them is the large-$N$ $SU(N)$ limit explored in \cite{SUN} which gives highly degenerate mean-field solutions, and a spin-Peierls order as the $1/N$ correction. The ordered state maximizes the number of fluctuating benzene-like hexagonal clusters, where three singlet bonds sitting on a hexagonal plaquette oscillate between two possible configurations. However, it remained unknown whether this ordered state would survive  higher-order corrections. Another explored limit is the large-$N$ $Sp(N)$ \cite{SpN}. This generalization of the spin-1/2 problem retains the ``spin-magnitude'' as an independent parameter. For large spins the ground-state possesses long-range magnetic order (in a pattern dubbed the $\sqrt{3}\times\sqrt{3}$ structure), and the elementary excitations are the spin waves associated with the broken spin rotational symmetry. For small spins, however, the ground-state does not break any symmetries, and the elementary spin carrying excitations are unconfined spin-1/2 bosonic spinons.

An alternative, less controlled, but perhaps more physical approach is to assume that the physics of the paramagnetic ground states of the Kagome magnet is described well by the short-ranged valence bonds, which can then be roughly approximated by the quantum dimers on the Kagome lattice. A quantum dimer model resulting from the overlap expansion of the valence-bond states was studied numerically by Zeng and Elser \cite{Elser}. Misguich and colaborators have considered two simple dimer models: one at the Rokhsar-Kivelson point \cite{KagRK}, and another offering more similarity to the overlap expansion (different resonant loops appear with different signs) \cite{KagSpecDimer}. The former was exactly solved by mapping to the triangular lattice Ising model, and both gave the dimer-liquid ground-states with large zero-temperature entropy. In a broader framework of frustrated quantum magnetism, Moessner, Sondhi, and various collaborators have studied various related problems on different lattices, and discussed connections between the dimer model and various gauge theories ~\cite{Z2MS}.

In this paper, we undertake a  set of calculations that could potentially provide considerable insight into the low energy singlet excitations of the Kagome Heisenberg antiferromagnet. It has been realized for some time \cite{Z2MS,SaJa,Z2a} that for a number of quantum antiferromagnets, the nature of the paramagnetic phases is closely related to (and essentially determined by) the properties of a much simpler model: the fully frustrated Ising model in a transverse field on a lattice that is dual to that of the original spin system. This connection (which we briefly review below) may be established in a number of different approaches. One is through a slave particle theory of such quantum paramagnets. The theory of fluctuations about the mean field state in such slave particle theories is a gauge theory which for frustrated magnets \cite{SaJa} is a \z2 gauge theory. This is then related by duality to the transverse field Ising model mentioned above. Another connection is through the quantum dimer description of the paramagnetic phases of the original spin system. As explained in Ref. \cite{Z2MS}, in appropriate limits, the quantum dimer model may be related to the fully frustrated transverse field Ising model.

The connection to the transverse field Ising model is particularly useful for issues related to the low energy singlet excitations of frustrated quantum paramagnets. First, by construction, the Ising model describes only the {\em singlet} sector of the original spin system, and is best to describe states below the spin gap. These are precisely the states one wishes to understand. Second, one can exploit the relative simplicity of the Ising model to perform calculations that would be extremely difficult for the original spin system. For instance, the Ising model can be simulated by Monte Carlo without a sign problem, unlike the original frustrated spin model. This has been nicely demonstrated in the extensive work of Moessner, Sondhi, and coworkers \cite{FrIsing2,FrIsing} on such Ising models.

It is therefore of extreme interest to study the particular Ising model that one may guess describes the singlet sector of the Kagome Heisenberg antiferromagnet. This is a fully frustrated transverse-field Ising magnet on the dual lattice (known as the dice lattice; see Fig.~\ref{KDGrid}b). This is an interesting model in itself, and appears not to have been examined before. As described in Appendix \ref{LGAp}, a Landau-Ginzburg analysis (which has been useful\cite{Berker} in analyzing analogous frustrated quantum Ising magnets on other lattices) shows an infinitely degenerate set of zero energy modes in the Gaussian approximation. This renders the Landau-Ginzburg analysis less useful for this Ising model. Thus studying the fully frustrated quantum Ising model may also contain lessons of its own for the theory of frustrated quantum systems.

The questions of interest are the following: Does this Ising model have a zero temperature ordered phase? We note that the combination of frustration and quantum effects could potentially completely destroy the ordered phase even at zero temperature (as happens for instance for the transverse field Ising antiferromagnet on the Kagome lattice \cite{FrIsing2,FrIsing}). Generally, the ordered states of the Ising model describe, in the original Heisenberg spin problem, ``confined'' paramagnets where the spin carrying excitations are (gapped) spin-$1$ magnons. The disordered states of the Ising model describe fractionalized paramagnets where there are (gapped) spin-$1/2$ spinons. Whatever the phase diagram of the dice Ising model, the nature of the excitations is an extremely appropriate question. Indeed, if this Ising model has a large number of low energy excitations, they would possibly correspond to the singlet excitations of the original Kagome lattice Heisenberg antiferromagnet.

\begin{figure}
\includegraphics[width=2.1in,angle=90]{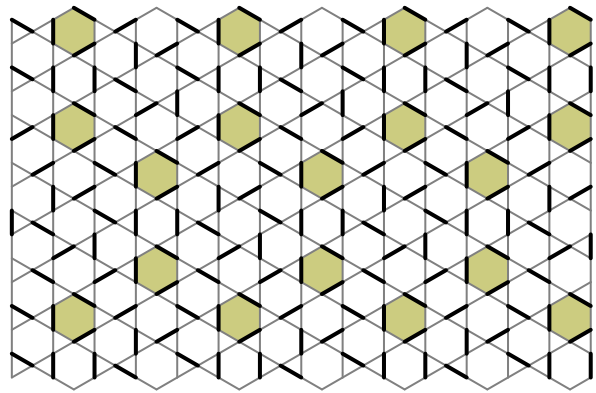}
\caption{\label{DimerGND}Description of the suggested ground-state of the Kagome quantum antiferromagnet. Dimers formally represent the frustrated bonds of the dual dice-lattice Ising model, but physically they represent singlet valence bonds. The translational symmetry is broken by a unit-cell with 36 sites. Dimers on the shaded hexagonal plaquettes resonate between the two possible configurations on the hexagon. Also, the six dimers around the central plaquette of every honeycomb cell resonate in a \emph{star} shaped configuration. The lowest lying excitations are various excited states of the \emph{stars}.}
\end{figure}

In this paper, we first argue that the fully frustrated transverse-field Ising model on the dice lattice supports (at least) two distinct phases. As usual, at large transverse fields (Section \ref{smallH}) there is a gapped Ising paramagnet. However, its excitations are peculiar since they are either localized or extremely heavy. At small transverse fields (Section \ref{largeH}) there is also an ordered phase which breaks translational symmetry (albeit in a rather complicated pattern with a large unit cell). The ordering pattern is best described in the language of the quantum dimer model on the Kagome lattice to which the model is equivalent at small transverse fields: the dimers crystallize in a honeycomb structure of benzene-like resonating hexagons. The elementary excitations have a tendency to be localized (large effective mass), and they have a non-zero finite gap which nevertheless is very small compared to the natural energy scales of the model. We suggest that this phase is what is actually realized in the original Kagome Heisenberg magnet (see Fig.~\ref{DimerGND}), and not a gapless spin-liquid. We argue that properties of this phase are consistent with results from the numerical studies, which in turn may suffer from serious finite-size effects. This then provides a qualitative picture of the singlet physics below the spin-gap observed by the exact diagonalization. We also performed a simple Monte-Carlo simulation (Section \ref{MC}) to show that the thermal fluctuations do not introduce order from disorder, and do not select the honeycomb structure out of many classically degenerate dimer coverings. In the appendices we provide a discussion of various peripheral matters and details of some calculations.

\bigskip

\section{Models}\label{models}

We will try to understand the low-energy physics of the Heisenberg model on the Kagome lattice:
\begin{equation}\label{HK}
H = J \sum_{\la ij \ra} \boldsymbol{S}_i \cdot \boldsymbol{S}_j
\end{equation}
where the sum runs over the nearest-neighbor sites. From the numerical exact diagonalization calculations we know that the ground state has no magnetic long range order. There is a non-zero gap for spin-carrying excitations. We will therefore focus on possible non-magnetically ordered phases of the model. A number of different arguments may be provided to relate the properties of such phases to those of the fully frustrated transverse field Ising model on the dice lattice.

The quantum dimer model is a popular way to capture some essential physics of quantum paramagnetic ground states of spin systems. Moessner and Sondhi ~\cite{Z2MS} have established a connection between the standard hard-core dimer model, and the fully frustrated Ising model in a (small) transverse field on the dual lattice. The connection is made by noting that in zero transverse field, the ground states of the classical fully frustrated Ising model are (up to a global spin flip) in one-to-one correspondence with hard-core dimer coverings of the dual lattice.  A small transverse field essentially introduces quantum resonances between dimer configurations, leading to the quantum dimer model. 

A closely related earlier approach is through a slave particle mean field theory of the quantum paramagnetic phases of the Heisenberg model. As argued in Ref. \cite{ReSaSpN}, for frustrated lattices the theory of fluctuations about the mean field theory is a \z2 gauge theory, where the gauge fields live on the links of the original (Kagome) lattice. For spin-$1/2$ systems, this gauge theory has non-trivial Berry phase terms (and was hence dubbed the ``odd'' \z2 gauge theory in Ref. \cite{Z2MS}). Under a duality transformation, this odd \z2 gauge theory maps onto the fully frustrated Ising model in a transverse field on the dual (dice) lattice. 

Finally, as discussed in \cite{Z2a,Z2h}, it is possible to directly formulate the Heisenberg spin model on any lattice as a theory of fermionic spinons coupled to an odd \z2 gauge theory.  In the spin-gapped quantum paramagnetic phases, the spinon fields may formally be integrated out, leaving a pure odd \z2 gauge theory as a description of physics below the spin gap. By the duality transformation alluded to above, one again obtains the fully frustrated transverse-field Ising model on the dual lattice. This approach gives a new flavor to it: the lattice does not have to be bipartite, and there is no need to rely on the dimer model from the beginning. We present a formal derivation and analysis in Appendix \ref{z2Ap}. A rough formal estimate for the relative importance of various coupling constants is also provided.

With these motivations, the starting point of this paper is the following frustrated Ising model in transverse field:
\begin{eqnarray}\label{ID}
H & = & -h \sum_{\la lm \ra} \epsilon_{lm} v^z_l v^z_m - K_3 \sum_{l_3} v^x_{l_3} -
K_6 \sum_{l_6} v^x_{l_6} -
\nonumber \\& & {} - 
K_{3+3} \sum_{(l_3 m_3)} v^x_{l_3} v^x_{m_3} - \cdots
\end{eqnarray}
This Hamiltonian is defined on the dice lattice, dual to Kagome, and the frustration is realized through the $\epsilon_{lm} = \pm 1$ factors, satisfying on each plaquette the following relation:
\begin{equation}\label{Frust}
\prod_{\diamondsuit} \epsilon_{lm} = -1
\end{equation}
The dice lattice, and a possible choice for $\epsilon_{lm}$ are shown in the Fig.~\ref{KDGrid}b. The $v^x$, $v^y$, and $v^z$ operators are the Pauli matrices of the fluctuating \z2 vortex field which corresponds to the singlet degrees of freedom. From now on we will drop all terms denoted by the ellipses, and keep only the lowest order ones involving the isolated 3-coordinated sites $l_3$, 6-coordinated sites $l_6$, and the pairs of next-nearest-neighbor 3-coordinated sites $(l_3 m_3)$. This theory is an effective theory of the Kagome Heisenberg magnet, describing the physics below the spin-gap. In the rest of the paper we study the properties of this model in different parameter regimes. As we formally argue in Appendix \ref{z2Ap}, when obtained from the original Heisenberg model, it is natural to consider the limit of large $h>> K_{3,6}$. This is also supported by comparison with the results of the numerical studies. Nevertheless we will study the model more generally and not just in the large-$h$ limit.

\section{Large-$h$ Limit: Valence Crystal Phase}\label{largeH}

Here we analyze the properties of the ground-state and excitation spectrum of the frustrated Ising model given by the Hamiltonian ~(\ref{ID}) in the $h \gg K_3, K_6, K_{3+3}, \dots $ limit. We will build the perturbative expansion for the exact low-energy effective theory.

The unperturbed Hamiltonian, with all $K_n$ equal to zero, is a classical Ising model on the fully frustrated dice lattice, and every spin configuration is an eigenstate. The ground-states have the minimal number of frustrated bonds (the bonds with positive energy: $ \epsilon_{lm} v^z_l v^z_m = -1 $). In trying to construct them, we have to leave at least one bond frustrated on every plaquette: otherwise, the product of four bond energies around a plaquette could not be negative, as required by the frustration condition ~(\ref{Frust}). It is convenient to return to a dimer representation of these states on the original Kagome lattice: we put a dimer on every Kagome bond which intersects a frustrated dice bond. This representation is accurate up to global spin flip of the Ising spins. There will be an odd number of dimers emanating from every Kagome site, and the ground-states will be given by the various hard-core dimer coverings.

\begin{figure}
\subfigure[{}]{\includegraphics[width=1.2in]{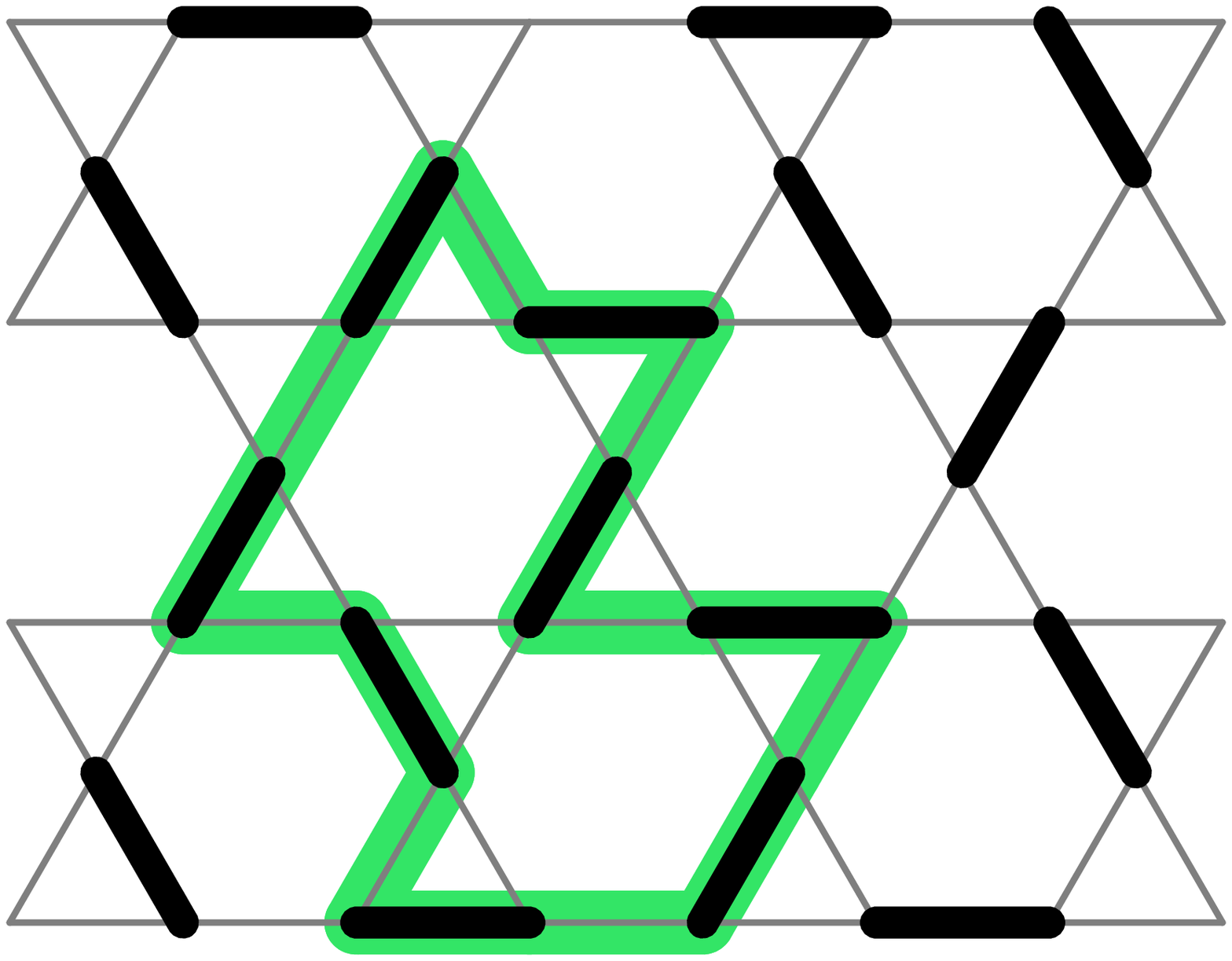}}
\subfigure[{}]{
  \includegraphics[width=0.6in]{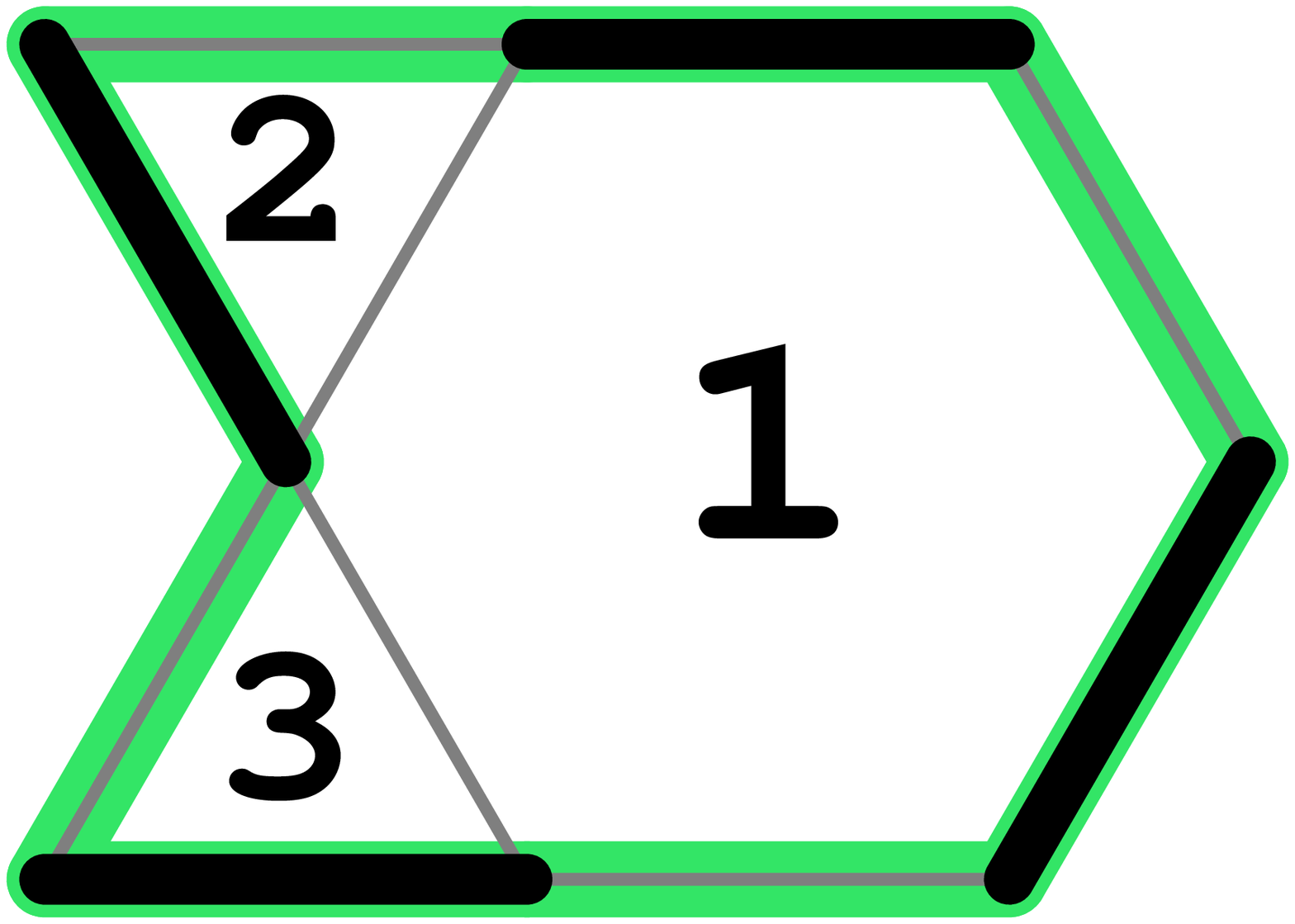}
  \includegraphics[width=0.6in]{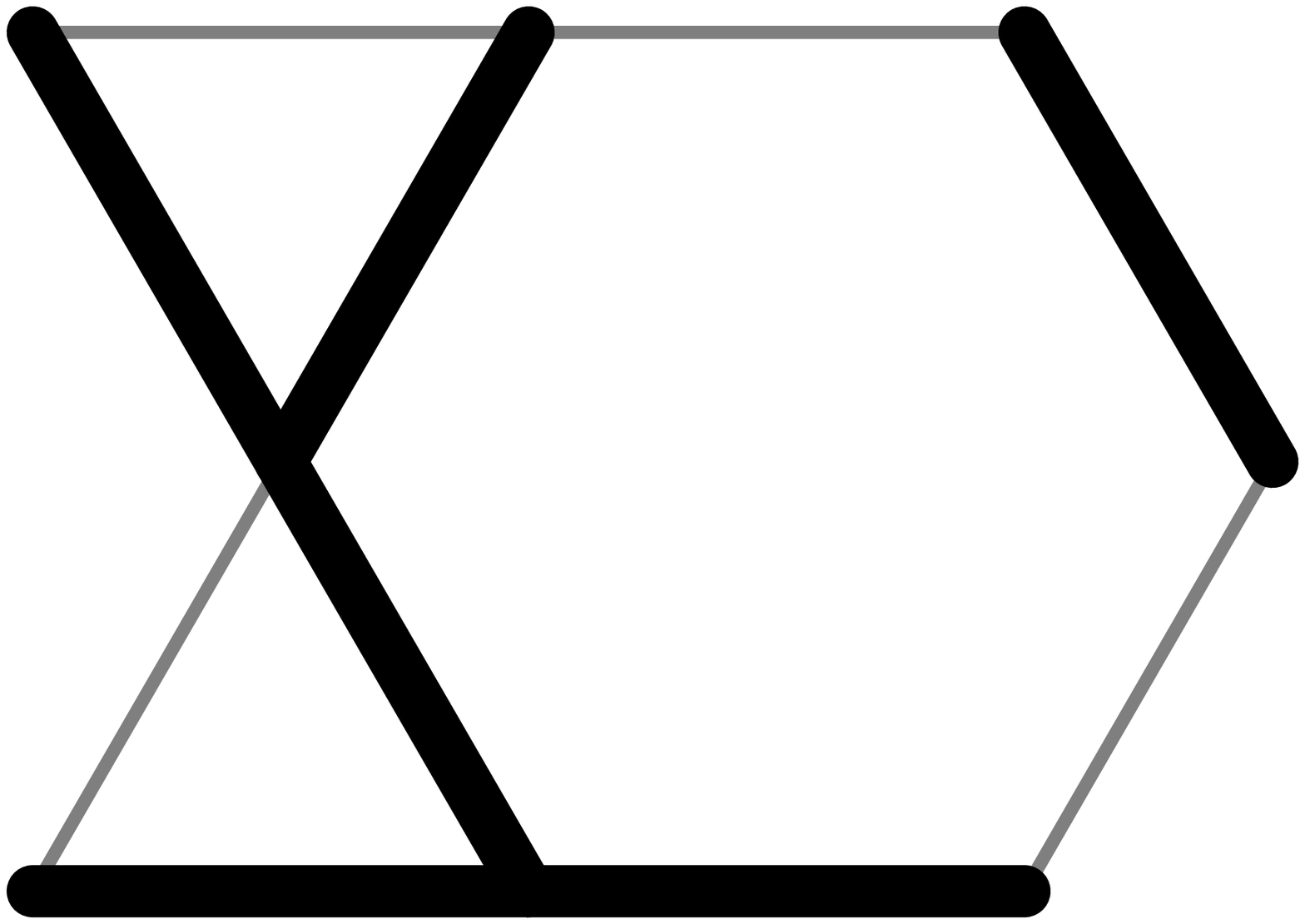}
  \includegraphics[width=0.6in]{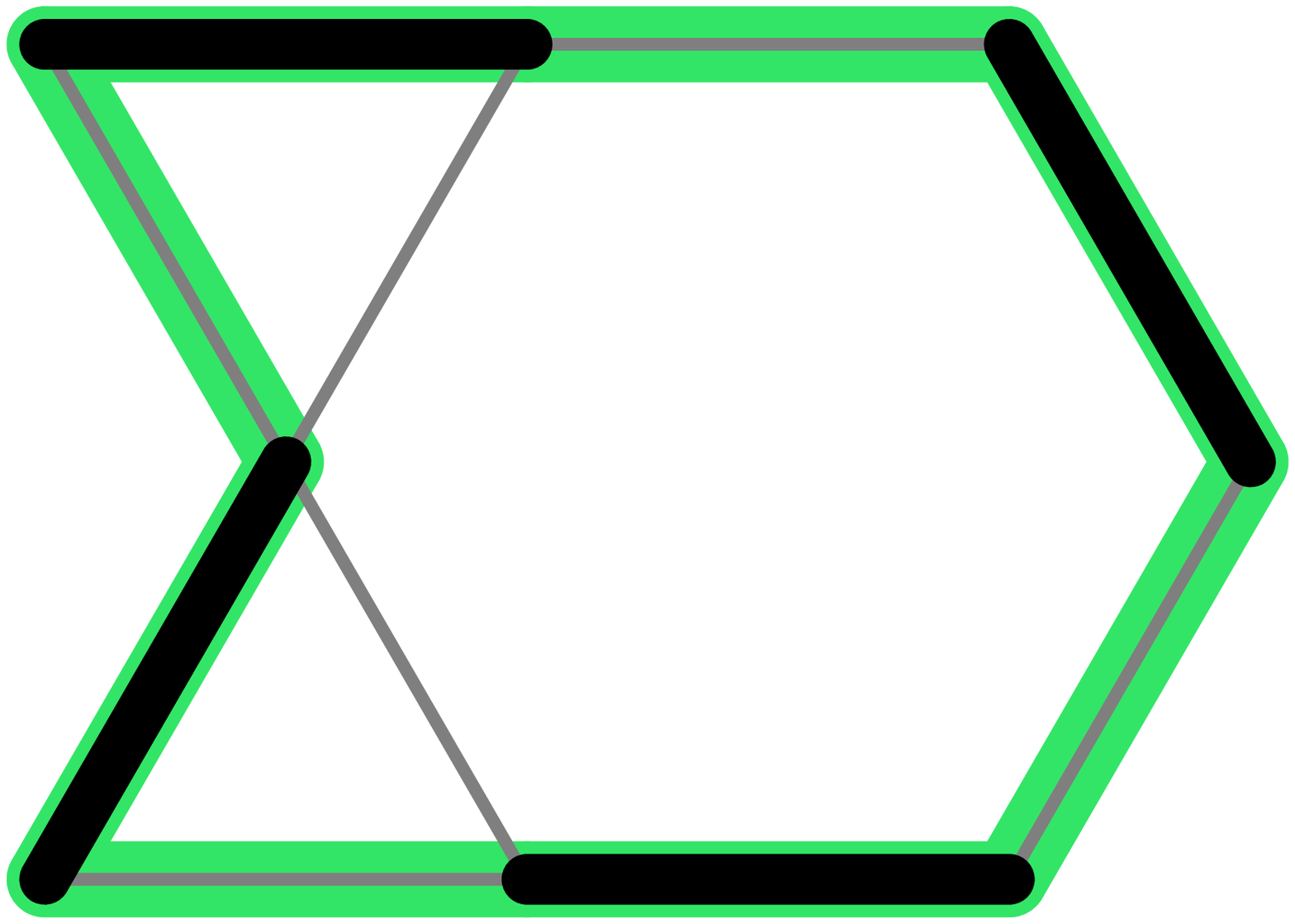}
}
\caption{\label{FlLoops0}(a) An example of a flippable loop. The flippable loops are tangential to the dimers, so that they cannot contain only one end of any dimer. (b) An example of a loop-flip between the two possible hard-core dimer arrangements on it. This loop contains a hexagon and a bowtie pair of triangles. The loop is flipped through a virtual state by the successive operation of $-K_6v^x_1$ and $-K_{3+3}v^x_2v^x_3$, where 1,2,3 refer to the dice lattice sites which correspond to the depicted Kagome plaquettes.}
\end{figure}

The perturbations $K_n$ introduce quantum fluctuations between different dimer coverings, and lift degeneracy of the unperturbed ground-states. The effect of a $v^x_l$ operator in the dimer picture is to toggle the dimer occupancy on all bonds of the Kagome plaquette which corresponds to the dice site $l$. Since we want to construct the effective theory, we need to find the combinations of $v^x_l$ operators which transform one hard-core dimer covering into another, possibly through some virtual states. Such processes can be described as dimer motion along the \emph{flippable} loops, and we will call them the \emph{loop flips} (see Fig.~\ref{FlLoops0} for explanation). It is convenient to introduce the following terminology: if a plaquette carries no dimers on its bonds, we will call it a \emph{defect} plaquette, and if it carries a flippable loop on its bonds, we will call it a \emph{perfect} plaquette. A plaquette can be perfect only if it has an even number of bonds. Of special interest are the elementary flippable loops which enclose only one hexagonal plaquette, and they are explained in Table \ref{FlLoops}.

\begin{table}[!b]
\begin{tabular}{|c||c|}
\hline
(a)
\includegraphics[width=1.5cm,height=1.5cm]{star6a.eps}
& (b)
\includegraphics[width=1.5cm,height=1.5cm]{star8Aa.eps}
\includegraphics[width=1.5cm,height=1.5cm]{star8Ba.eps}
\includegraphics[width=1.5cm,height=1.5cm]{star8Ca.eps} \\
\hline
(c)
\includegraphics[width=1.5cm,height=1.5cm]{star12a.eps}
& (d)
\includegraphics[width=1.5cm,height=1.5cm]{star10Aa.eps}
\includegraphics[width=1.5cm,height=1.5cm]{star10Ba.eps}
\includegraphics[width=1.5cm,height=1.5cm]{star10Ca.eps} \\
\hline
\end{tabular}
\caption{\label{FlLoops}The elementary flippable loops:
(a) flippable hexagon, (b) flippable 8-bond loops (rhombus, arrow and trapeze, from left to right),
(c) the star, (d) flippable 10-bond loops. The elementary flippable loops by definition enclose only one hexagon. For arbitrary dimer covering, a unique elementary flippable loop can be found on every hosting hexagonal plaquette: it goes through all the hexagon sites, and includes those surrounding triangles which hold a dimer on a bond that does not belong to the hexagon. The hexagon in (a) is \emph{perfect}, while the hexagon in (c) is a \emph{defect} hexagon. The length of the elementary loop is directly related to the number of dimers sitting on the hexagon, which is equal to the number of \emph{defect} triangles around the hexagon: 3 in (a), 2 in (b), 1 in (d), and none in (c).}
\end{table}

\begin{figure}
\subfigure[{}]{\includegraphics[width=0.9in]{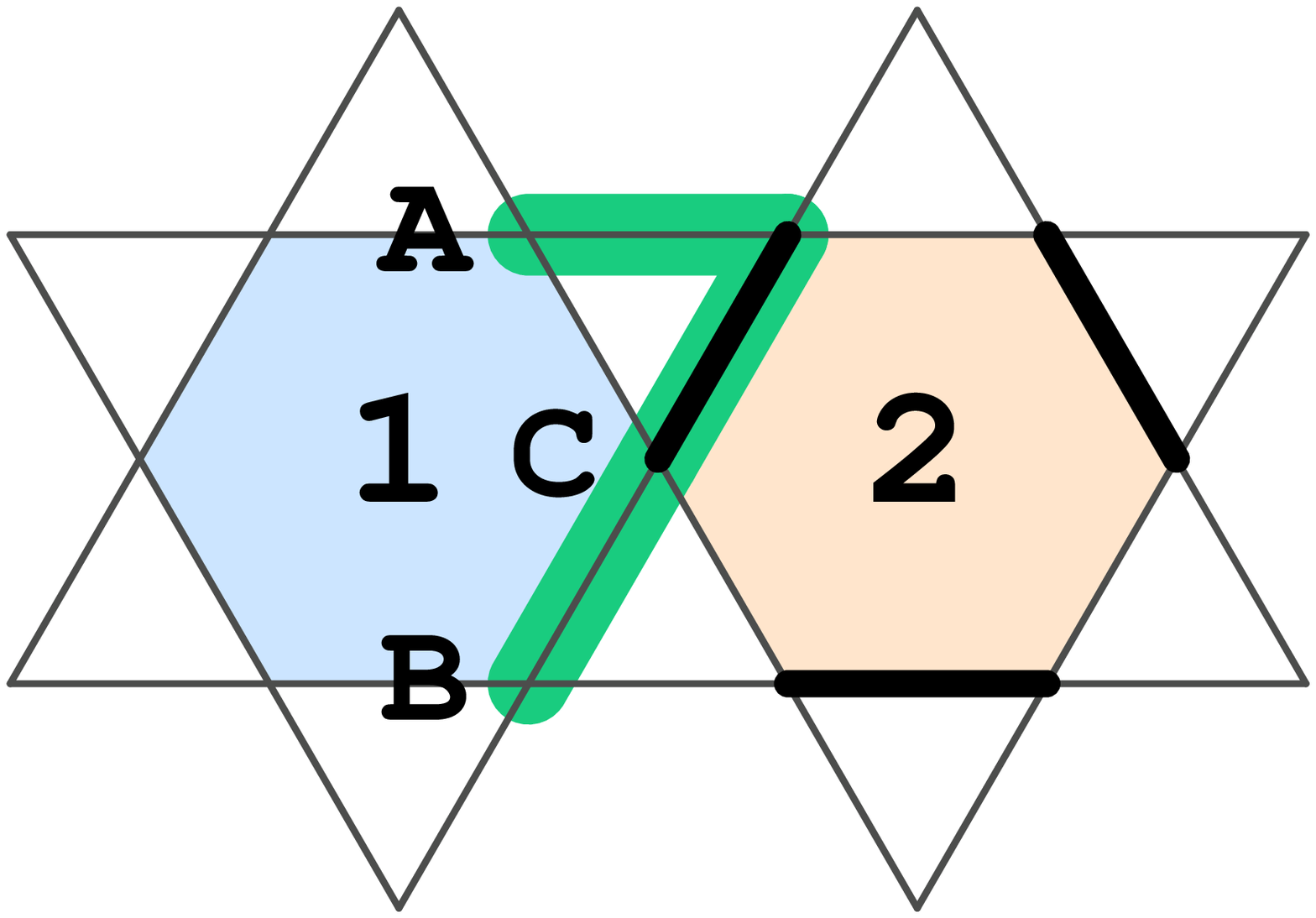}}
\subfigure[{}]{\includegraphics[width=0.9in]{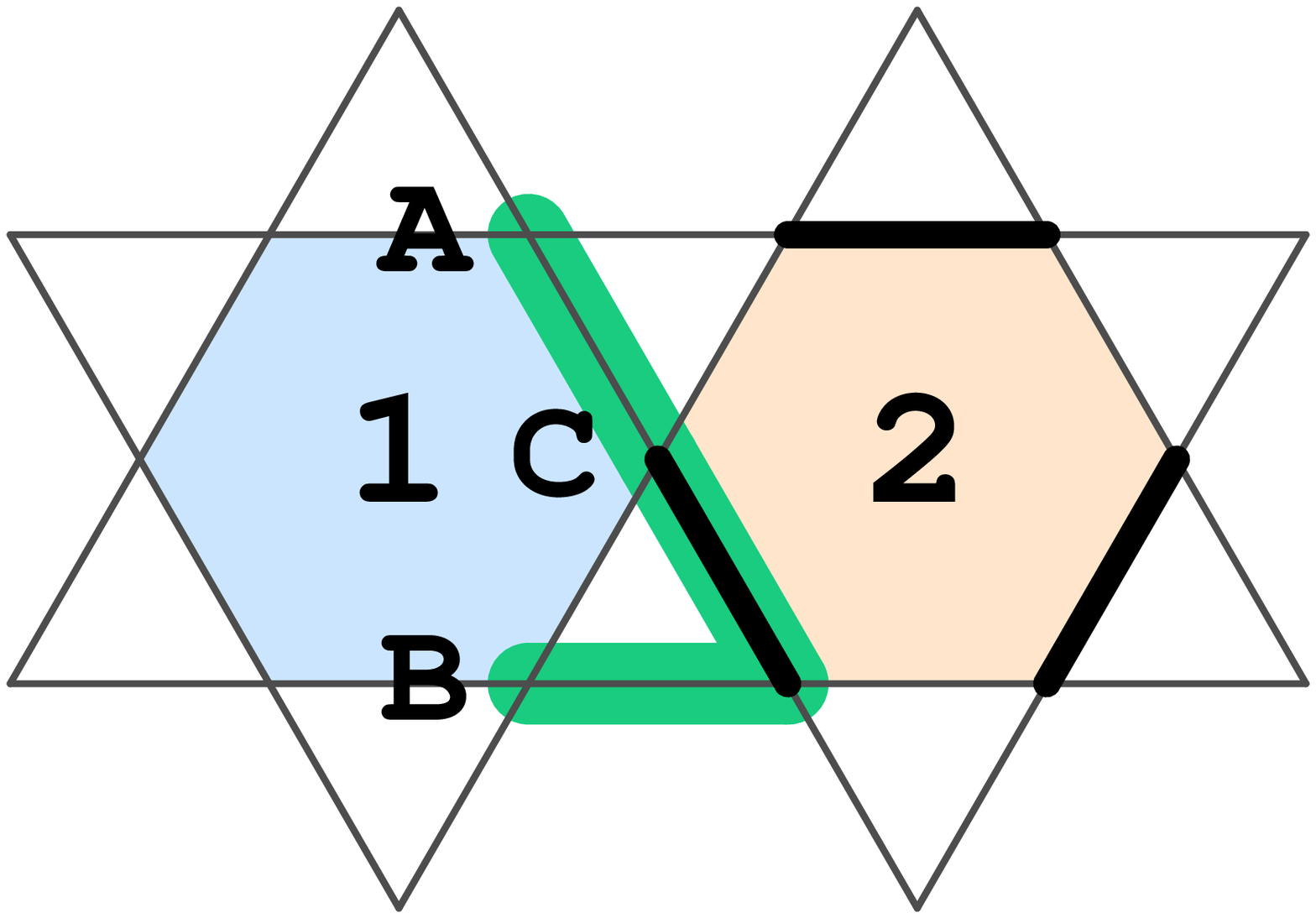}}
\vskip -2mm
\caption{\label{UWCommute}Two perfect hexagons cannot be nearest neighbors, and the perfect hexagon flip does not affect length of any other elementary flippable loop. Consider a perfect hexagon 2 being flipped from the configuration (a) to (b). The flippable loop on the hexagon 1 has to go through the sites A, B, and C, and therefore pass through the shaded bonds in order to be tangential to the dimers. As a consequence, it has to include at least one triangle, so that it cannot be a perfect hexagon, and also, its length is not affected by the flip on the hexagon 2.}
\end{figure}

\begin{figure}
\subfigure[{}]{\includegraphics[width=2.1in,angle=90]{kagome-honeycomb.eps}}
\subfigure[{}]{\includegraphics[width=2.1in,angle=90]{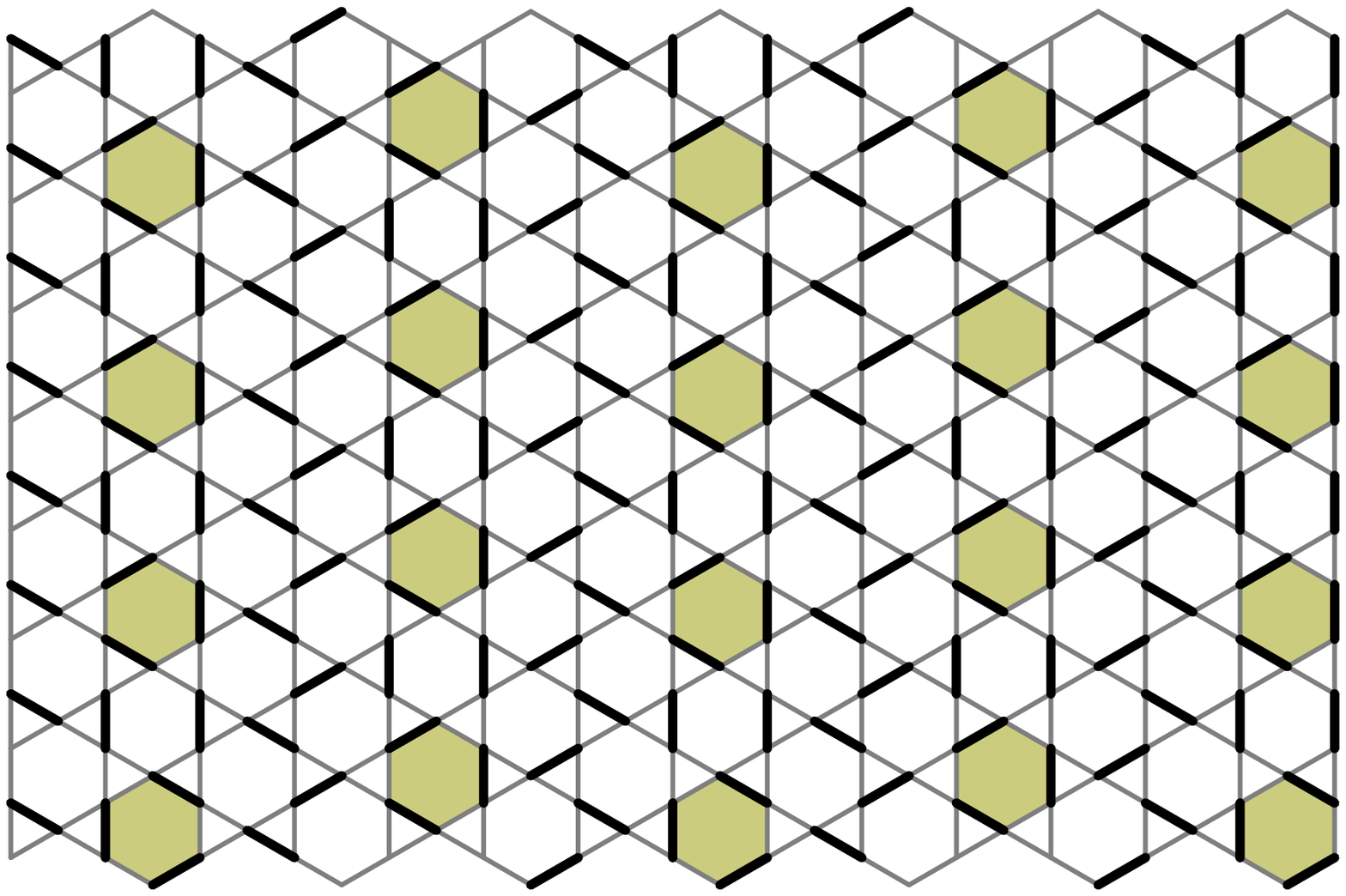}}
\vskip -2mm
\caption{\label{HoneyStripe}Two hard-core dimer patterns that maximize the number of perfect hexagons: (a) honeycomb pattern; (b) stripe pattern. The perfect hexagons are shaded to guide the eye. Note that the 8-bond flippable loops appear only as ``connections'' between the perfect hexagons, and the 10-bond flippable loops touch exactly one perfect hexagon. The honeycomb pattern has the 12-bond flippable loops, the stars: they sit inside the honeycomb cells.}
\end{figure}

The lowest order terms in the perturbation theory are the single plaquette flips. The $K_3$ and $K_{3+3}$ terms cannot connect between two hard-core dimer coverings, so that to the lowest order the effective theory is:
\begin{equation}\label{Dimer0}
\widetilde{H} = - N h - K_6 W_6
\end{equation}
where $N$ is the number of Kagome sites, and $W_6$ is the \emph{kinetic energy} operator of the perfect hexagons:
\begin{equation}\label{FHKin}
W_6 = \sum_{\hexagon} \left(
\ket{\sixRLau}\bra{\sixRLad} + \ket{\sixRLad}\bra{\sixRLau} \right)
\end{equation}
The sum runs over all hexagonal plaquettes of the Kagome lattice. The quantum fluctuations created by the kinetic term yield the ground states in which the number of resonating perfect hexagons is maximized. This is an exact statement, since all perfect hexagon flips commute with one another: as illustrated in Fig.~\ref{UWCommute}, the perfect hexagons can never be nearest neighbors, and therefore cannot affect one another. In order to gain more insight about these states, we want to recall some observations from References \cite{SUN} and \cite{KagHe}. The total number of dimers in a hard-core dimer covering on the Kagome lattice is $N_2=N/2$, and the number of triangular plaquettes is $ N_{\triangle} = 2N/3 $, so that $ N_2 = 3N_{\triangle}/4 = (1-1/4)N_{\triangle} $. Since a triangle can carry at most one dimer, we see that one quarter of all triangles are the \emph{defects} in any hard-core dimer covering: $ N_{\triangle d} = N_{\triangle}/4 $. Next, we note that every perfect hexagon has exactly three neighboring \emph{defect} triangles, and since no two perfect hexagons can be neighbors, those defect triangles are not shared between them. It follows from this that the total number of perfect hexagons has an upper bound: $ N_{\hexagon p} \leqslant N_{\triangle d}/3 = N_{\triangle}/12 = N_{\hexagon}/6 $. The maximum possible density of perfect hexagons is one per six hexagonal plaquettes, and it can be achieved in a variety of ways. In Fig.~\ref{HoneyStripe} we show two characteristic possibilities, (a) the \emph{honeycomb}, and (b) the \emph{stripe} state. In general, these states are constructed by placing the perfect hexagons as close as possible to each other. The closest they can be is the next nearest neighbors, provided that between them is another hexagonal plaquette, and not a bowtie pair of triangles ($\bowtie$), because in the latter case there would have been a site (the center of the bowtie) involved in no dimers. This rule allows one to arrange perfect hexagons in strings which may be straight, bent at an angle of $120^o$, or forked into two new strings at the $120^o$ angles. The \emph{stripe} state is an example of straight strings, while the \emph{honeycomb} state has strings forking at each perfect hexagon.

\begin{figure}
\subfigure[{}]{\includegraphics[width=2.1in]{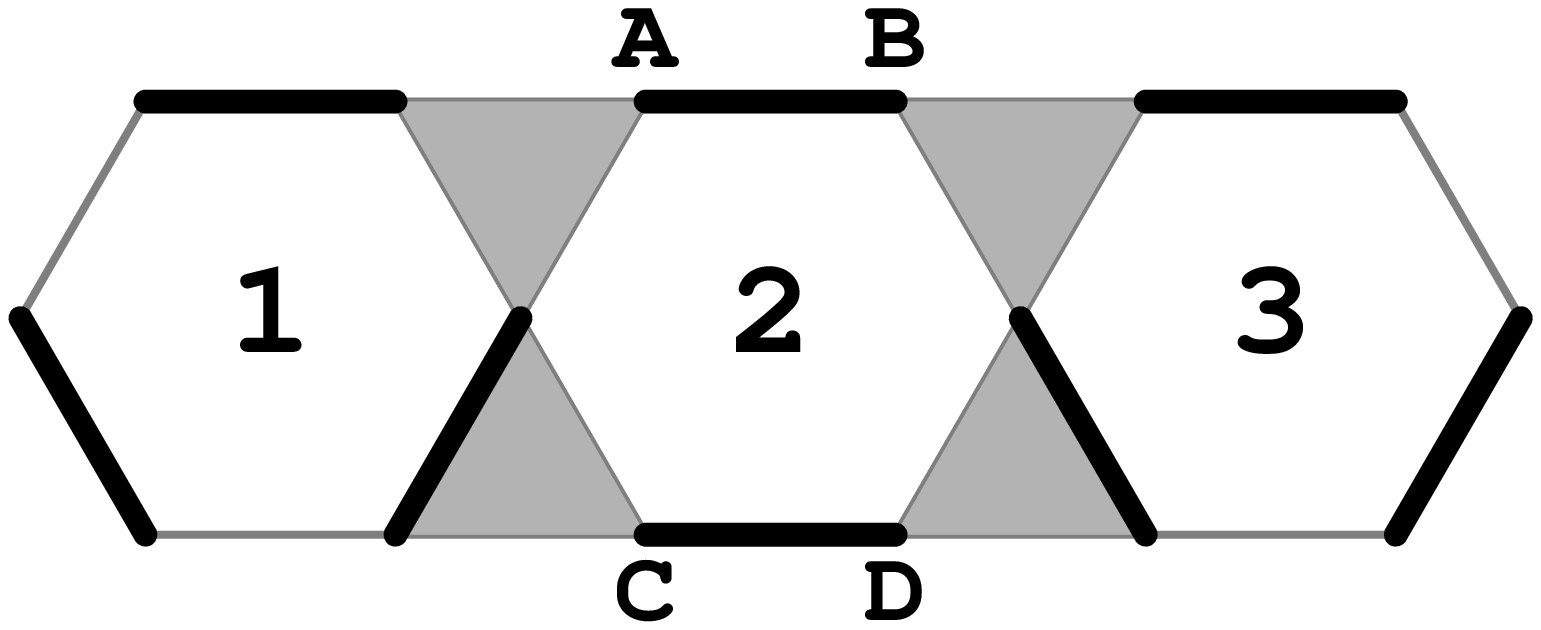}}
\subfigure[{}]{\includegraphics[width=2.1in]{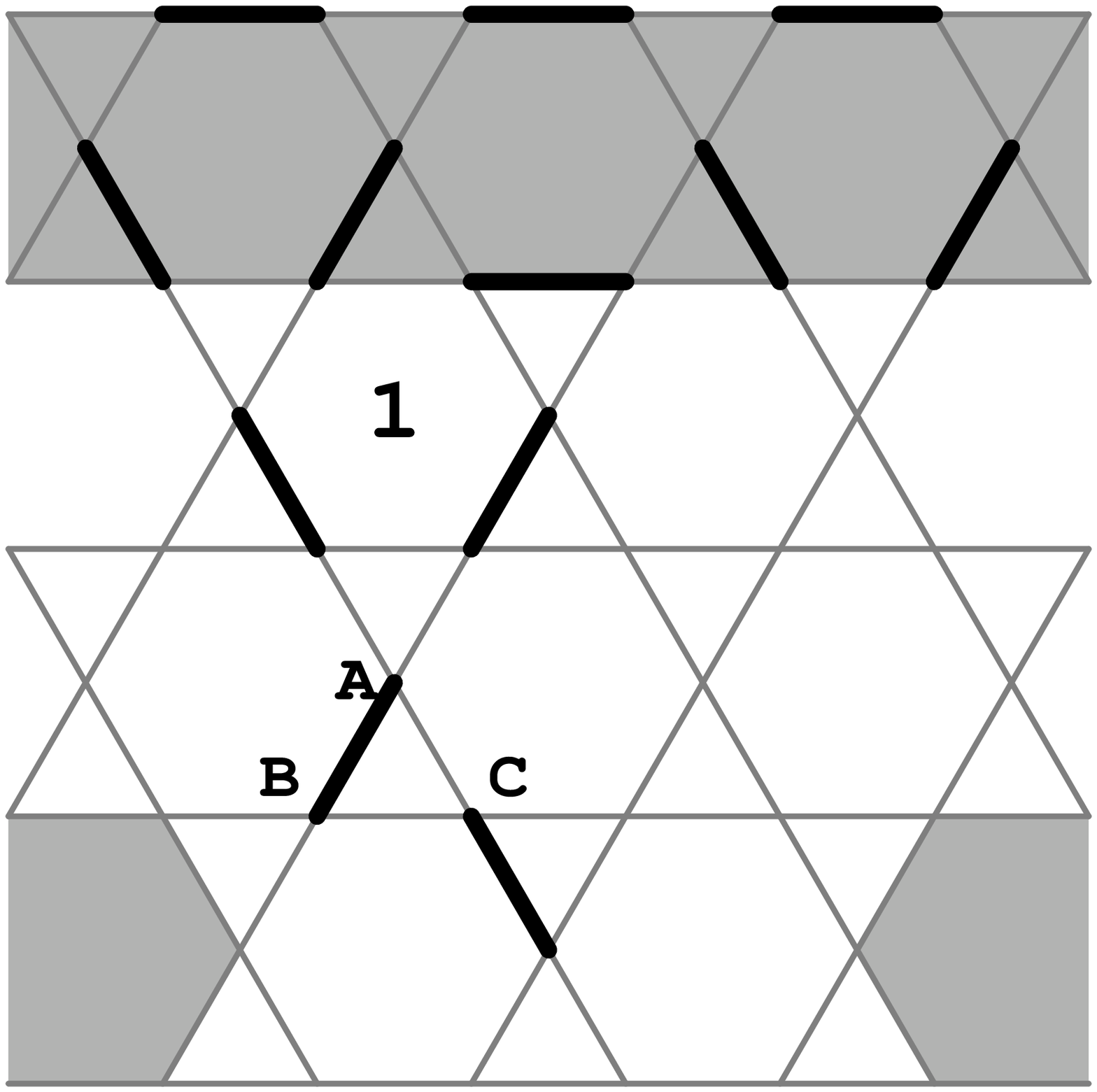}}
\caption{\label{FlLoops1} \\
(a) The hexagonal plaquette between a pair of perfect hexagons always hosts an 8-bond flippable loop. Once the perfect hexagons 1 and 3 are placed, the only way for all of the sites A,B,C, and D to be involved in dimers is to pair A,B and C,D. Then, the hexagonal plaquette 2 carries two dimers on its bonds, and therefore hosts an 8-bond flippable loop. However, it cannot be an \emph{arrow}-shaped loop (see Table \ref{FlLoops}), because both shaded bowties must have one normal and one \emph{defect} triangle due to the perfect hexagons, and an \emph{arrow}-shaped loop must contain a bowtie with two normal triangles. \\
(b) Only 10-bond flippable loops surround the perfect hexagon strings. Consider a string, and a neighboring hexagon 1. One cannot put there a 6-bond flippable loop, because this would be a perfect hexagon next to another perfect hexagon from the string. A \emph{star}-shaped 12-bond flippable loop is not an option either, because this hexagon always has one neighboring \emph{defect} triangle from the string. There is only one way to place two dimers on the bonds of hexagon 1 in attempt to create an 8-bond flippable loop on it. However, this requires the site A to pair with either B or C, which in turn makes it impossible for another string of perfect hexagons to take its normal place. At least a spot is lost for a perfect hexagon, and this costs energy $K_6$! Similar is true for the corners of the outward bending strings, while this issue does not occur for the inward bending, as in the case of the \emph{honeycomb} state, since 10-bond loops are the only choice there. \\[2mm]
Relative positions of various kinds of elementary flippable loops can also be established by using the relation between their length and the number of \emph{defect} triangles around them (Table \ref{FlLoops}). When the number of \emph{perfect} hexagons is maximized, all \emph{defects} are placed around them ($N_{\triangle d}=3N_{\hexagon p}$), and the length of any flippable loop depends on the number of its perfect hexagon neighbors.}
\end{figure}

Before we proceed with the next order of the perturbation theory, we need to make some additional remarks. If we look at the elementary flippable loops realized on the various hexagonal plaquettes in Fig.~\ref{HoneyStripe}, we observe that between every two closest perfect hexagons there is an 8-bond flippable loop, right on the sides of the strings there are only 10-bond flippable loops, and in the case of the \emph{honeycomb} state, there is a 12-bond \emph{star}-shaped flippable loop sitting at the center of every honeycomb cell. Also, one never finds an \emph{arrow}-shaped 8-bond flippable loop between two perfect hexagons. These are quite general features of the states with the maximum number of perfect hexagons, which we explain in more detail in the Fig.~\ref{FlLoops1}.

At the second order of perturbation theory, we need to include the combinations of two $K_3$, $K_6$, and $K_{3+3}$ flips. One of them is the flip on the \emph{arrow} shaped 8-bond flippable loop (a hexagon, and a bowtie: see Fig.\ref{FlLoops0}b):
\begin{eqnarray}\label{ArrowContr}
\Delta H^{(1)}_{\emph{ar.}} & = & - \frac{K_6 K_{3+3}}{2h} W^{(ar.)}_{8} \\
W^{(ar.)}_{8} & = & \sum_{\hexagon} \left(
\ket{\eightRLbu}\bra{\eightRLbd} + \ket{\eightRLbd}\bra{\eightRLbu}
+ \emph{rotat.} \right)
\end{eqnarray}
All other allowed combinations flip the same plaquette twice, leaving the dimer configuration unchanged. A hexagonal plaquette can be flipped twice if it is not a flippable loop: it has to be the part of an elementary flippable loop with 8, 10, or 12 bonds. As the potential energy associated with the flip is different in each case, we need to know the numbers $U_n$ of the $n$-bond flippable loops in the dimer configuration in order to write the appropriate contribution in the effective theory:
\begin{equation}\label{HexagonContr}
\Delta H^{(1)}_{\hexagon \times 2} = - \frac{K_6^2}{4h} U_8 -
\frac{K_6^2}{8h} U_{10} - \frac{K_6^2}{12h} U_{12}
\end{equation}
The number operators $U_n$ are explicitly written in the Table \ref{PotEnergy}.
\begin{table*}
\begin{eqnarray}
U_{6} & = & \sum_{\hexagon} \left(\ket{\sixRLau}\bra{\sixRLau} +
 \ket{\sixRLad}\bra{\sixRLad} \right) \nonumber \\
U_{8} & = & \sum_{\hexagon} \left(
 \ket{\eightRLau}\bra{\eightRLau} + \ket{\eightRLad}\bra{\eightRLad} +
 \ket{\eightRLbu}\bra{\eightRLbu} + \ket{\eightRLbd}\bra{\eightRLbd} +
 \ket{\eightRLcu}\bra{\eightRLcu} + \ket{\eightRLcd}\bra{\eightRLcd} +
 \emph{rotations} \right) \nonumber \\
U_{10} & = & \sum_{\hexagon} \left(
 \ket{\tenRLau}\bra{\tenRLau} + \ket{\tenRLad}\bra{\tenRLad} +
 \ket{\tenRLbu}\bra{\tenRLbu} + \ket{\tenRLbd}\bra{\tenRLbd} + 
 \ket{\tenRLcu}\bra{\tenRLcu} + \ket{\tenRLcd}\bra{\tenRLcd} +
 \emph{rotations} \right) \nonumber \\
U_{12} & = &\sum_{\hexagon} \left(\ket{\twelveRLau}\bra{\twelveRLau} +
 \ket{\twelveRLad}\bra{\twelveRLad} \right) \nonumber
\end{eqnarray}
\caption{\label{PotEnergy}The flippable loop number operators: $U_n$ counts the number of $n$-bond flippable loops in the dimer configuration. They correspond to the \emph{potential energy} in the dimer model.}
\end{table*}

We treat the triangular plaquette double flips similarly. Every triangle can be either a \emph{defect}, or normal in any dimer covering. The \emph{defects} contain no dimers, while the normal triangles contain one dimer. If their numbers are $N_{\triangle d}$ and $N_{\triangle}$ respectively,  then the contribution of the triangle double flips is:
\begin{equation}\label{TriangleContr}
\Delta H^{(1)}_{\triangle \times 2} = 
  - \frac{K_3^2}{6h} N_{\triangle d} - \frac{K_3^2}{2h} (N_{\triangle}-N_{\triangle d})        
  = - \frac{K_3^2}{h} \frac{5N}{18}
\end{equation}
We have used the identities $N_{\triangle d}=N_{\triangle}/4$, and $N_{\triangle}=2N/3$, where $N$ is the number of Kagome sites. Finally, it remains to consider the double flips of bowties $(\bowtie)$. They can contain either one \emph{defect} triangle or none. For every \emph{defect} there are three bowties containing it, while the total number of bowties in the lattice is equal to the number of Kagome sites $N$. Therefore, the contribution of the bowties is:
\begin{eqnarray}\label{BowtieContr}
\Delta H^{(1)}_{\bowtie \times 2} & = &
  - \frac{K_{3+3}^2}{8h} 3N_{\triangle d} - \frac{K_{3+3}^2}{4h} (N-3N_{\triangle d}) =
  \nonumber \\ & = & - \frac{K_{3+3}^2}{h} \frac{3N}{16}
\end{eqnarray}
We see that the triangle and bowtie contributions reduce to mere constants for arbitrary dimer covering. Then, it pays to investigate closer the contribution of double hexagon flips. First, we want to express the number of defect triangles in terms of the $U_n$ operators. This can be achieved by noting that every type of elementary flippable loop has a fixed number of both kinds of triangles (see Table \ref{FlLoops}): $N_{\triangle d} = (3 U_6 + 2 U_8 + U_{10})/3$, where the factor of $1/3$ corrects the over-counting of triangles shared between the elementary flippable loops. Using this, we can eliminate $U_8$ and $U_{10}$ from ~(\ref{HexagonContr}):
\begin{equation}\label{HexagonContr2}
\Delta H^{(1)}_{\hexagon \times 2} = - \frac{K_6^2}{12h} U_{12} + \frac{3K_6^2}{8h} (U_6-N_{\triangle d})
\end{equation}
Now, we can add ~(\ref{ArrowContr}, \ref{HexagonContr2}, \ref{TriangleContr}, \ref{BowtieContr}), and write the effective theory to the second order:
\begin{eqnarray}\label{Dimer1}
\widetilde{H} & = & - N h - K_6 W_6 - \frac{K_6 K_{3+3}}{2h} W^{(ar.)}_{8} +
  \nonumber \\
  & & + \Delta H^{(1)}_{\triangle \times 2} + \Delta H^{(1)}_{\hexagon \times 2}
  + \Delta H^{(1)}_{\bowtie \times 2}
\end{eqnarray}
The ground-state of this Hamiltonian can be obtained by the \emph{second level} perturbation theory in which the unperturbed Hamiltonian is given by ~(\ref{Dimer0}). The $W^{(ar.)}_{8}$ term slightly spreads the ground-state wavefunction from the sharp \emph{honeycomb} or a \emph{stripe}-like state, but the correction to the ground-state energy due to this term appears only at the higher orders: both the \emph{honeycomb} and \emph{stripe} states cannot contain the \emph{arrow}-shaped 8-bond flippable loops, and even if they could, the flip would destroy the two neighboring perfect hexagons, so that $\la 0|W^{(ar.)}_{8}|0 \ra = 0$. Therefore, the energy shift is dominated by the \emph{potential energy} part of ~(\ref{Dimer1}) involving $U_n$ operators. This part fully commutes with ~(\ref{Dimer0}): the perfect hexagon flips $W_6$ cannot change the length of any elementary flippable loop (see Fig.~\ref{UWCommute}), and the $U$ operators simply count the number of flippable loops with the given length. Only the  double hexagon flips ~(\ref{HexagonContr2} select the actual ground state: when the number of perfect hexagons is maximized, the $U_6$ operator behaves as a number, taking the value $N_{\hexagon}/6=N/18$, so that the potential energy is controlled by the number of the 12-bond \emph{star}-shaped flippable loops. The \emph{honeycomb} state (Fig.~\ref{HoneyStripe}a) maximizes their number, and therefore the quantum fluctuations select it as the ground-state. The degeneracy of this ground-states is exponentially large in the system size, at this order of the perturbation theory. It comes from the freedom to flip any \emph{star}-loops without energy cost.

Various new kinetic and potential terms appear in the effective dimer model at the higher orders. They further spread the ground-state wavefunction, but its main component remains the \emph{honeycomb} structure. The ground-state degeneracy is finally lifted when the \emph{star} kinetic terms appear (only the star-flips do not destroy the perfect hexagons):
\begin{equation}\label{StarKin}
W_{12} = \sum_{\hexagon} \left(
\ket{\twelveRLau}\bra{\twelveRLad} + \ket{\twelveRLad}\bra{\twelveRLau} \right)
\end{equation}
In terms of $K_n/h$, this first happens at the fourth order by combining the flips on one hexagon and three bowties:
\begin{equation}\label{StarKin33}
\Delta H^{(3)}_{\emph{star}} \propto - \frac{K_6 K_{3+3}^3}{h^3} W_{12}
\end{equation}
However, since in principle it might happen that $K_{3+3} \ll K_3^2/h$, the dominant \emph{star} kinetic term may occur when a hexagon flip is combined with six triangle flips:
\begin{equation}\label{StarKin3}
\Delta H^{(6)}_{\emph{star}} \propto - \frac{K_6 K_3^6}{h^6} W_{12}
\end{equation}
In any case, the ground-state will be the \emph{honeycomb} state with resonating perfect hexagons and stars, while the lowest lying excitations will be gapped. The gap is very small, of the order of $K_6 K_{3+3}^3 / h^3$ or $K_6 K_3^6 / h^6$. The number of such excitations is exponential in the system size. The only remaining degeneracy is 12-fold, due to the broken translational symmetry (the honeycomb unit-cell has 12 hexagonal plaquettes).

It is also interesting to understand this ground-state in the dual Ising picture. As explained at the beginning of this section, a dimer on the Kagome lattice represents a frustrated bond of the dual Ising model on the dice lattice. Every Kagome dimer covering corresponds to two Ising spin configurations on the dice lattice, related to each other by the global spin-flip. If the Kagome dimers are flipped along a flippable loop, the corresponding effect in the dual picture is the simultaneous flip of all dice Ising spins which sit ``inside'' the Kagome plaquettes enclosed by that flippable loop. Now we can translate the description of the ground-state. Every resonating perfect hexagon is a \emph{flippable} spin on a 6-coordinated site of the dice lattice, in the state of equal superposition of ``up'' and ``down``. Every resonating star-shaped flippable loop is a \emph{flippable} cluster of seven Ising spins coherently fluctuating between two states of defined spin orientation (one spin is on a 6-coordinated site, and the other six are on the surrounding 3-coordinated sites). However, certain dimers are static in the ground-state: two of them reside between every pair of neighboring perfect hexagons. In order to describe them in the dual language, it is sufficient to arrange the corresponding Ising spins on the dice lattice in some appropriate static configuration. Therefore, the ground-state breaks the global spin-flip symmetry of the dual Ising model, with $3/4$ of all spins assuming a fixed orientation, and $1/4$ of spins fluctuating. The translational symmetry is broken only by the arrangement of frustrated bonds, and locations of the fluctuating spins; there need not be actual long-range order in terms of the orientation of non-fluctuating spins, since this depends on a relatively arbitrary assignment of $\epsilon_{lm}=\pm 1$ in ~(\ref{ID}). In the dual language, this state is formed first by minimizing the number of frustrated bonds, and then by maximizing the number of flippable spins which gain the kinetic energy from the transverse field. 

Since this is a valence-bond crystal phase which breaks the spin-flip symmetry of the dual Ising model, it is also a confined phase. The elementary spinful excitations are  gapped spin-1 magnons. We also note that this phase is stable against the weak fluctuations of the large loops that we have ignored from the Hamiltonian ~(\ref{ID}). The stability has been demonstrated in the perturbation theory. 

At the end, we ask what changes if the system becomes finite. As we have seen, the infinite system breaks translational symmetry. The ground-states are 12-fold degenerate, and related to one another by translations. Roughly speaking, we can even regard the lowest lying excited states as 12-fold degenerate. If the lattice becomes finite, this remaining degeneracy will be lifted, and no symmetry will be spontaneously broken. Let us apply the perturbation theory developed in this Section to a finite sample with $N$ sites. The 12-fold degenerate ground (and excited) states, shaped at the low orders, can be eventually mixed only by high order perturbations, which are essentially just translations. In the dimer language, they correspond to dimer flips on large flippable loops which enclose a finite fraction of all plaquettes of the sample. Therefore, their order scales as $N$, and the energy scale associated with them is exponentially small. This is enough to completely mix the states with the same energy, but there will be no mixtures of the states with different energies. As a consequence, the spectrum of the finite system will be very similar to that of the infinite system, and just slightly more dispersed. For example, let us consider a sample with $N=36$ sites. It takes enclosing at least $\sim 16$ plaquettes by the translational flippable loops, so that the translational degeneracy is lifted at $\sim 16^{\textrm{th}}$ order. The exact count of the lowest lying excitations depends on the boundary conditions, but we can still demonstrate that this number is not small. If the boundary conditions allow the \emph{honeycomb} pattern, then there will be at most one star and two perfect hexagons for $N=36$. Some basic excited states can be obtained by exciting the star and two perfect hexagons in various combinations ($2^3$ total, including the ground-state), and each is 12-fold ``degenerate'' in the thermodynamic limit. The other can be obtained by departing from the honeycomb pattern, which may require removing one or both perfect hexagons. The ``degeneracy'' is lifted in this finite system, and these states form bands. Their number is greater than $2^3 \times 12=96$. This is not inconsistent with the number obtained from numerics: $1.15^N \sim 153$ below the spin-gap. We also note that $N=36$ is small enough to miss the spinon confinement. Confinement is noticeable only at distances much larger than size of the \emph{honeycomb} pattern unit-cell, so that seeing it would require at least $N=72$ sites.

\section{Small-$h$ Limit: Fractionalized Spin-Liquid Phase}\label{smallH}

The limit $h \ll K_3, K_{3+3}, K_6$\dots is convenient to analyze directly in the frustrated Ising model. If $h$ vanishes, then all Ising spins in ~(\ref{ID}) are aligned with the transverse fields in the $x$-direction, making the ground-state completely disordered and uncorrelated in terms of $v^z_l$. An elementary excitation is created when one spin is flipped against the transverse field. These excitations are \z2 vortices, or visons according to ~(\ref{VisonX}), and they are localized and gapped. For finite $h$, the visons can in principle hop between sites and lower their gap by acquiring kinetic energy. However, for small $h$ the gap is guarantied to persist, and the ground-state remains disordered, and unique.

In the following we will assume that $K_3$ and $K_6$ are positive, and we will ignore $K_{3+3}$ and other terms for simplicity. One way to study the properties of the excited states is to consider perturbatively the effective theory for one vison. At the lowest order, this effective theory is simply the nearest neighbor hopping Hamiltonian:
\begin{eqnarray}\label{VisonHop1}
\widetilde{H} & = & -h \sum_{\la lm \ra} 
\Bigl( |l \ra \epsilon_{lm} \la m| + h.c. \Bigr) + \nonumber \\
& & + 2 K_3 \sum_{l_3} |l_3 \ra \la l_3|
+ 2 K_6 \sum_{l_6} |l_6 \ra \la l_6|
\end{eqnarray}
where $l_3$ and $l_6$ are the 3 and 6-coordinated sites of the dice lattice, and $|l\ra = v_l^z |0\ra$. It can be easily diagonalized in the momentum space by working with the 6-site elementary cells of the tile in Fig.~\ref{KDGrid}. It was shown in the reference \cite{DiceAB} that for $K_3=K_6$ the energy spectrum of this model is completely dispersionless and divided into three macroscopically degenerate levels. Remarkably, this remains true for arbitrary values of $K_3$ and $K_6$:
\begin{eqnarray}\label{VisonBands1}
E_1 & = & 2 K_6 \nonumber \\
E_2 & = & 2 K_3 - \sqrt{6} h \\
E_3 & = & 2 K_3 + \sqrt{6} h \nonumber
\end{eqnarray}
Therefore, the visons are localized at the lowest order of perturbation theory, in the similar way to a single electron in the magnetic field (this analogy does not generally hold). A natural question to ask is whether this localization persists to higher, or maybe all orders of perturbation theory. We try to find an answer by considering the time-ordered Green's function in the interaction picture:
\begin{equation}\label{VisonGreen}
i G(l,m;t_l,t_m) = \frac{\la 0|T v_l^z(t_l) v_m^z(t_m) S|0 \ra}{\la 0|S|0 \ra}
\end{equation}
where $T$ is the time-ordering operator, and:
\begin{equation}
S  = T \exp \Bigl( -i \int_{-\infty}^{\infty} \dd t H'(t) \Bigr) \nonumber
\end{equation}
\begin{equation}\label{IntPicture}
v^z(t) = e^{i H_0 t} v^z(0) e^{-i H_0 t}
\end{equation}
The perturbation $H'$ is the vison-hopping part of ~(\ref{ID}), while $H_0$ is the vison potential energy. The expansion of $S$ in ~(\ref{VisonGreen}) generates many vacuum expectation values of the products of $v^z$ operators, which are then integrated over the internal time variables. These vacuum expectations are easy to calculate for any given set of time moments: the $v^z(t)$ operators can be evolved to $t=0$ when they simply flip a spin, while the evolution parts $e^{\pm i H_0 t}$ only introduce a phase factor, since they always act on a $H_0$ eigenstate. We immediately see that a vacuum expectation which appears in ~(\ref{VisonGreen}) can be non-zero only if the product of $v^z$ is made from a path of bonds bridging between the sites $l$ and $m$. We only need to turn our attention to the \emph{connected} clusters of bonds.

\begin{figure}
\subfigure[{}]{\includegraphics[width=0.5in]{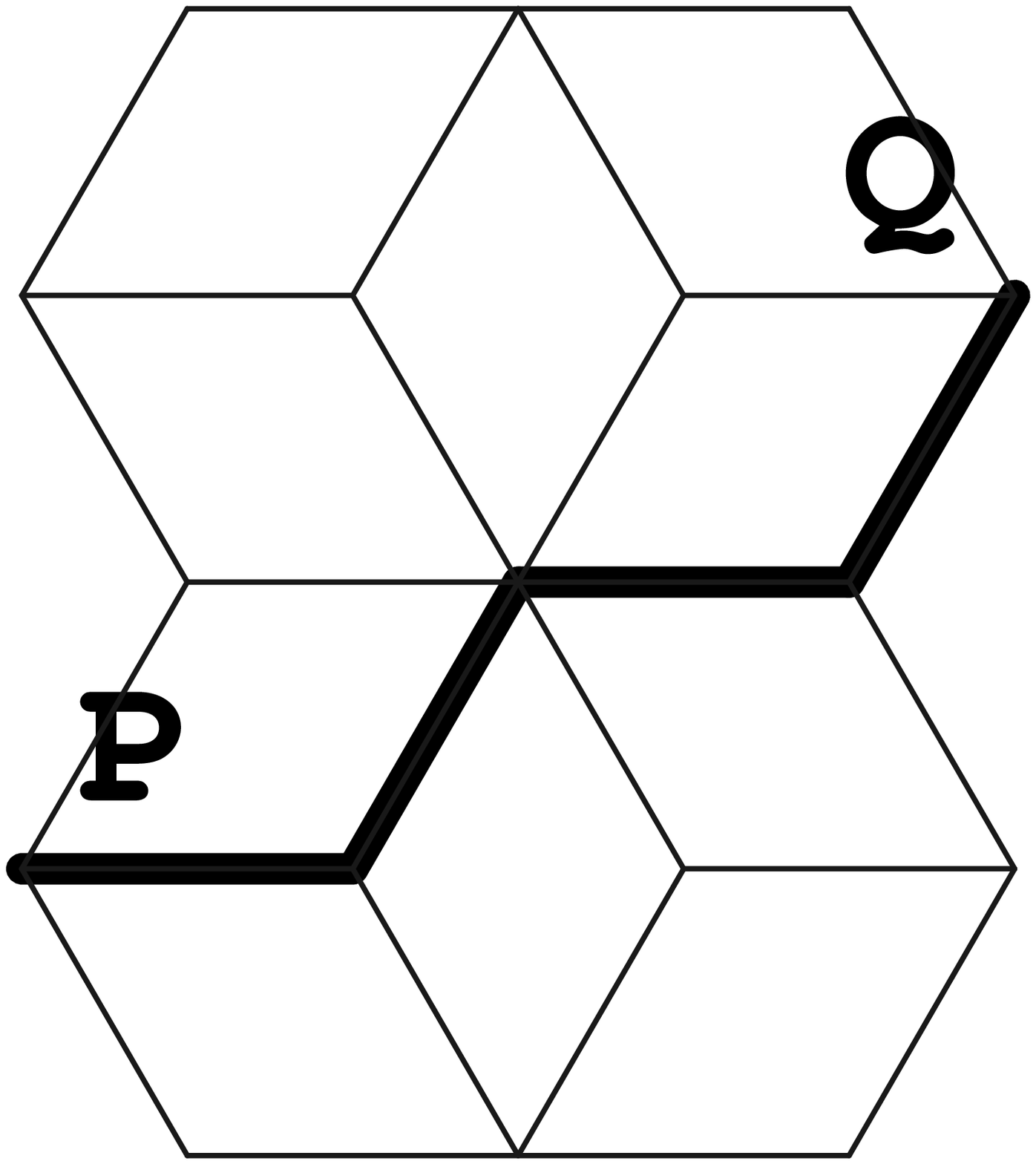}}
\subfigure[{}]{\includegraphics[width=0.5in]{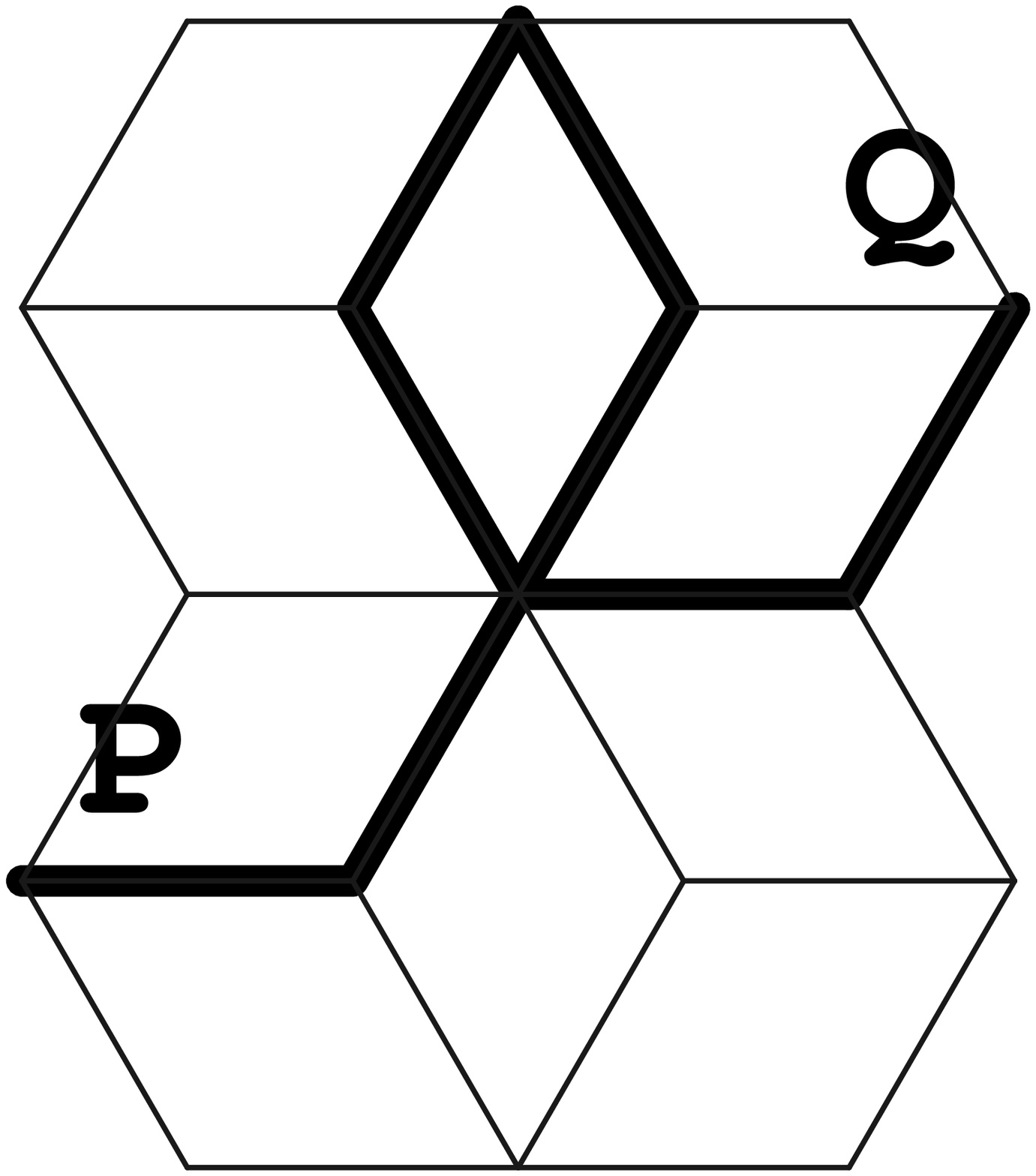}}
\subfigure[{}]{\includegraphics[width=0.5in]{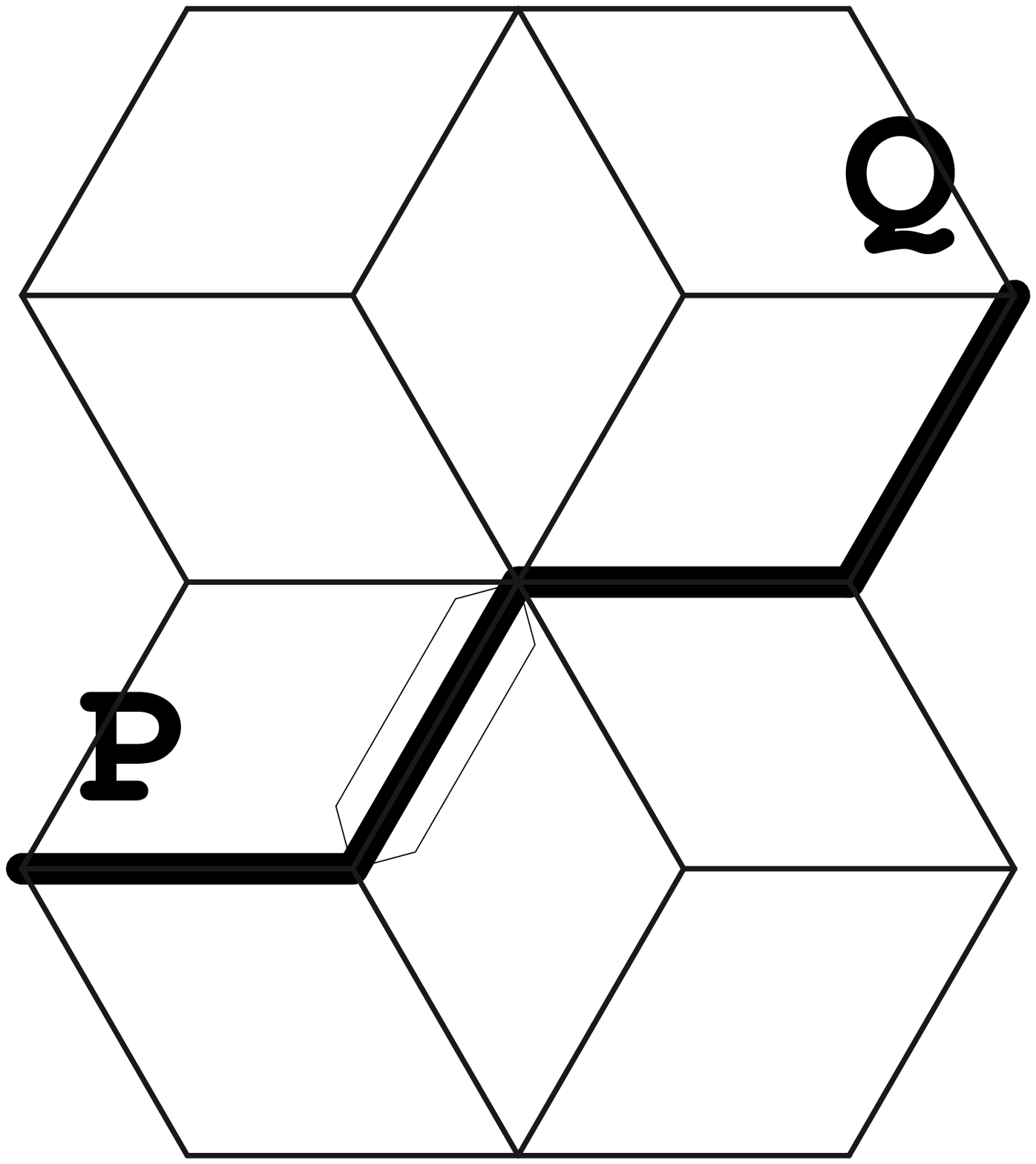}}
\subfigure[{}]{\includegraphics[width=0.5in]{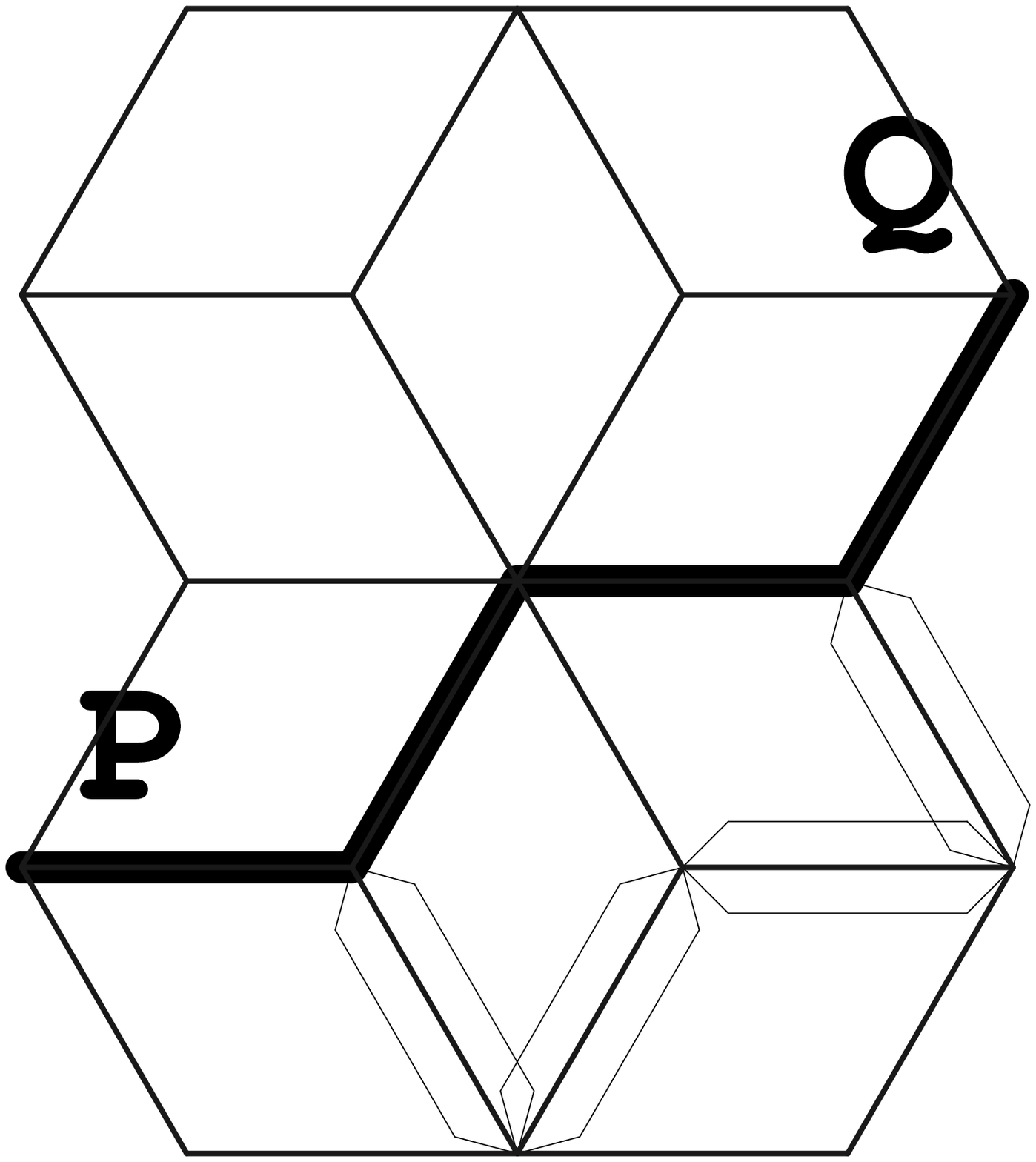}}
\vskip -2mm
\caption{\label{DicePaths}Examples of the paths between the sites $P$ and $Q$: (a) simple path; (b)-(d)  complex paths. The thin line-doublets around the bonds in (c) and (d) mean that there are loops going through those bonds twice.}
\end{figure}

Consider a particular pair of sites $P$ and $Q$, and the $n^{\textrm{th}}$ order term in the perturbative expansion of $iG(P,Q;t_P,t_Q)$. Its value, proportional to $h^n$, is contributed separately by all paths of the length $n$ between $P$ and $Q$. The two paths cancel out each other if they enclose an odd number of rhombic plaquettes: the total ``flux'' through the loop formed by them is $\pi$, that is, $\prod_{ \textrm{loop} } \epsilon_{lm} = -1$. Let us distinguish the two types of paths:  \emph{simple} and \emph{complex}. The \emph{simple} paths never visit any site more than once, while the \emph{complex} paths visit at least one site more than once. The \emph{complex} paths can be obtained from the \emph{simple} ones by adding the loop segments to them, where a loop may take some bonds an even number of times. We illustrate this in the Fig.~\ref{DicePaths}.

\begin{figure}
\subfigure[{}]{\includegraphics[width=1.2in]{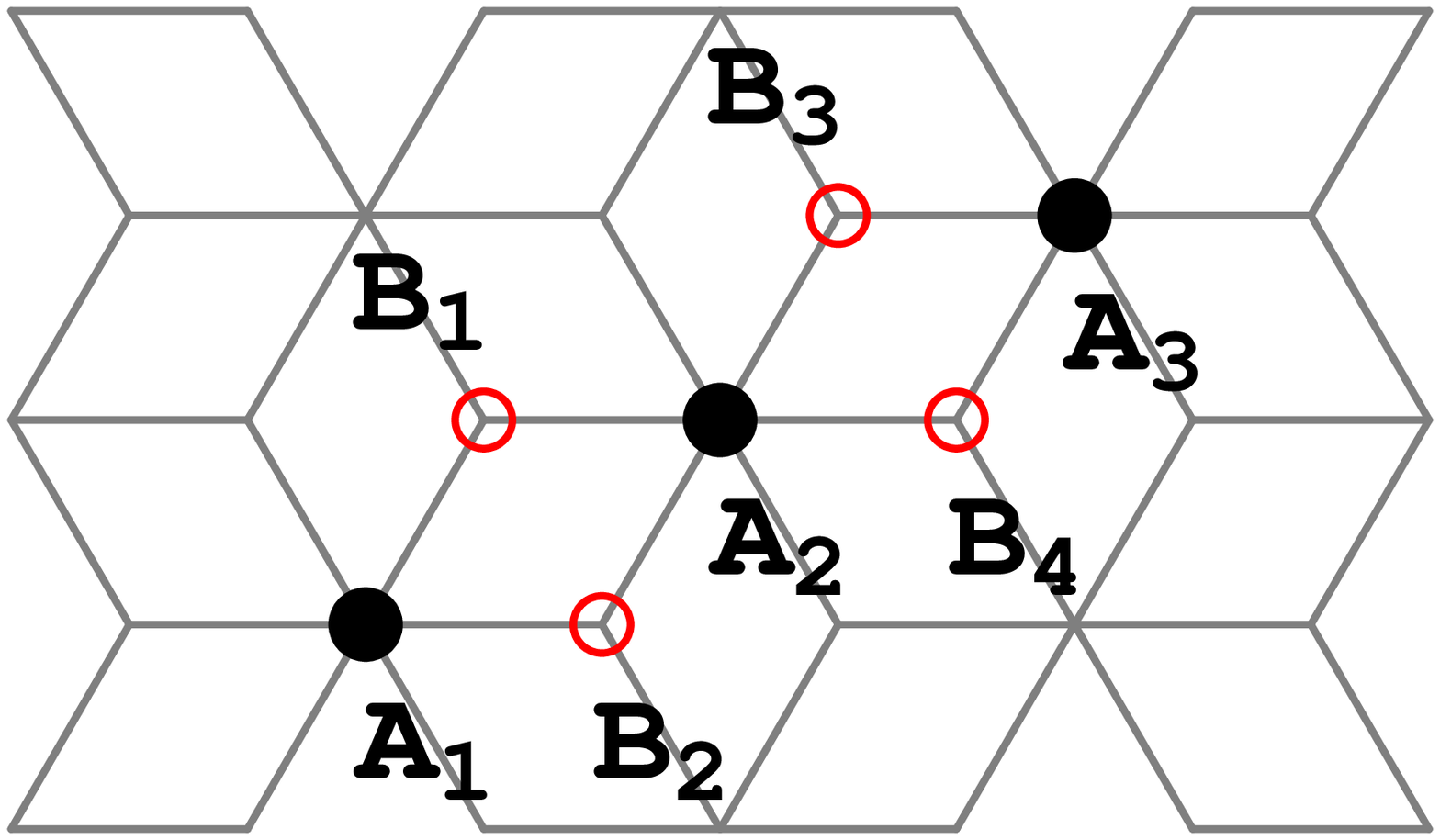}}
\subfigure[{}]{\includegraphics[width=1.3in]{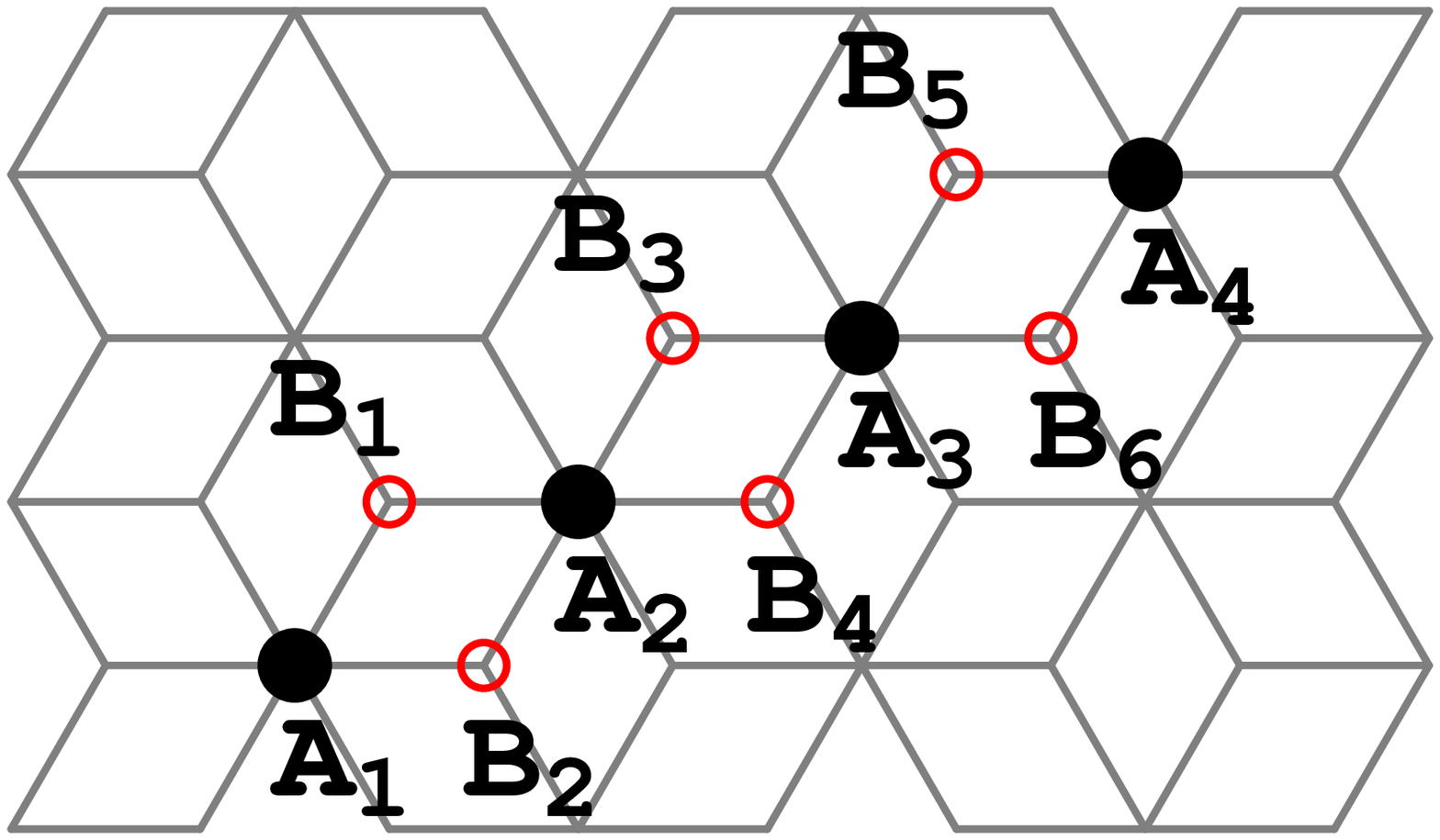}}
\subfigure[{}]{\includegraphics[width=0.7in]{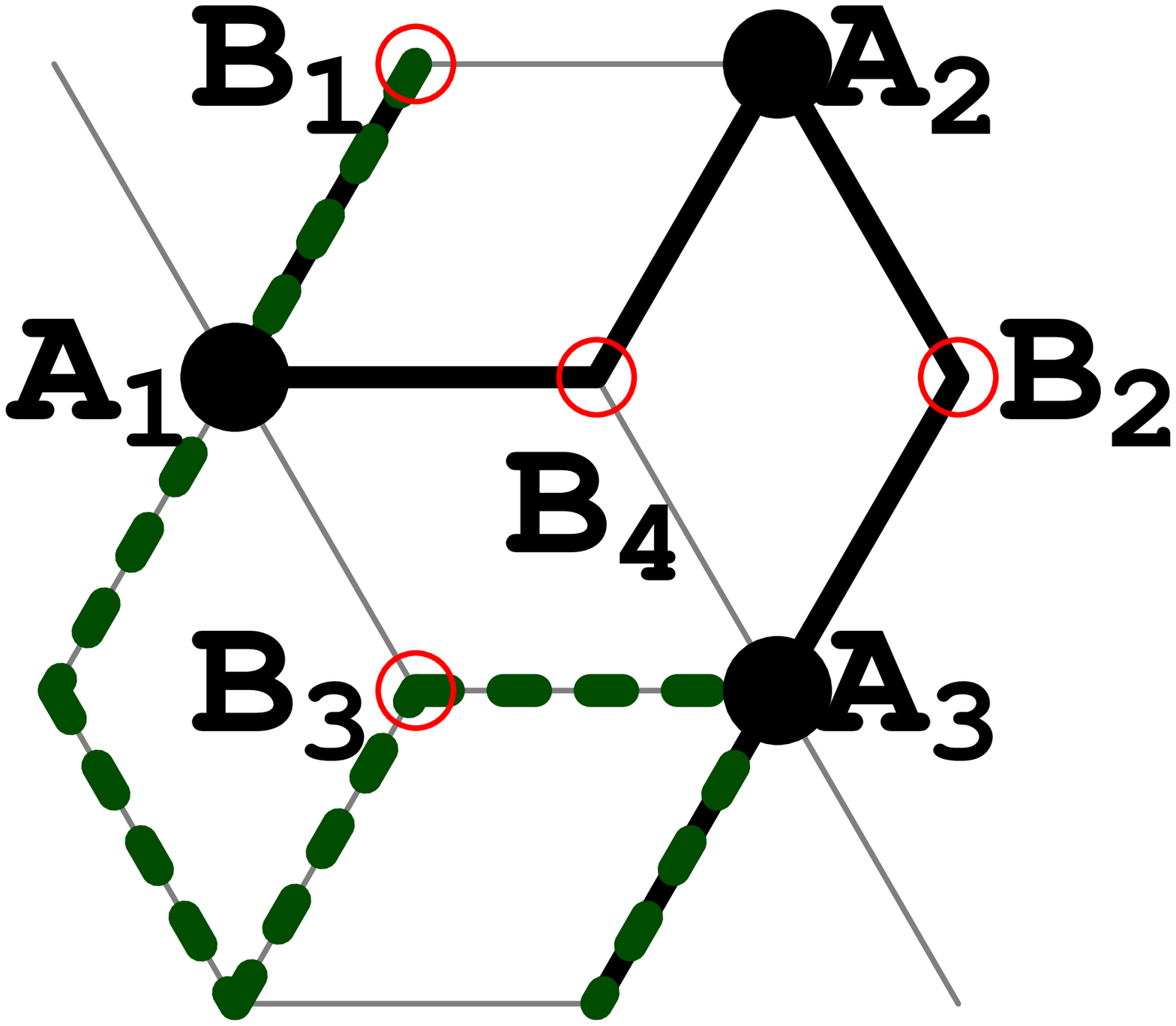}}
\vskip -2mm
\caption{\label{CancelPaths}If the paths contain only the marked 6-coordinated sites, they would have to contain also all of the marked 3-coordinated sites in order not to be cancelable (see the text).}
\end{figure}

For any given \emph{simple} path between some two \emph{distant} sites $P$ and $Q$, we can construct a path that cancels it out. In demonstrating this, we also explore what the sufficient distance between the $P$ and $Q$ is. (a) Consider first the simple paths which contain only the three 6-coordinated sites depicted in the Fig.~\ref{CancelPaths}a. If the path goes through the sites $A_1-B_1-A_2$, and never visits the site $B_2$, then it can be canceled by the path which goes through $A_1-B_2-A_2$, and continues beyond as the original one. The only way to avoid the path cancellation is to visit all the depicted $B$-sites, and this can be done only by choosing the path's end points among them (otherwise, more 6-coordinated sites would belong to the path). We see that in this case the path's end-points are the next-nearest-neighbors. (b) Next, consider the paths which contain the four 6-coordinated sites depicted in the Fig.~\ref{CancelPaths}b. For the same reason as before, all the $B$-sites must be included in the path, otherwise there will be another path which cancels it out. But this time, in attempting to do so we would have to introduce another 6-coordinated site into the path (the remaining neighbors of the depicted $B$-sites - analyzed as the case c), since the path-ends can take only two out of three $B$-sites left ``outside'' of the straight segment of the path. We can already see that all the simple paths with more than three 6-coordinated sites arranged in a chain cancel out. This chain may, of course, bend, but somewhat special situation occurs when a triplet of 6-coordinated sites $(A_1,A_2,A_3)$ of the path sits on the three touching plaquettes, as shown in the Fig.~\ref{CancelPaths}c. (c) Again, all the $B$-sites must be visited, or the path is canceled out, and one of them must be a path's end-point if no other 6-coordinated sites may belong to the path. Since all $B$-sites are now the next-nearest-neighbors, we may try to choose another 3-coordinated site further away as the path's end-point, and construct the path like the one plotted with the solid line in the Fig.~\ref{CancelPaths}c. But, again there is a path which cancels it out, and it is plotted with the dashed line. By trying to add more 6-coordinated sites to the case c) we would only have too many $B$-sites to cover as before. In conclusion, all the simple paths whose end-points are separated more than the next-nearest-neighbors must cancel out.

Many complex paths are also canceled - namely, whenever we can construct another path of the same topology, and between the same end-points, which encloses an odd number of plaquettes with the original one. However, for some paths this construction is not possible, and a different approach is needed in order to see whether they are canceled out or not. An example of such a path which appears at the $12^{\textrm{th}}$ order of the perturbation theory is shown in the Fig.~\ref{DicePaths}d. The complex paths between the sites further away than the next-nearest-neighbors appear only above the $6^{\textrm{th}}$ order, so that below this order the Green's function between distant sites (and arbitrary times) strictly vanishes, and the visons appear strictly localized.

At present, we cannot tell whether the vison localization persists to all orders. If not, they would have an extremely large inertia in the small-$h$ limit, and this would be a novel mechanism for creation of the large quasi-particle effective mass. The possibility of vison localization is also hinted by the Monte Carlo calculations presented in the Section \ref{MC}. Namely, the spatial vison-vison correlations effectively vanish at separations beyond the next-nearest neighbors. We would like to contrast this with another case of a fractionalized spin-liquid on the Kagome lattice: in a particular easy-axis limit explored in Ref.~\cite{KagEA} the vison correlations are short-ranged, but they exhibit a clear exponential decay over a significant range of separations.

Finally, we note that the phase obtained here is also the unconfined phase. The elementary singlet excitations are gapped visons, while the elementary spin-carrying excitations are gapped $S=1/2$ spinons.

\section{Monte-Carlo Results}\label{MC}

In order to gain some more insight into the frustrated transverse field Ising model, we attempted to study it numerically using the classical Monte Carlo technique. The quantum Ising model ~(\ref{ID}) translates to the 3-dimensional classical Ising model on the time-stacked dice-lattice layers: the corresponding Boltzmann weights are all positive and hence the model can be simulated by Monte Carlo without a sign problem. We reliably measured the spin-spin correlations in the disordered phase, and answered the question of whether the thermal fluctuations alone can yield order-from-disorder like the quantum fluctuations. However, we could not truly explore the complete phase diagram and the nature of the phase transition with this simple classical technique; for those purposes, a more sophisticated Monte Carlo technique is needed.

\begin{figure}
\includegraphics[width=2.1in]{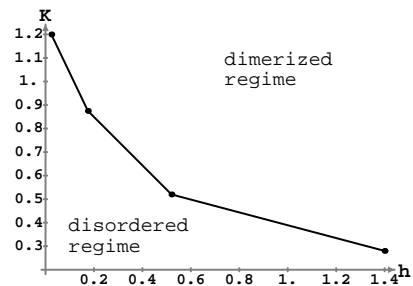}
\vskip -2mm
\caption{\label{Z2Regimes}The regimes of the (2+1)D classical Ising model on the fully frustrated dice lattice in ~(\ref{MCAction}). $\widetilde{K} \equiv \widetilde{K}_3 = \widetilde{K}_6$. The crossover line was obtained by joining the points $(\widetilde{h} , \widetilde{K})$ at which the heat-capacity is peaked for several fixed ratios $\widetilde{K} / \widetilde{h}$. In defining the heat-capacity $\widetilde{h}$ was treated as the inverse temperature $\beta$.}
\end{figure}

The model we study is given by the action analogous to ~(\ref{ID}):
\begin{eqnarray}\label{MCAction}
S & = & \sum_{\tau}\Bigl[-\widetilde{h}\sum_{lm}\epsilon_{lm} v_{l,\tau} v_{m,\tau} -
  \nonumber \\ & & -\widetilde{K}_3\sum_{l_3} v_{l_3,\tau} v_{l_3,\tau + 1}
  -\widetilde{K}_6\sum_{l_6} v_{l_6,\tau} v_{l_6,\tau + 1} \Bigr]
\end{eqnarray}
We have to regard this problem as a classical one, because the quantum limit $\widetilde{h} \to 0$, and $\widetilde{K}_{3,6} \to \infty$ is not accessible. If $\widetilde{K}_3 = \widetilde{K}_6 = \widetilde{K}$, two regimes emerge, depicted in the Fig.~\ref{Z2Regimes}. In order to obtain this diagram, we scan along $\widetilde{K} \propto \widetilde{h}$ lines, and measure the heat-capacity, using $h$ as the inverse ``temperature'' $\beta$; then we record position of the heat-capacity peak, and join the peak positions $(\widetilde{h} , \widetilde{K})$ from different scans into a crossover line between the two regimes.

\begin{figure}
\subfigure[{}]{\includegraphics[width=2.4in]{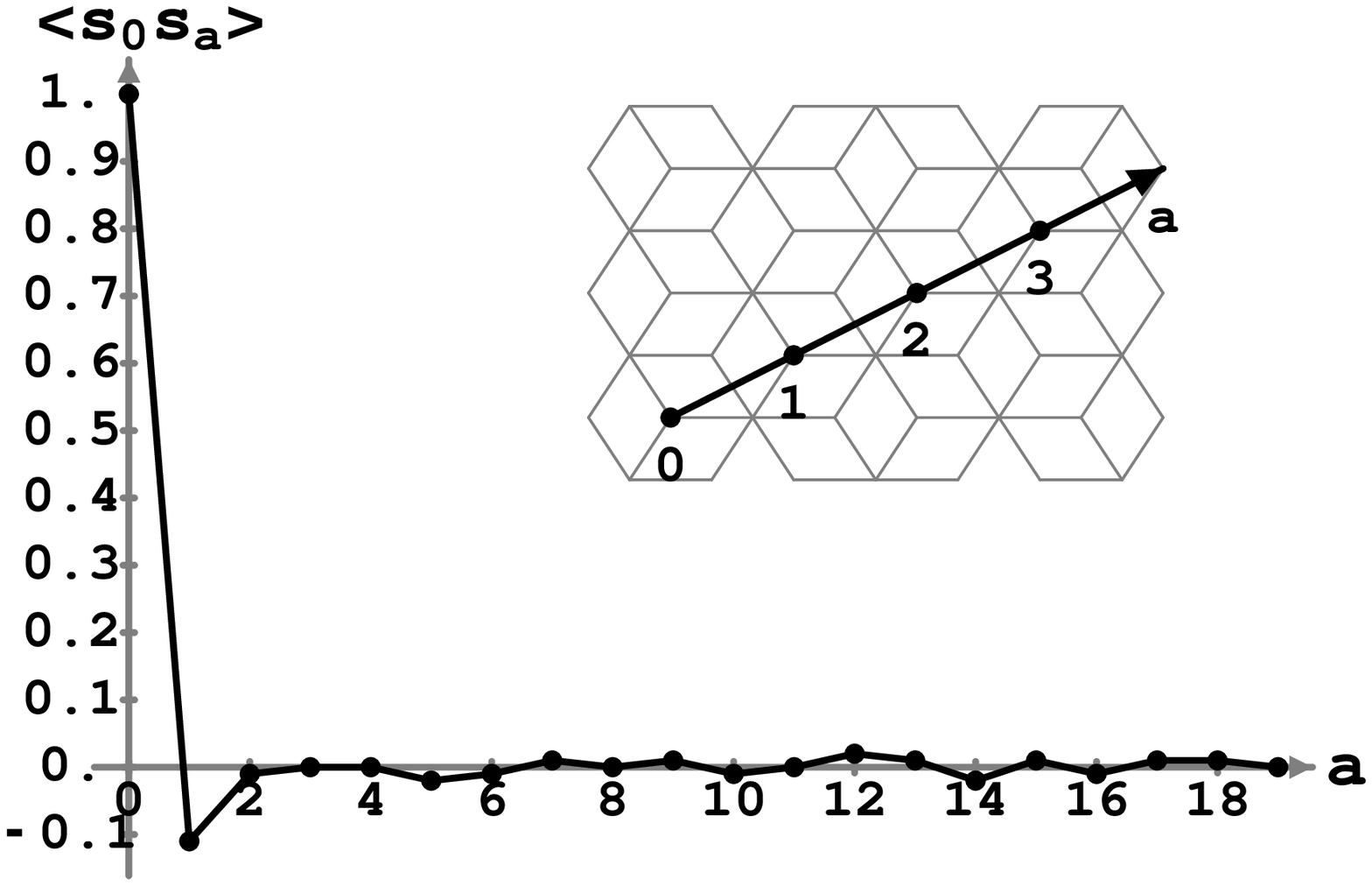}}
\vskip -2.3mm
\subfigure[{}]{\includegraphics[width=2.4in]{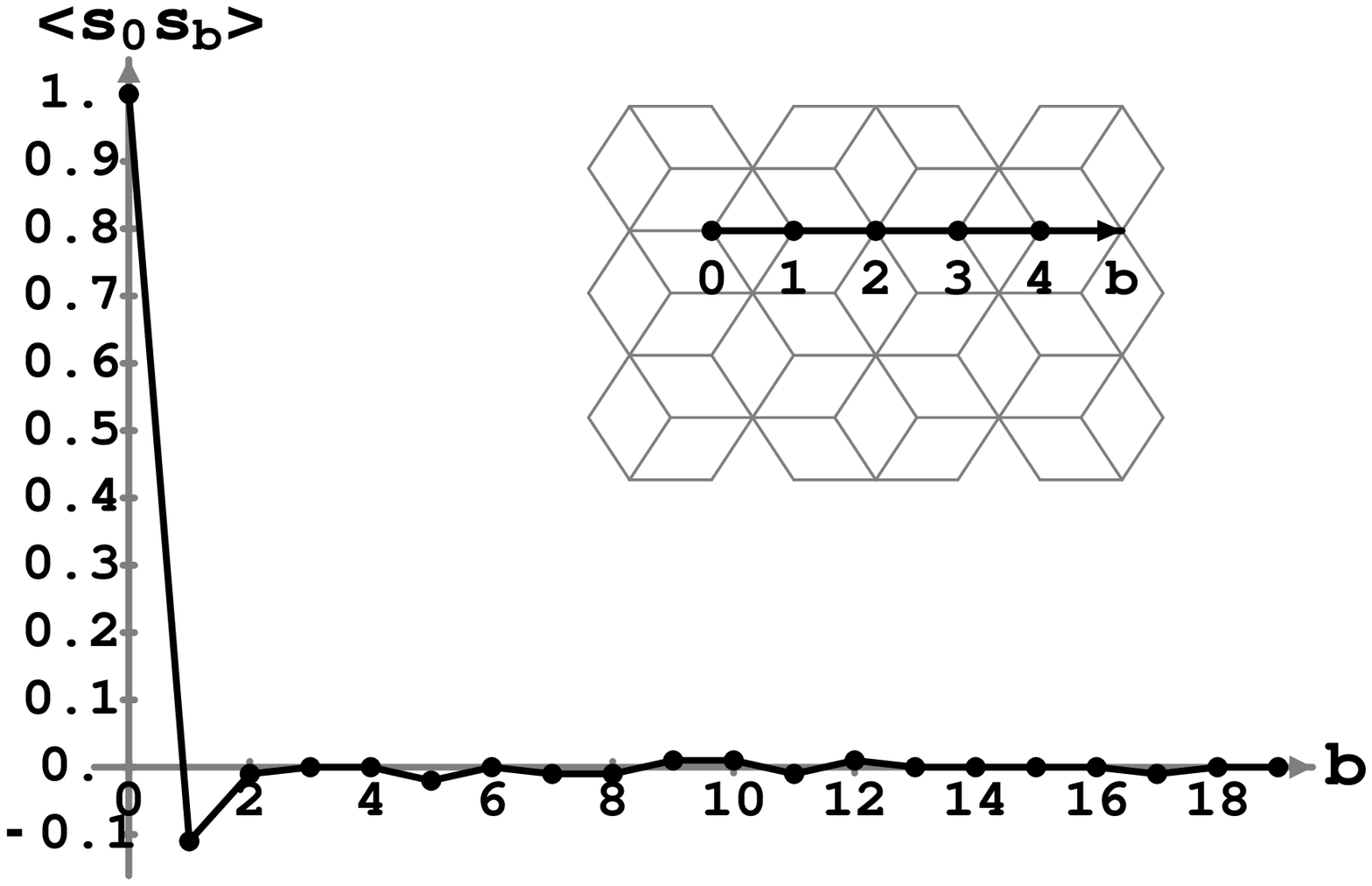}}
\vskip -2.3mm
\subfigure[{}]{\includegraphics[width=2.4in]{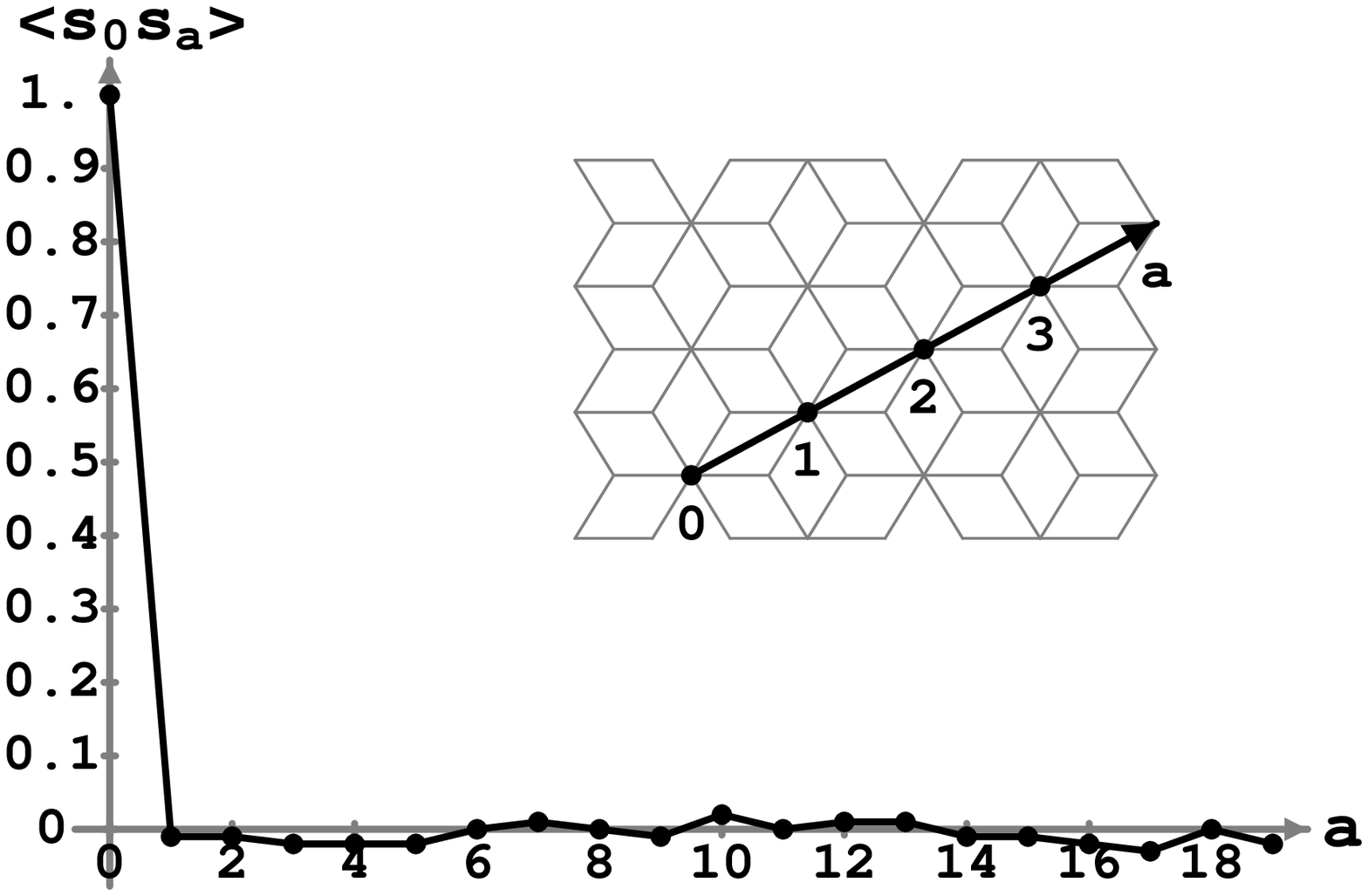}}
\vskip -2.3mm
\subfigure[{}]{\includegraphics[width=2.4in]{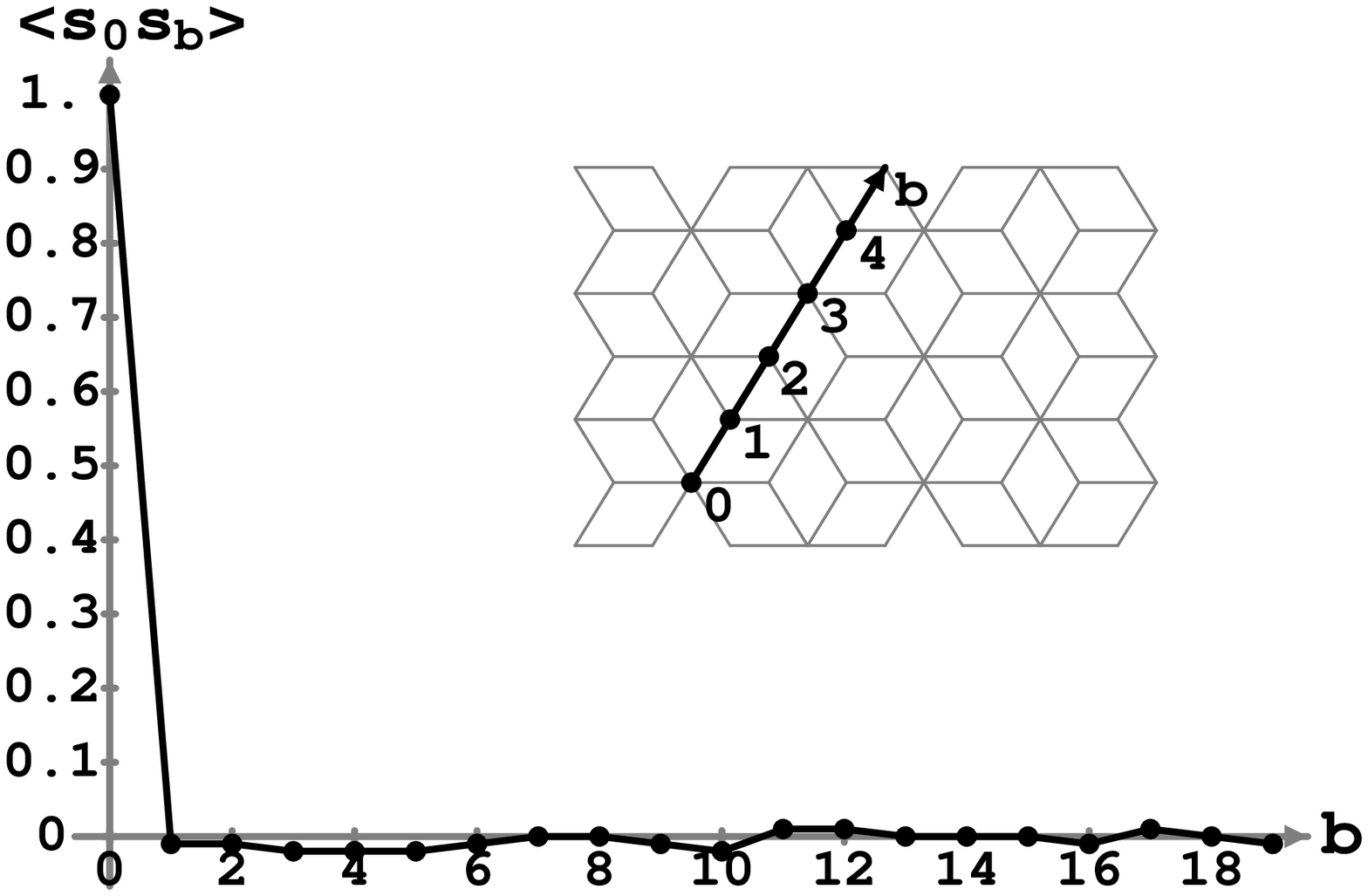}}
\vskip -2.3mm
\caption{\label{MCCorr}Monte-Carlo measured spatial spin-spin correlations in the disordered phase. The sample size was $20 \times 18 \times 20$ unit-cells (each with 6 sites), with $\widetilde{h} = 0.15$, and $\widetilde{K}_3 = \widetilde{K}_6 = 0.75$. The type of the site at the origin, and the spatial direction are shown in the legend.}
\end{figure}

For small $\widetilde{h}$ and $\widetilde{K}$ a disordered regime occurs. It is characterized by seemingly vanishing spatial spin-spin correlations beyond the nearest-neighbor sites, as shown in Fig.~\ref{MCCorr}, while the time correlations decay exponentially. This regime corresponds to the disordered phase and the small-$h$ limit of ~(\ref{ID}). The numerical result for the spatial correlations is consistent with the statement made in the Section \ref{smallH} that the visons may be localized in the vicinity of the next-nearest-neighbor sites.

For large $\widetilde{h}$ and $\widetilde{K}$, a ``dimerized'' regime occurs: if the frustrated dice bonds (with positive bond energy) are plotted as dimers on the Kagome lattice (wherever a Kagome bond intersects a frustrated dice bond), then a hard-core dimer covering is obtained. Our present method was not able to establish whether the crossover line between these two regimes is actually a phase transition line or not. However, based on our analytical study we expect that this is the phase transition. Since large coupling constants yield the dimerized regime in which all dimer coverings have the same action, the remaining phase properties can be decided only by entropy, which is a manifestation of quantum fluctuations. In the Ising language, once the action is fixed to its lowest value, the maximum entropy state has the largest number of \emph{flippable} clusters of spins (free to fluctuate without changing the action). In the dimer language, the \emph{honeycomb} pattern will be selected, in a mechanism essentially analogous to that obtained in the Section \ref{largeH}. Indeed, every perfect hexagon of dimers represents a \emph{flippable} spin on the dice-lattice. Once their number is maximized by the entropy, the only other elementary flippable loops which could possibly resonate without action cost are the \emph{stars}, and to maximize their number, the perfect hexagons must arrange in the \emph{honeycomb} pattern. Unfortunately, our simulation could not reach equilibrium in this regime, at least due to very slow dynamics at large coupling constants.

Since the quantum fluctuations are capable of producing the order-from-disorder, it is natural to ask whether the thermal fluctuations are capable too. If we set the transverse field couplings in ~(\ref{ID}) to zero, we obtain the classical two-dimensional Ising model on the fully frustrated dice lattice. In this case, the Monte Carlo easily reveals that there are no phase transitions, measured through the ``heat capacity'', and the model is always in its disordered phase. Therefore, the thermal fluctuations alone cannot introduce a long-range order, even when $\widetilde{h}$ becomes large and the ``dimerized'' regime occurs. The presence of the transverse fields is crucial for the entropical selection of a long-range ordered state in the Monte Carlo.

\section{Discussion: Implications for the Kagome quantum Heisenberg magnet}\label{dis}

The analysis in the preceding sections has established some properties of the fully frustrated transverse field Ising model on the dice lattice. Specifically, we analyzed the two limits of small and large transverse fields $K_n$ in ~(\ref{ID}), and described the two distinct phases that obtain in those limits. In this Section, we consider the lessons for the original Kagome Heisenberg magnet. The disordered phase of the Ising model ($h \ll K_n$) has a unique ground-state and a clear gap in its spectrum. This will describe spin liquid phases of the original Kagome magnet. On the other hand, the \emph{honeycomb} dimer phase ($h \gg K_n$) breaks translational symmetry, but with a very large unit cell. This phase has a number of interesting properties. First it clearly describes a valence bond crystal phase of the original Kagome magnet with a large unit cell. The spinful excitations will be gapped spin-$1$ magnons. However, below the spin gap this state has a large number of low-energy excitations with a very small but non-zero gap. The number of low-lying excitations that we can account for is exponential in the system size. However, their exact number below the spin-gap is hard to estimate in our approach. The numerical calculations \cite{KNum} on the Kagome Heisenberg magnet revealed a large number of seemingly gapless singlet excitations in the 36-site sample of the Kagome quantum magnet, and a finite-size scaling analysis suggested that in the thermodynamic limit the spectrum could be gapless. We suggest that the actual ground state in the thermodynamic limit displays the honeycomb dimer ordering present in the dual Ising model at small transverse fields. At the end of Section \ref{largeH} we argued that our suggestion is consistent with the numerical data. This ground-state breaks translational symmetry with a unit-cell containing 36 sites, while 36 sites was the largest sample size that could be studied by the exact diagonalizations so far. Therefore, the numerics might suffer from some serious finite-size effects. In particular, the large unit-cell dimer ordering, if it exists, will be missed in the current numerical calculations, as well as the spinon confinement into magnons. However, some gross features of the low-lying spectrum, namely the presence of a large number of singlet excitations below the spin-gap, are not too sensitive to the finite sample sizes used in the numerical studies. (Whether or not these singlet excitations have a true gap or are gapless in the thermodynamic limit is of course more sensitive to finite size). We emphasize that the specific valence bond crystal state proposed in this paper is in a confined phase where there are no spinon excitations. This is in contrast to naive expectations based on the apparent translation symmetric ground state obtained in the numerical calculations.

Quite generally, we expect that analysis similar to that presented in this paper is applicable to a variety of spin models in paramagnetic phases. These include the checkerboard lattice, and various other lattices frustrated by the next-nearest neighbor exchange. Then, if we were to generalize our hypothesis to other models with near neighbor exchange interactions, we would expect that the $h \gg K_n$ limit always describes the effective theory for singlet excitations. This theory is a quantum dimer model dominated by the kinetic energy. A naive guess for the ground-state is that the smallest resonating flippable loops assume the maximum possible density, typically leading to a \emph{plaquette} long range order. Indeed, such \emph{plaquette} phases have been observed in the Heisenberg model on the checkerboard lattice \cite{CbNum}, and the phase diagrams of the quantum dimer model on the square \cite{SPlaq}, and honeycomb \cite{HcDimer} lattices, while a dimer RVB phase was found on the triangular lattice \cite{TDimer}. In this paper we demonstrated analytically a plaquette phase in the Kagome lattice Heisenberg model. The peculiarity of the Kagome lattice is that the maximum density of smallest resonating flippable loops is not sufficient to lift the macroscopic degeneracy of the ground-states. This is why the singlet gap is extremely small. In other lattices, the checkerboard for example, the degeneracy is easily lifted, and the singlet gap is fairly significant.

One of the remaining fundamental questions is whether the spin-liquids with gapless singlet excitations (Type II) really exist in the quantum spin models, as suggested in the Ref.~\cite{Lhlong,Lhshort}. The only known examples which numerically exhibit such properties are the Kagome lattice Heisenberg model \cite{KNum}, and the triangular lattice antiferromagnet with ring-exchange \cite{TNum}. In the latter case, a small ring-exchange coupling destroys the Neel-order and creates a spin-liquid, while at a slightly larger coupling a singlet-gap is gradually opened. Armed with our experience on the Kagome lattice, we speculate that the case of the triangular lattice is also not a Type II spin-liquid with gapless singlet states. Probably this phase will not be found in other quantum spin models either.

Another interesting feature that emerged from our calculations is a rather small (possibly zero) dispersion of the excitations in either phase on the dice lattice. The elementary excitations of both the ordered and disordered phases appear dispersionless at least to the sixth order of perturbation theory. Although at present we don't know whether this persists to all orders, we speculate that it might be the case, at least in the disordered phase. Indeed, a simple Gaussian Landau-Ginzburg analysis gives a dispersionless spectrum. In the Appendix \ref{LGAp} we study the $O(N)$ generalization of the Ising model on the fully frustrated dice lattice, and recover a  dispersionless spectrum in the large-$N$ limit. In addition, we show that in this approximation, the model is always in the paramagnetic phase. It is clear that both the Landau-Ginzburg approach and the large-$N$ approximation have fundamental limitations in this problem. As we have argued, the correct phase diagram includes an ordered phase, albeit of a somewhat unconventional kind.

\section{Conclusion}\label{concl}

In this paper we have studied various aspects of the fully frustrated transverse field Ising model on a dice lattice. This model is useful as a description of the physics of singlet excitations below the spin gap of the quantum Kagome Heisenberg antiferromagnet. We argued that this transverse field Ising model supports at least two quantum phases: a \emph{plaquette} phase where the ground state breaks both global spin flip and translational symmetries, and a disordered phase with short-ranged (Ising) spin correlations. In terms of the original Kagome Heisenberg magnet, the former describes a state with a \emph{honeycomb} pattern of resonating benzene-like arrangements of singlet bonds on the hexagonal plaquettes, with a 36-site unit-cell. The elementary singlet excitations are gapped with a very small gap and localized inside the cells of the honeycomb structure. Based on comparison with the numerical studies, we suggested that this specific large-unit-cell valence bond crystalline order describes the ground state of the Kagome Heisenberg quantum magnet. The elementary spin-carrying excitations are gapped spin-$1$ magnons. We hope that our work stimulates attempts to look for this ordering pattern in future studies of the Kagome magnet. 

In addition, we argued that due to frustration and unusual geometry the fully frustrated dice lattice transverse field Ising model possesses a strong tendency to localize its elementary excitations: they are either completely localized, or they have an extremely large effective mass (narrow energy bands).

We are grateful to O.~I.~Motrunich for several useful discussions and comments on the manuscript. This work was supported by the MRSEC Program of the NSF under grant DMR-0213282, and by a grant from the NEC Corporation. TS also acknowledges support from the Alfred P. Sloan Foundation and the hospitality of Harvard University where part of this work was done.

\appendix

\section{\z2 Gauge description of the Kagome Quantum Magnet}\label{z2Ap}

Here we briefly derive a representation of 
 the Kagome Heisenberg quantum magnet as a \z2 gauge theory. For more in-depth discussion, see references \cite{Z2a,Z2h,Z2t}. Our starting point is the Heisenberg model on the Kagome lattice:
\begin{equation}\label{HK}
H = J \sum_{\la ij \ra} \boldsymbol{S}_i \cdot \boldsymbol{S}_j
\end{equation}
where the sum runs over the nearest-neighbor sites. The Heisenberg model can be viewed as a special limit of another model. Consider a theory of hopping spinons $f_{\alpha i}$ coupled to the \z2 gauge field $\sigma^z_{ij}$ which lives on the bonds of the Kagome lattice:
\begin{eqnarray}\label{SZK}
H & = & -h_0 \sum_{\la ij \ra} \sigma^x_{ij} -
\sum_{\la ij \ra} \sigma^z_{ij} \Bigl[ t_{ij}
\sum_{\alpha = \uparrow \downarrow}
   (f^{\dagger}_{\alpha i} f^{\phantom{\dagger}}_{\alpha j} + h.c.) +
\nonumber \\ & & {} + 
\Delta_{ij} (f^{\dagger}_{\uparrow i} f^{\dagger}_{\downarrow j} -
f^{\dagger}_{\downarrow i} f^{\dagger}_{\uparrow j} + h.c.) \Bigr]
\end{eqnarray}
The spinon operators obey fermion anticommutation relations, and the gauge-field operators are Pauli matrices. This theory possesses a local \z2 gauge symmetry: if the signs of $f_{\alpha i}$ at a particular site $i$ and $\sigma^z_{ii'}$ on all bonds emanating from that site are simultaneously changed, the Hamiltonian ~(\ref{SZK}) remains invariant. The appearance of the local symmetry means that the Hilbert phase is larger than that of the Heisenberg model, and a constraint per site is needed to project back to the physical Hilbert space. This is achieved through a gauge requirement on the physical states:
\begin{equation}\label{PhysGauge}
G_i = \prod_{i' \in i} \sigma^x_{ii'} (-1)^{\sum \limits_{\alpha = \uparrow \downarrow} 
f^{\dagger}_{\alpha i} f^{\phantom{\dagger}}_{\alpha i}} = -1
\end{equation}
The operators $G_i$ are generators of the local gauge transformations at sites $i$, and the product is over all bonds $\la ii' \ra$ emanating from the site $i$. In the $h_0 \rightarrow \infty$ limit of ~(\ref{SZK}), all gauge spins tend to align in the x-direction, so that the gauge requirement ~(\ref{PhysGauge}) fixes the number of spinons to be one-per-site. The spinon gauge theory ~(\ref{SZK}) then reduces to the Heisenberg model at the second order of perturbation theory, with $J \sim t_{ij}^2/h_0, \Delta_{ij}^2/h_0$.

The next step assumes that the spin-carrying excitations in the original Heisenberg model are gapped. We rely on the experimental and numerical studies which show such a gap in the spectrum of the Kagome magnet. Thus, the spinon degrees of freedom can be ``integrated-out'' in ~(\ref{SZK}), leaving behind a pure $\mathbb{Z}_2$ effective gauge theory with local terms (we will come back to this step later):
\begin{eqnarray}\label{ZK}
H' & = & -h \sum_{\la ij \ra} \sigma^x_{ij} 
- K_3 \sum_{\triangle} \prod_{\triangle} \sigma^z_{ij} 
- K_6 \sum_{\hexagon} \prod_{\hexagon} \sigma^z_{ij} -
\nonumber \\& & {} - 
K_{3+3} \sum_{\bowtie} \prod_{\bowtie} \sigma^z_{ij} - \cdots
\end{eqnarray}
The terms involving $\sigma^z$ operators are the sums over all plaquette products. The elementary plaquettes appear at the lowest order, but in principle, all connected clusters of bond-loops can appear at higher orders. This form of the effective Hamiltonian is required by the \z2 gauge symmetry: the sign change of $\sigma^z_{ii'}$ on all bonds emanating from any site $i$ must leave the Hamiltonian invariant. The spin-gap results with a local and perturbative nature of ~(\ref{ZK}): the coupling constants $K_n$ of the $n$-bond clusters are of the order of $\epsilon x^n$, where $\epsilon$ is some energy scale, and $x \ll 1$. The spinon ``integration'' also modifies the gauge requirement ~(\ref{PhysGauge}) to:
\begin{equation}\label{PhysGaugeZ}
G'_i = \prod_{i' \in i} \sigma^x_{ii'} = -1
\end{equation}
The expressions ~(\ref{ZK}) and ~(\ref{PhysGaugeZ}) define a pure \z2 gauge theory on the Kagome lattice. We now express this theory in its dual form. Let $v^{x,y,z}_l$ be the Pauli operators defined on the sites of the lattice dual to Kagome, the dice lattice. Their relation to the \z2 gauge field of the Kagome lattice is the following:
\begin{eqnarray}
\sigma^x_{ij} = \epsilon_{lm} v^z_l v^z_m \label{VisonZ} \\
\prod_{\textrm{\emph{plaq.}}} \sigma^z_{ij} = v^x_l \label{VisonX}
\end{eqnarray}
In ~(\ref{VisonZ}) the dual dice bond $<\!lm\!>$ intersects the Kagome bond $<\!\!ij\!\!>$, while in ~(\ref{VisonX}) the dual dice site $l$ sits inside the Kagome elementary plaquette appearing in the product on the left-hand side. The numbers $\epsilon_{lm}$ are fixed to $+1$ and $-1$ in such a way that on every dice elementary plaquette the condition of full frustration holds:
\begin{equation}\label{Frust2}
\prod_{\diamondsuit} \epsilon_{lm} = -1
\end{equation}
This is the consequence of the gauge requirement ~(\ref{PhysGaugeZ}). One choice of sign-arrangement for $\epsilon_{lm}$ is depicted in Fig.~\ref{KDGrid}. The pure \z2 gauge theory ~(\ref{ZK}) can now be rewritten as the transverse field quantum Ising model on the fully frustrated dice lattice:
\begin{eqnarray}\label{ID2}
H' & = & -h \sum_{\la lm \ra} \epsilon_{lm} v^z_l v^z_m - K_3 \sum_{l_3} v^x_{l_3} -
K_6 \sum_{l_6} v^x_{l_6} -
\nonumber \\& & {} - 
K_{3+3} \sum_{(l_3 m_3)} v^x_{l_3} v^x_{m_3} - \cdots
\end{eqnarray}
The ``kinetic energy'' terms contain in general the products of $v^x_l$ on various clusters of the dice lattice sites. 

The Hamiltonian ~(\ref{ID2}) captures only the form of the effective theory below the spin gap. The coupling constants are not known, because the path-integration of spinons in ~(\ref{SZK}) cannot be carried out analytically. However, we can at least attempt to estimate their relative values. For these purposes, we express the original Heisenberg model in the slave-particle path-integral, as it was done in the Reference \cite{Z2a}. The action involves the spinon Grassmann field $f_{\alpha i}$ living on the sites $i$ of the (2+1)D lattice, and the gauge Ising-like field $\sigma_{ij}$ living on the lattice bonds:
\begin{eqnarray}\label{Z2Action}
S & = & -\sum_{\la ij \ra} \sigma_{ij} \Bigl[
  \widetilde{t}_{ij} (\overline{f}_{\alpha i}f_{\alpha j} + c.c.) +  \\
  & & + \widetilde{\Delta}_{ij} (\overline{f}_{\uparrow i}\overline{f}_{\downarrow j} -
     \overline{f}_{\downarrow i}\overline{f}_{\uparrow j} + c.c.) \Bigr] -
  \sum_i \overline{f}_{\alpha i}f_{\alpha i} + S_B \nonumber
\end{eqnarray}
where the Berry's phase $S_B$, realizing the projection to the physical Hilbert space,  is given by:
\begin{equation}\label{Z2Berry}
e^{-S_B} = \prod_{i,j=i-\widehat{\tau}} \sigma_{ij}
\end{equation}
The action ~(\ref{Z2Action}) is an exact rewriting of the Heisenberg model in the limit $ \widetilde{t}_{ij}, \widetilde{\Delta}_{ij} \ll 1 $ \cite{Z2a}. We obtain it upon integrating out $\sigma_{ij}$, and $ J\Delta\tau \sim (\widetilde{t}_{ij}^2, \widetilde{\Delta}_{ij}^2) $, where $\Delta\tau \to 0$ is the imaginary-time lattice spacing used in the path-integral (of the Heisenberg model). We can relate the action coupling constants $\widetilde{t}_{ij}$ and $\widetilde{\Delta}_{ij}$ to the Hamiltonian ~(\ref{SZK}) coupling constants, $t_{ij}$ and $\Delta_{ij}$. In order to simplify notation, from now on we will use one symbol $\widetilde{t}$ to represent all the couplings in the action, and $t$ for all the couplings in the Hamiltonian. The connection is:
\begin{equation}\label{CConst1}
\widetilde{t}_{ij} \sim \bigg \lbrace
\begin{array}{l@{,}l}
$1 \ $
  & \textrm{$\ $ on the temporal links $\la ij \ra$} \\
  t\delta $\ $  & 
  \textrm{$\ $ on the spatial links $\la ij \ra$}
\end{array} \nonumber
\end{equation}
where $\delta = \sqrt{\Delta\tau / h_0} \to 0$. Now we integrate out the spinon field, and obtain a pure \z2 gauge theory with the action: 
\begin{eqnarray}\label{PureZ2Action}
S & = & -\log\det A(\sigma_{ij};\widetilde{t}_{ij}) + S_B 
  \nonumber \\
  & = & - \sum_{\Box}\widetilde{K}_{\Box}\prod_{\Box}\sigma_{ij} + S_B
\end{eqnarray}
The matrix $A$ above is the matrix which couples the spinons in ~(\ref{Z2Action}). In the expanded form, this action must contain various gauge-invariant products of $\sigma_{ij}$ on the closed loops, with the loop-dependent coupling constants $\widetilde{K}_{\Box}$. The very assumption that this expansion is possible and convergent is an essential one, and it requires a large spin-gap (of the order of $J$), determined at a microscopic scale. A large spin-gap guaranties that the effective theory will have only local terms, and significance of various closed loops in ~(\ref{PureZ2Action}) will rapidly decrease with increasing loop size, making the expansion convergent. At this point we are not providing any formal estimate of the spin-gap. We simply use some external sources to obtain information about it, for example the numerics.

Using the fact that a $\sigma_{ij}$ factor always comes together with a $\widetilde{t}_{ij}$ factor in ~(\ref{Z2Action}), we can make an estimate for the coupling constants $\widetilde{K}_{\Box}$ (to the lowest order):
\begin{equation}\label{CConst2}
\widetilde{K}_{\Box} \sim \prod_{\Box} \widetilde{t}_{ij}
\end{equation}
and this implies that $\widetilde{K}_{\Box}$ can be treated as small numbers, scaling as $\delta^n$ with $n$ being the number of spatial links in the loop. In order to formally derive the connection between this action and the Hamiltonians ~(\ref{ZK}) and ~(\ref{ID2}), we neglect at this point all the higher order loop terms (enclosing more than one elementary plaquette). If the action contains only the elementary plaquettes, we can readily construct the \emph{dual} theory, as outlined in \cite{Z2a}. It takes the form of the Ising model on the fully frustrated dual lattice:
\begin{equation}\label{DualAction}
S = -\sum_{\la lm \ra} \widetilde{K}_{lm}\epsilon_{lm} v_l v_m
\end{equation}
The Ising gauge-field $\epsilon_{lm}$ living on the links of the dual lattice is \emph{frozen} such that its product over any dual elementary plaquette is $-1$. The only fluctuating field is the vison Ising field $v_l$. The duality transformation establishes connection between the coupling constants $\widetilde{K}_{\Box}$ defined for a plaquette and $\widetilde{K}_{lm}$ defined for the dual bond piercing that plaquette:
\begin{equation}\label{Duality}
\tanh\widetilde{K}_{\Box} = e^{-2\widetilde{K}_{lm}} \quad , \quad
\tanh\widetilde{K}_{lm} = e^{-2\widetilde{K}_{\Box}}
\end{equation}
For simplicity, assume that there is only one kind of elementary plaquettes. Then, the  Hamiltonian describing the dual theory has the form:
\begin{equation}\label{DualHamiltonian}
H = -h \sum_{\la lm \ra} \epsilon_{lm} v_i^z v_j^z - K \sum_{l} v_l^x
\end{equation}
The connection between the action ~(\ref{DualAction}) and Hamiltonian above is:
\begin{equation}\label{ActHam}
\begin{array}{l@{,}l}
K\Delta\tau = e^{-2\widetilde{K}_{lm}} $\ $
  & \textrm{$\ $ on the temporal dual links $\la lm \ra$} \\
h\Delta\tau = \widetilde{K}_{lm} $\ $  
  & \textrm{$\ $ on the spatial dual links $\la lm \ra$}
\end{array} \nonumber
\end{equation}
Now we can estimate the values of the Hamiltonian coupling constants. The temporal elementary plaquettes always have 4 bonds (two spatial and two temporal), while we assume that the spatial elementary plaquettes have $n$ bonds. It follows that:
\begin{eqnarray}\label{kdt}
K\Delta\tau & = & e^{-2\widetilde{K}_{lm}\textrm{(dual temporal link)}} \nonumber \\
  & = & \tanh \widetilde{K}_{\Box}\textrm{(spatial plaq.)} \\
  & & \xrightarrow{\Delta\tau \to 0} t^n 
  \bigl( \frac{\Delta\tau}{h_0} \bigr)^{n/2} \nonumber
\end{eqnarray}
\begin{eqnarray}\label{hdt}
h\Delta\tau & = & \widetilde{K}_{lm}\textrm{(dual spatial link)} = \nonumber \\
  & = & -\frac{1}{2}\log\tanh\widetilde{K}_{\Box}\textrm{(temporal plaq.)} \\
  & & \xrightarrow{\Delta\tau \to 0} 
  -\log\bigl[ t \sqrt{\Delta\tau / h_0} \bigr] \nonumber
\end{eqnarray}
Taking the ratio of ~(\ref{kdt}) and ~(\ref{hdt}), and noting that the Heisenberg model exchange coupling is $ J \sim t^2/h_0 $, we obtain:
\begin{equation}\label{KoverH}
\frac{K}{h} = \frac{2(J\Delta\tau)^{n/2}}{\vert \log (J\Delta\tau)\vert}
\end{equation}
The energy scale of $J$ is also a measure of the spin-gap. Since the spinons have to be integrated out, the starting path-integral must accurately represent the energy scales above the spin-gap. The energy cut-off must be much larger than the spin-gap, and hence $ J\Delta\tau \ll 1 $. This is also compatible with request $ \widetilde{t}_{ij}, \widetilde{\Delta}_{ij} \ll 1 $, needed for the path-integral with action ~(\ref{Z2Action}) to reduce to the Heisenberg model. Consequently, $h \gg K$ (for arbitrary $n \geqslant 0$).

This simple analysis suggests that the large-$h$ limit of ~(\ref{ID2}) is a good effective theory for the singlet physics below the spin-gap. It also seems natural to expect from ~(\ref{KoverH}) that the higher-order loop terms, present in the more accurate effective Hamiltonian, have the coupling constants $K_n$ which decay with the loop length $n$ as $K_n \propto x^n$, where $x=\sqrt{J\Delta\tau} \ll 1$.

And finally, we can ask whether this approach is useful for extensions of the Heisenberg model. The next-nearest and further neighbor exchange interactions are reflected in the action ~(\ref{Z2Action}) as additional spinon hopping and pairing terms mediated by the \z2 gauge field. The spinons can still be integrated out (provided that there is a large spin-gap), and the pure \z2 gauge theory ~(\ref{PureZ2Action}) would effectively live on the more complicated lattice which has bonds between all pairs of sites that host an exchange coupling. The dual theory would, in general, be more complicated than the frustrated Ising model (the effective lattice may not even be planar), but we could still write an effective Hamiltonian like ~(\ref{ZK}) together with the gauge condition ~(\ref{PhysGaugeZ}), and expect a large-$h$ limit to be realized. Similarly, the ring exchange interactions can be implemented in the action ~(\ref{Z2Action}) by explicitly adding the loop products of the gauge field to it. This would yield larger values for the loop couplings $K_n$ in the pure \z2 gauge theory, possibly making them comparable with $h$. Having in mind the results of this paper in the small-$h$ limit, this hints that the multiple-spin exchange is a good candidate to yield truly fractionalized spin liquids.

\section{$O(N)$ Large-$N$ Limit}\label{LGAp}

A relatively simple way to gain some first intuition for the quantum Ising model on the frustrated dice lattice ~(\ref{ID}) is to generalize it to a limit where the semiclassical approximations become exact. Consider the $O(N)$ generalization of spins in the large-$N$ limit. We will demonstrate that this Gaussian theory has a dispersionless spectrum, and supports only a disordered phase for all values of the coupling constants.

\begin{figure}
\includegraphics[width=1in]{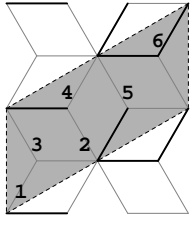}
\caption{\label{FrustDiceCell}A choice of the unit cell on the frustrated dice lattice. The site labeling corresponds to the internal indices of the matrix in Table \ref{HMatrix}.}
\end{figure}

We write the imaginary time path-integral and impose the spin-magnitude constraint using a field $\lambda$ of Lagrange multipliers. The action obtained from ~(\ref{ID}) by keeping only the elementary transverse field terms is:
\begin{eqnarray}\label{ONAction}
S_{\lambda} & = & \int \dd \tau \Bigl\lbrace
  -h \sum_{\la lm \ra} \epsilon_{lm} \boldsymbol{S}_{l\tau} \boldsymbol{S}_{m\tau}
  - \frac{K_3}{2} \sum_{l_3} \boldsymbol{S}_{l_3\tau} 
  \frac{\partial^2\boldsymbol{S}_{l_3\tau}}{\partial\tau^2} \nonumber \\
& & - \frac{K_6}{2} \sum_{l_6} \boldsymbol{S}_{l_6\tau} 
  \frac{\partial^2\boldsymbol{S}_{l_6\tau}}{\partial\tau^2}
  - \sum_l i \lambda_{l\tau}(\boldsymbol{S}_{l\tau}^2-1) \Bigr\rbrace
\end{eqnarray}
In the saddle-point approximation, the Lagrange multiplier field $\lambda$ becomes ``homogeneous'', with only two independent values $\lambda_3$ and $\lambda_6$ corresponding to the 3 and 6-coordinated sites. This is the consequence of the \z2 gauge-invariance, namely the arbitrariness of the choice of $\epsilon_{lm}$, subject to the condition ~(\ref{Frust}). If $\epsilon_{lm}$ are chosen to be negative on the thick bonds in the Fig.~\ref{KDGrid}(b), then the action can be diagonalized on the six by six matrices in the frequency-momentum space, and the spin degrees of freedom can be easily integrated out:
\begin{equation}\label{ONLagAction}
\widetilde{S}_{\lambda} = N \Bigl[ \frac{1}{2} \sum_{\omega} \sum_{q_x q_y}
\log\det(H_{q_x q_y \omega}+i\Lambda) + \mathcal{N}\textrm{tr} (i\Lambda) \Bigr]
\end{equation}
Here we used the elementary cell in Fig.~\ref{FrustDiceCell}, $\mathcal{N}$ is the number of space-time sites, $\Lambda = \textrm{diag} (\lambda_6, \lambda_6, \lambda_3, \lambda_3, \lambda_3, \lambda_3) $, and the matrix $H_{q_x q_y \omega}$ is given in the Table \ref{HMatrix}.
\begin{table*}
\begin{equation}
H_{q_x q_y \omega} = \left(
\begin{array}{cccccc}
-\frac{K_6}{2}\omega^2 & 0 & \frac{h}{2}(1+e^{-iq_y}) &
  -\frac{h}{2}e^{-iq_y} & \frac{h}{2}e^{-iq_x} & \frac{h}{2}e^{-iq_x}(1-e^{-iq_y}) \\
0 & -\frac{K_6}{2}\omega^2 & \frac{h}{2} &
   \frac{h}{2}(1+e^{-iq_y}) & \frac{h}{2}(-1+e^{-iq_y}) & -\frac{h}{2}e^{-iq_y} \\
\frac{h}{2}(1+e^{iq_y}) & \frac{h}{2} & -\frac{K_3}{2}\omega^2 & 0 & 0 & 0 \\
-\frac{h}{2}e^{iq_y} & \frac{h}{2}(1+e^{iq_y}) & 0 & -\frac{K_3}{2}\omega^2 & 0 & 0 \\
\frac{h}{2}e^{iq_x} & \frac{h}{2}(-1+e^{iq_y}) & 0 & 0 & -\frac{K_3}{2}\omega^2 & 0 \\
\frac{h}{2}e^{iq_x}(1-e^{iq_y}) & -\frac{h}{2}e^{iq_y} & 0 &
  0 & 0 & -\frac{K_3}{2}\omega^2
\end{array}
\right)
\end{equation}
\caption{\label{HMatrix}The matrix describing the spin-spin couplings in  ~(\ref{ONAction})}
\end{table*}
The matrix under the determinant in ~(\ref{ONLagAction}) has three two-fold degenerate eigenvalues that have no dependence on the momentum. If we relabel $i\lambda_3=-m_3^2$ and $i\lambda_6=-m_6^2$, they are:
\begin{eqnarray}
v_3 & = & \frac{K_3}{2}\omega^2 + m_3^2 \nonumber \\
v_{1,2} & = &\frac{1}{2} \Bigg[ \frac{K_3+K_6}{2}\omega^2 + m_3^2 + m_6^2 \pm
  \nonumber \\
  & & \sqrt{6h^2 + \Bigl( \frac{K_3-K_6}{2}\omega^2 + m_3^2-m_6^2 \Bigr)^2} \Bigg] \end{eqnarray}
The action then becomes:
\begin{equation}\label{ONLagAction2}
\widetilde{S}_{\lambda} = N \mathcal{N} \left[ 
 \int \frac{\dd\omega}{2\pi} \log(v_1 v_2 v_3) - (2m_6^2+4m_3^2) \right]
\end{equation}
In order for this action to correspond to a physical hermitian Hamiltonian, both $m_3^2$ and $m_6^2$ must be real numbers. In addition, the path-integral converges only when all eigenvalues $v_i$ are real and positive for all frequencies. These requirements are satisfied if:
\begin{equation}\label{Conv}
m_3^2 > 0 \quad , \quad m_6^2 > 0 \quad , \quad
  4 m_3^2 m_6^2 > 6h^2
\end{equation}
The failure to satisfy these conditions by the saddle-point values of $m_3$ and $m_6$ would signify an instability.

Now we vary $m_3^2$ and $m_6^2$ to find the minimal action, integrate out the frequency, and obtain the following saddle-point equations:
\begin{eqnarray}\label{SPE}
\frac{1}{\sqrt{2 K_3 m_3^2}} 
  + K_3K_6 \frac{(K_3+K_6) + \frac{2(m_3^2+m_6^2)}{\sqrt{\alpha_1 \alpha_2}}}
    {\sqrt{\alpha_1}+\sqrt{\alpha_2}} & = & 6
  \nonumber \\
\frac{1}{\sqrt{2 K_3 m_3^2}} 
  - K_3K_6 \frac{(K_3-K_6) + \frac{2(m_3^2-m_6^2)}{\sqrt{\alpha_1 \alpha_2}}}
    {\sqrt{\alpha_1}+\sqrt{\alpha_2}} & = & 2
\end{eqnarray}
where:
\begin{equation}\label{Alphas}
\alpha_{1,2} = \Bigg(\frac{m_3^2}{K_3}+\frac{m_6^2}{K_6}\Bigg) \pm
  \sqrt{\bigg(\frac{m_3^2}{K_3}-\frac{m_6^2}{K_6}\bigg)^2 + \frac{6h^2}{K_3 K_6}}
\end{equation}
The conditions ~(\ref{Conv}) directly translate to $\alpha_{1,2} > 0$. The saddle-point equations are too complicated to be solved exactly. If the solution that meets the conditions ~(\ref{Conv}) exists for every finite non-zero $K_3$, $K_6$, and $h$, then the system is always in the paramagnetic phase. In order to prove that such solutions always exist, we can solve the inverse problem: assume that $m_3^2$ and $m_6^2$ are known, real and positive, fix arbitrary values for $K_3$ and $K_6$, and find the value of $h$ that satisfies the saddle-point equations.

A couple of straight-forward algebraic manipulations reduce the saddle-point equations ~(\ref{SPE}) to:
\begin{eqnarray}
\frac{K_6\alpha+2m_6^2}{K_3\alpha+2m_3^2} & = & 2 - \frac{1}{2\sqrt{2K_3m_3^2}}
  \nonumber \\
\Bigl(\alpha+\frac{2m_3^2}{K_3}\Bigr)^2 & = & 8K_6^2 \alpha^2 
  \Bigl(\frac{m_3^2}{K_3} + \frac{m_6^2}{K_6} + \alpha\Bigr)
\end{eqnarray}
and $\alpha = \sqrt{\alpha_1 \alpha_2}$. It follows that:
\begin{equation}\label{M3}
\frac{m_3^2}{K_3} = \frac{x^2}{2}
\end{equation}
\begin{equation}\label{M6}
\frac{m_6^2}{K_6} = \frac{(\alpha + x^2)^2}{8K_6^2 \alpha^2} -
  \frac{x^2}{2} - \alpha
\end{equation}
\begin{equation}\label{XEqu}
x^3 + \alpha \bigl[1-4K_6(2K_3+K_6)\alpha\bigr] x + 2K_6\alpha^2 = 0
\end{equation}
Treating ~(\ref{XEqu}) as a quadratic equation for $\alpha$, we find the real and positive solution:
\begin{eqnarray}\label{AlphaX}
\alpha & = & x \frac{1+\sqrt{1+8K_6[2(2K_3+K_6)x-1]x}}{4K-6[2(2K_3+K_6)x-1]}
  \nonumber \\ & & \textrm{for} \quad x>\frac{1}{2(2K_3+K_6)}
\end{eqnarray}
We can now study the dependence of $\beta=6h^2/K_3K_6$ on $x$. From ~(\ref{Alphas}) we obtain:
\begin{equation}
\beta = 4\frac{m_3^2}{K_3}\frac{m_6^2}{K_6}-\alpha^2
\end{equation}
and using ~(\ref{M3}), ~(\ref{M6}), and ~(\ref{AlphaX}) we express $\beta$ as a function of $x$. One can easily check that $\beta=0$ for $m_3^2=1/8K_3$, $m_6^2=1/8K_6$, and $\beta \to \infty$ for $m_3^2=h\sqrt{3}/2$, $m_6^2=h\sqrt{3}$. In between these limiting cases, $\beta(x)$ is a continuous, monotonously growing function in the region $x>1/K_3$, for all values of $K_3$ and $K_6$. Due to a very complicated dependence of $\beta$ on $x$, we verify this by numerical plotting for a variety of $K_3$ and $K_6$ values, including all important limits. The fact that $\beta(x)$ is a monotonous function means that for every value of $h$ there is a unique solution for $x$, i.~e. ~for the saddle-point equations. The ``squared masses'' $m_3^2$ and $m_6^2$ are real and positive for all $h$, $K_3$, and $K_6$, and vary continuously with the coupling constants. Therefore, this theory does not support a phase transition, and always remains in the paramagnetic phase.

\newpage


\end{document}